\documentclass[preprint,3p,times]{elsarticle}

\usepackage{amsmath,amssymb,amsthm,amsfonts}
\usepackage{mathabx}
\usepackage{mathrsfs}
\usepackage{bm}
\usepackage{graphicx}
\usepackage{float}
\usepackage{subfig}
\usepackage{color,xcolor}
\usepackage{tabularx}
\biboptions{numbers,sort&compress}
\usepackage{epsfig,psfrag}
\usepackage{tikz}
\usepackage{todonotes}
\usepackage{comment}
\usepackage{graphicx,psfrag}
\usepackage{hyperref}
\usepackage{enumitem}
\usetikzlibrary{shapes}

\setcitestyle{authoryear,round}

\usepackage[commandnameprefix=always,commentmarkup=footnote]{changes}
\makeatletter
\@namedef{Changes@AuthorColor}{blue}
\colorlet{Changes@Color}{blue}
\makeatother

\makeatletter
\hypersetup{colorlinks=true}
\AtBeginDocument{\@ifpackageloaded{hyperref}
	{\def\@linkcolor{magenta}
		\def\@anchorcolor{black}
		\def\@citecolor{teal}
		\def\@filecolor{cyan}
		\def\@urlcolor{magenta}
		\def\@menucolor{red}
		\def\@pagecolor{cyan}
		\begingroup
		\@makeother\`%
		\@makeother\=%
		\edef\x{%
			\edef\noexpand\x{%
				\endgroup
				\noexpand\toks@{%
					\catcode 96=\noexpand\the\catcode`\noexpand\`\relax
					\catcode 61=\noexpand\the\catcode`\noexpand\=\relax
				}%
			}%
			\noexpand\x
		}%
		\x
		\@makeother\`
		\@makeother\=
	}{}}
\makeatother

\makeatletter \oddsidemargin 0.0cm \evensidemargin 0.0cm
\newcommand{\be}{\begin{equation}}
\newcommand{\en}{\end{equation}}

\def\bm#1{\mbox{\boldmath{$#1$}}}

\numberwithin{equation}{section}
\theoremstyle{plain}

\newtheorem{theorem*}{Theorem}

\theoremstyle{definition}

\newcommand{\rd}{\mathrm{d}}
\newcommand{\ri}{\mathrm{i}}

\DeclareMathOperator{\tr}{tr}
\DeclareMathOperator{\Div}{Div}
\DeclareMathOperator{\ddiv}{div}

\DeclareMathOperator{\cof}{cof}

\journal{Journal of the Mechanics and Physics of Solids}

\begin{document}

\newcommand{\beq}{\begin{equation}}
\newcommand{\beql}[1]{\begin{equation}\label{#1}}
\newcommand{\eeq}{\end{equation}}
\newcommand{\bea}{\begin{eqnarray}}
\newcommand{\beal}[1]{\begin{eqnarray}\label{#1}}
\newcommand{\eea}{\end{eqnarray}}
\newcommand{\bean}{\begin{eqnarray*}}
\newcommand{\eean}{\end{eqnarray*}}

\newcommand{\del}{\delta}

\newcommand{\mbf}[1]{\mathbf{#1}}
\newcommand{\sbf}[1]{\boldsymbol{#1}}

\newcommand{\intL}{\int_{\cal L}}

\newcommand{\beps}{\sbf{\varepsilon}}
\newcommand{\bsig}{\sbf{\sigma}}
\newcommand{\blam}{\sbf{\lambda}}
\newcommand{\bmu}{\sbf{\mu}}
\newcommand{\ba}{\mbf{a}}
\newcommand{\bi}{\mbf{i}}
\newcommand{\bB}{\mbf{B}}
\newcommand{\bS}{\mbf{S}}
\newcommand{\bV}{\mbf{V}}
\newcommand{\bA}{\mbf{A}}
\newcommand{\bE}{\mbf{E}}
\newcommand{\bU}{\mbf{U}}
\newcommand{\bR}{\mbf{R}}
\newcommand{\bX}{\mbf{X}}
\newcommand{\bK}{\mbf{K}}
\newcommand{\bd}{\mbf{d}}
\newcommand{\bs}{\mbf{s}}
\newcommand{\bff}{\mbf{f}}
\newcommand{\bF}{\mbf{F}}
\newcommand{\bg}{\mbf{g}}
\newcommand{\bG}{\mbf{G}}
\newcommand{\bL}{\mbf{L}}
\newcommand{\bM}{\mbf{M}}
\newcommand{\bH}{\mbf{H}}
\newcommand{\bkron}{\mbf{1}}
\newcommand{\sigm}{\sigma_{\rm m}}
\newcommand{\epsm}{\varepsilon_{\rm m}}
\newcommand{\bI}{\mbf{I}}
\newcommand{\bIv}{\mbf{I}_{\rm K}}
\newcommand{\bId}{\mbf{I}_{\rm D}}
\newcommand{\bIs}{\mbf{I}_{\rm S}}
\newcommand{\bCe}{\mbf{C}_{\rm e}}
\newcommand{\bDe}{\mbf{D}_{\rm e}}
\newcommand{\bC}{\mbf{C}}
\newcommand{\bD}{\mbf{D}}
\newcommand{\bu}{\mbf{u}}
\newcommand{\bv}{\mbf{v}}
\newcommand{\bw}{\mbf{w}}
\newcommand{\bz}{\mbf{z}}
\newcommand{\bT}{\mbf{T}}
\newcommand{\bt}{\mbf{t}}
\newcommand{\bn}{\mbf{n}}
\newcommand{\bN}{\mbf{N}}
\newcommand{\bx}{\mbf{x}}
\newcommand{\bq}{\mbf{q}}
\newcommand{\bh}{\mbf{h}}

\newcommand{\cE}{{\cal E}}
\newcommand{\cH}{{\cal H}}
\newcommand{\cA}{{\cal A}}
\newcommand{\cD}{{\cal D}}

\newcommand{\G}{{\rm I\!G}}

\newcommand{\bnab}{\sbf{\nabla}}
\newcommand{\bnabs}{\sbf{\nabla}_{\rm s}}

\newcommand{\half}{\mbox{$\frac 1 2$}}
\newcommand{\dV}{\,\mbox{d}V}
\newcommand{\dx}{\,\mbox{d}x}
\newcommand{\dxi}{\,\mbox{d}\xi}
\newcommand{\deta}{\,{\rm d}\eta}
\newcommand{\dS}{\,\mbox{d}S}
\newcommand{\ds}{\,\mbox{d}s}
\newcommand{\dA}{\,\mbox{d}A}
\newcommand{\dt}{\,\mbox{d}t}
\newcommand{\pard}[2]{\frac{\partial #1}{\partial #2}}
\newcommand{\parder}[2]{\frac{\partial #1}{\partial #2}}

\begin{frontmatter}

\title{{\bf Incremental equations in curvature-dependent surface elasticity}}

\author[mymainaddress]{Xiang Yu}

\author[mysecondaryaddress]{Michal \v{S}mejkal}

\author[mysecondaryaddress,myteritaryaddress]{Martin Hor\'{a}k}

\address[mymainaddress]{Department of Mathematics, School of Computer Science and Technology, Dongguan University of Technology, Dongguan, 523808, China}

\address[mysecondaryaddress]{Czech Technical University in Prague, Faculty of Civil Engineering, Department of Mechanics, Th\'{a}kurova 2077/7, 166 29 Prague 6, Czechia}

\address[myteritaryaddress]{Czech Academy of Sciences, Institute of Information Theory and Automation, Pod vod\'{a}renskou v\v{e}\v{z}\'{i} 4, 182 00 Prague 8, Czechia}

\begin{abstract}
We develop a general incremental framework for hyperelastic solids whose surfaces exhibit both stretch-dependent and curvature-dependent elastic behavior. Building upon a variational formulation of curvature-dependent surface elasticity, we derive compact governing equations expressed in a coordinate-free Lagrangian setting that remain valid for arbitrary geometries. Linearization about an arbitrarily large finite deformation yields incremental bulk and surface balance laws that closely resemble the classical small-on-large theory, but are now extended to include surface-curvature-induced stresses. The applicability of the general theory is demonstrated by analyzing the onset of periodic beading in a soft cylindrical substrate coated with a surface layer exhibiting stretching- or curvature-dependent behavior, illustrating how surface stretching and bending effects influence instability thresholds for both compressible and incompressible bulk. This unified formulation thus provides a foundation for studying stability phenomena in elasto-capillary systems where surface curvature plays a critical mechanical role.
\end{abstract}

\begin{keyword}
Surface elasticity\sep Curvature-dependence\sep Incremental theory \sep Periodic beading
\end{keyword}
\end{frontmatter}

\section{Introduction}

Surface elasticity plays an important role in understanding the mechanical behavior of materials at small scales, such as micro/nano materials, thin films, and soft solids, where surface effects dominate due to a high surface-to-volume ratio \citep{mogilevskaya2010effects,style2013surface,long2016effects,chen2022recent,roudbari2022review}. The theoretical framework for surface elasticity originates from the pioneering work of Gurtin and Murdoch \citep{gurtin1975continuum,gurtin1978surface}, who developed a phenomenological surface elasticity theory that treats the surface of a solid as a zero-thickness membrane endowed with its own free energy, distinct from that of the bulk. The model accounts for surface stress and strain through the surface energy, thereby providing an intrinsic explanation for surface effects \citep{lu2021modified,chen2022gurtin}. With the rapid advancements of nanoscience and nanotechnology, the Gurtin-Murdoch theory has found broad applications in analyzing the elastic properties of nanomaterials and nanostructures, including nanowires \citep{chen2006size,wang2009surface}, nanofilms \citep{cammarata1994surface}, and composites with nanoscale heterogeneities \citep{sharma2003effect,duan2005size,duan2005eshelby,mogilevskaya2021use}.

Despite its success in various applications, the Gurtin–Murdoch model accounts only for the contribution of surface strain to the surface energy and neglects curvature effects that become important in problems such as nanostructures or biological membranes; see \citep{mogilevskaya2021fiber,zhang2024novel}. To address this limitation, Steigmann and Ogden \citep{steigmann1999elastic} proposed a generalized theory of surface elasticity that incorporates both curvature dependence and finite-strain effects, thereby extending the classical Gurtin-Murdoch model. The formulation treats the surface as a two-dimensional shell with inherent bending resistance; thus, the surface energy depends on both the surface strain and surface curvature. Linearized versions have been widely explored for analytical tractability, particularly in plane-strain and cylindrical geometries \citep{chhapadia2011curvature,dai2023discussion}. For instance, in nanoindentation, the model predicts size-dependent hardness by accounting for curvature-induced stresses at the contact interface \citep{li2019nanoindentation}. Similarly, in fracture mechanics, it has been applied to model curvilinear cracks and multiple interacting fractures, yielding bounded stresses at crack tips in contrast to the singularities predicted by classical theories \citep{zemlyanova2018frictionless,zemlyanova2012modeling}.

The present work is motivated by the absence of a general incremental formulation for curvature-dependent surface elasticity, which hinders the stability analysis of elasto-capillary problems involving curvature effects. This study can be viewed as a sequel to the prior study \cite{yu2025incremental}, which established the incremental equations for strain-dependent surface elasticity without the effect of surface curvature. Although some progress has been made in deriving incremental equations for curvature-dependent surface elasticity \citep{ogden1997effect,dryburgh1999bifurcation,ogden2002plane}, these derivations are restricted to the plane-strain case. As a result, past investigations have often relied on ingenious but {\it ad hoc} methods tailored to specific geometries and surface energies. For example,  \cite{taffetani2024curvature} recently analyzed periodic beading in a soft elastic cylinder with a thin coating described by the Helfrich free energy; see \cite{helfrich1973elastic}. Their procedure involved deriving the governing equations directly from a variational principle, enforcing incompressibility through a stream-function formulation, and then linearizing the system for bifurcation analysis. While effective, this approach requires repeating the variation and linearization process for each problem, which is cumbersome and problem-specific. In this paper, we address this gap by deriving the incremental equations for curvature-dependent surface elasticity in the mathematical framework described by \cite{biot1939xliii} and \cite{ogden1984non}.  Our main contribution is the derivation of a general form of the corresponding {\it incremental} equations for curvature-dependent surface elasticity in a compact, coordinate-free Lagrangian formulation, thereby facilitating their applicability to arbitrary constitutive models and geometries. It should be emphasized that the coordinate-free formulation for curvature-dependent surfaces was developed by \cite{vsilhavy2013direct} and \cite{tomassetti2024coordinate}.

The rest of the paper is divided into six sections as follows. Section \ref{sec:2} reviews the kinematics of deformable surfaces. Section \ref{sec:VarFormulation} presents a coordinate-free derivation of governing equations in curvature-dependent surface elasticity. The incremental equations are developed in Section \ref{sec:incremental} in a coordinate-free manner. Furthermore, expressions of the incremental equations in arbitrary curvilinear coordinates are presented in section \ref{sec:coords}.
Section \ref{sec:app} illustrates the theory by analyzing periodic beading of a soft-coated cylinder. Finally, concluding remarks are provided in Section \ref{sec:con}.

\section{Kinematics}\label{sec:2}
In this section, we introduce the necessary definitions and preliminaries for describing deformations of elastic surfaces. The presentation follows \cite{HorakSurface25} and \cite{yu2025incremental}; further details can be found in \cite{steigmann1999elastic}, \cite{vsilhavy2013direct}, \cite{steinmann2008boundary}, and \cite{gao2014curvature}.

We consider a bounded reference domain $\Omega \subset \mathbb{R}^3$ with sufficiently regular boundary $\Gamma = \partial \Omega$. A portion of the boundary, denoted by $S \subset \Gamma$, is assumed to carry surface energy, and its boundary $\partial S$ is taken to be smooth. Let $\bm{X}$ denote the position of a representative material point in the reference configuration. The material point $\bm{X}$ is mapped to new position $\bm{y}$ in the deformed configuration  $\bm{\phi}(\Omega)$ under the deformation  $\bm{\phi}:{\Omega} \to  \mathbb{R}^3$,
\begin{align}
\bm{y}=\bm{\phi}(\bm{X}),\quad \bm{X}\in\Omega.
\end{align}
The deformation $\bm{\phi}$ is assumed to be sufficiently smooth, injective, and orientation-preserving. Moreover, $\bm{\phi}$ admits a continuous extension to the closure $\overline{\Omega}$, and we shall not distinguish between $\bm{\phi}$ and this extension.    The associated surface deformation  ${\bm{\phi}}_s:\Gamma \to \bm{\phi}(\Gamma)$ is given by the restriction of $\bm{\phi}$ to the boundary, i.e.,
\begin{align}
{\bm{\phi}}_s(\bm{X}) = {\bm{\phi}}(\bm{X}), \quad \bm{X} \in \Gamma.
\end{align}

\subsection{Bulk deformation}  
The  deformation gradient of the bulk is defined by
\begin{align}
\bm{F} = \nabla {\bm{\phi}}=\frac{\partial\bm{y}}{\partial\bm{X}}, 
\end{align}
where $\nabla$ denotes the gradient with respect to the reference coordinates.  It maps infinitesimal vectors from the reference to the deformed configuration. The Jacobian determinant, denoted as $J = \det(\bm{F})$, transforms the reference volume element $\mathrm{d}V$ into current volume element $\mathrm{d}v$, such that $\mathrm{d}v = J \,\mathrm{d}V$. Orientation preservation requires $J > 0$.  

Having established the mappings of line and volume elements, we now turn to surface elements. Nanson’s formula relates the oriented area element $\bm{N}\,\mathrm{d}A$ in the reference configuration to its deformed counterpart $\bm{n}\,\mathrm{d}a$ by 
\begin{align}\label{eq:cof}
\bm{n} \,\rd a = \cof(\bm{F})\bm{N}\,\rd A, 
\end{align}
where $\bm{N}$ and $\bm{n}$ denote the unit outward normals in the reference and deformed configurations, respectively. The cofactor tensor of $\bm{F}$ is defined  via the universal property 
\begin{align}\label{eq:cofF}
\bm{F}\bm{u}\times\bm{F}\bm{v}=\cof(\bm{F})(\bm{u}\times\bm{v}),\qquad \forall\ \bm{u},\bm{v}\in\mathbb{R}^3,
\end{align}
and is given explicitly by $\cof(\bm{F})= J \bm{F}^{-T}$ when $\bm{F}$ is invertiable, where $\times $ denotes the cross product.

\subsection{Surface deformation}\label{surfDef}
To define deformation measures relevant for the boundary surface, we first establish how the tensor fields defined on the boundary surface are differentiated, following the general definition of the differential of a function defined on a manifold.

Let  $\bm{D}_s: \Gamma \to V$ be a tensor field from the boundary surface $\Gamma$ to a finite-dimensional vector space $V$.  For our purposes, $V = \mathbb{R}$, $\mathbb{R}^3$, or $\mathbb{R}^3 \otimes \mathbb{R}^3$, so that $\bm{D}_s$ may represent a scalar, vector, or second-order tensor field, respectively.  The {\it surface gradient} of $\bm{D}_s$ at $\bm{X}$ is defined as a linear transformation $\nabla_s\bm{D}_s(\bm{X}): \mathbb{R}^3 \to V$ such that
\begin{equation}
\lim_{\bm{Y}\to \bm{X},{\bm{Y}\in \Gamma} }
\frac{| \bm{D}_s(\bm{Y}) - \bm{D}_s(\bm{X}) - \nabla_s \bm{D}_s(\bm{X})(\bm{X}-\bm{Y}) |}{|\bm{X}-\bm{Y}|} = 0,
\end{equation}
and 
\begin{equation}
\nabla_s \bm{D}_s(\bm{X}) \bm{I}_s = \nabla_s \bm{D}_s(\bm{X}),
\label{eq26}
\end{equation}
where $\bm{I}_s = \bm{I} - \bm{N}\otimes \bm{N}$ is the orthogonal projection onto the tangent space $T_{{X}}\Gamma$ of the surface $\Gamma$ at $\bm{X}$, with $\bm{I}$ the identity tensor in $\mathbb{R}^3$, and $|\cdot|$ denotes the Euclidean norm. The first condition generalizes the derivative of a scalar-valued function of one variable to that of a vector-valued function of multiple variables, while the second extends the domain of the derivative map from the tangent space $T_X \Gamma$ to $\mathbb{R}^3$ by setting $\nabla_s \bm{D}_s(\bm{X})\bm{N}=0$. This convention follows \cite{vsilhavy2013direct} and differs slightly from alternative formulations (e.g., \cite{gurtin1975continuum}), in which the surface gradient is restricted directly to the tangent space.  
In the present setting, the surface gradient acts as a linear map on $\mathbb{R}^3$, which unifies its domain at different points on the surface and offers the advantage of being manipulated easily. In particular, if $\bm{D}_s$ is the restriction of a smooth mapping ${\bm{D}}:\Omega \to V$  to the boundary surface $\Gamma$, the surface gradient can be expressed as the projection of the bulk gradient onto the tangent space,  
\begin{equation}
\nabla_s \bm{D}_s =(\nabla \bm{D})\bm{I}_s.
\end{equation}

Another important operator for tensor fields defined on the boundary surface is the \emph{surface divergence}. Let 
$\bm{R}_s:\Gamma \to V \otimes \mathbb{R}^3$ 
be a tensor field defined on the boundary surface, where typically $V=\mathbb{R}$ or $\mathbb{R}^3$. The surface divergence of $\bm{R}_s$ is a unique field $\Div_s(\bm{R}_s):\Gamma\to V$ satisfying
\begin{equation}
\Div_s (\bm R_s)\cdot\bm{v}=\tr\big(\nabla_s(\bm R_s^T \bm v)\big)
\qquad \forall\,\bm v\in V, \label{eq:divs_def_Silhavy}
\end{equation}
where $\bm{R}_s^T:\Gamma \to \mathbb{R}^3 \otimes V$ denotes the transpose of $\bm{R}_s$. For example, when $V=\mathbb{R}$ and $\bm{R}_s:\Gamma\to \mathbb{R}\otimes\mathbb{R}^3\cong\mathbb{R}^3$ is a vector field on $\Gamma$, the surface divergence takes the familiar form $\Div_s (\bm R_s)=\tr(\nabla_s\bm{R}_s)$. With this definition, one also has the product (Leibniz) rule: for any \(\bm a:\Gamma\to V\),
\begin{equation}
\Div_s(\bm R_s^T\bm a)=\Div_s (\bm R_s)\cdot \bm{a}+ \bm R_s \bm\cdot \nabla_s \bm a,
\label{eq:divs_product_Silhavy}
\end{equation}
where $\bm R_s \cdot \nabla_s \bm a :=\tr (\bm R_s^T\nabla_s \bm a)$.

\subsection{Surface deformation gradient}
Applying the above to the surface deformation map $\bm{\phi}_s$ (with $\bm{D}_s = \bm{\phi}_s$ and $V = \mathbb{R}^3$), the surface deformation gradient $\bm{F}_s:\mathbb{R}^3 \to \mathbb{R}^3$ is defined as
\begin{equation}
\bm{F}_s = \nabla_s \bm{\phi}_s. \nonumber
\end{equation}
It maps tangent vectors in the reference configuration to tangent vectors in the deformed configuration. Since $\bm{F}_s$ is rank-deficient, we introduce its {\it pseudo-inverse} $\bm{F}_s^{-1}:\mathbb{R}^3 \to \mathbb{R}^3$, which satisfies
\begin{equation}
\bm{F}_s^{-1}\bm{F}_s = \bm{I}_s, 
\qquad
\bm{F}_s \bm{F}_s^{-1} = \bm{i}_s,\qquad \bm{F}_s^{-1}\bm{i}_s=\bm{F}_s^{-1}
\label{eq39}
\end{equation}
where $\bm{i}_s = \bm{I} - \bm{n}\otimes \bm{n}$ defines the projection onto the tangent space of the deformed boundary surface $\bm{\phi}(\Gamma)$ with $\bm{n}$ being the current normal vector, and the last equality impose a condition similar to \eqref{eq26} which ensures that $\bm{F}_s^{-1}$ acts only on the tangent vectors of the deformed surface. By definition, the following identities are immediate:
\begin{equation}\label{eq:FIs}
\bm{F}_s \bm{I}_s = \bm{F}_s, \qquad
\bm{i}_s \bm{F}_s = \bm{F}_s, \qquad
\bm{F}_s^{-1} \bm{i}_s = \bm{F}_s^{-1}, \qquad
\bm{I}_s \bm{F}_s^{-1} = \bm{F}_s^{-1}.
\end{equation}

If $\bm{\phi}_s$ admits a smooth extension $\bm{\phi}$ to a neighborhood of the boundary, the surface deformation gradient and its pseudo-inverse can be expressed in terms of the bulk deformation gradient:
\begin{equation}
\bm{F}_s = \bm{F} \bm{I}_s, 
\qquad 
\bm{F}_s^{-1} = \bm{F}^{-1} \bm{i}_s.
\label{eq41}
\end{equation}
Finally, as a direct consequence of \eqref{eq:FIs}, we note that
\begin{equation}\label{eq:FsTn}
\bm{F}_s \bm{N} = \bm{0}, \qquad \bm{F}^T_s \bm{n} = \bm{0},
\end{equation}
i.e., the surface deformation gradient maps the referential normal $\bm{N}$ and, in transpose form, the current normal $\bm{n}$ to zero, reflecting its tangential nature. The second equality plays an important role in determining the various relations between the deformation of the surface and its unit normal.

The change in infinitesimal surface area can be described solely using the surface deformation gradient by introducing the concept of  \emph{surface Jacobian determinant}, defined as the ratio of the deformed area of an infinitesimal surface element to its corresponding original (reference) area. In view of \eqref{eq:cof}, this quantity can be expressed using the surface deformation gradient $\bm{F}_s$ as
\begin{equation}
{J}_s:= |\cof(\bm{F}_s) \bm{N}|.
\end{equation}
It is straightforward to check that $J_s$ satisfies an equality analogous to \eqref{eq:cofF} for vectors lying in the tangent space:
\begin{align}
|\bm{F}_s\bm{u}\times \bm{F}_s\bm{v}|=J_s|\bm{u}\times\bm{v}|,\qquad \forall\ \bm{u},\bm{v}\in T_X\Gamma,
\end{align}
which may serve as an alternative definition for the surface Jacobian determinant from the perspective of exterior algebra \citep{winitzki2009linear}.

In order to express some of the equations in the actual configuration, we apply the Piola transformation
\bea \label{eq:piola_t}
\Div_s (\bm R_s) =J_s \ddiv_s( J_s^{-1} \bm R_s \bm{F}_s^T),
\eea
where $\ddiv_s$ denotes surface divergence operator on the deformed surface $\bm{\phi}(\Gamma)$ and $\bm R_s:\Gamma \to V \otimes \mathbb{R}^3$ is a superficial tensor field on $\Gamma$, meaning it satisfies $\bm R_s = \bm R_s \bm{I}_s$.

\subsection{Surface second gradient and curvature}
Higher-order deformation measures require differentiation of surface gradients.  
For a smooth field $\bm{D}_s:\Gamma \to V$, the \emph{surface second gradient} is defined as
\begin{equation}
\nabla_s^2 \bm{D}_s = \nabla_s ( \nabla_s \bm{D}_s ), 
\end{equation}
which is viewed as a bilinear transformation from $\mathbb{R}^3 \times \mathbb{R}^3$ to $V$.  As a bilinear map, $\nabla^2\bm{D}_s$ is generally nonsymmetric; it is convenient to introduce a symmetric version by restriction to $T_X\Gamma\times T_X\Gamma$. In particular, for the surface deformation map $\bm{\phi}_s$, we define a symmetric bilinear map that accounts for the second gradient effect by
\begin{equation}
\bm{H}_s = \nabla_s^2 \bm{\phi}_s\circ (\bm{I}_s\times \bm{I}_s), 
\end{equation}
where $\bm{I}_s\times \bm{I}_s:\mathbb{R}^3\times\mathbb{R}^3\to T_X\Gamma\times T_X\Gamma$ denotes the restriction map satisfying
 $(\bm{I}_s\times\bm{I}_s)(\bm{u},\bm{v}):=(\bm{I}_s\bm{u},\bm{I}_s\bm{v})$ for all $\bm{u},\bm{v}\in\mathbb{R}^3$, and $\circ$ stands for function composition.

The geometry of the surface is characterized by its curvature.  The \emph{surface curvature tensor} is defined as
\begin{equation}
\bm{b}= -\nabla^{{y}}_s \bm{n},
\end{equation}
which acts as a symmetric endomorphism on the tangent space $T_{{y}}\Gamma$ and $\nabla^{{y}}_s$ denotes actual (spatial) surface gradient.  
The principal curvatures are the eigenvalues of the surface curvature tensor $\bm{b}$. 
The mean curvature $H$ and the Gaussian curvature $K$ are defined as $H=\frac{1}{2}\tr(\bm{b})$ and $K=\frac{1}{2}[(\tr \bm{b})^2-\tr(\bm{b}^2)]$. Thus, $H$ and $K$ correspond to the first and second invariants of the curvature tensor $\bm{b}$, respectively. It is also useful to introduce the \emph{relative curvature} tensor $\bm{\kappa}$, that is defined as
\begin{equation}
\bm{\kappa} = -\bm{n} \cdot \bm{H}_s.
\end{equation}
By applying the surface gradient to the second equation in \eqref{eq:FsTn}, we obtain
\begin{align}
\bm{n}\cdot \bm{H}_s-\bm{F}_s^T\bm{b}\bm{F}_s=0.
\end{align}
Thus, the relative curvature tensor $\bm{\kappa}$ is related to the surface curvature tensor $\bm{b}$ by
\begin{equation}
\bm{\kappa}=-\bm{F}_s^T\bm{b}\bm{F}_s. 
\end{equation}
Note that, by construction, the relative curvature acts only tangentially, namely $\bm{\kappa} = \bm{I}_s \bm{\kappa}  \bm{I}_s$.

\subsection{Curve deformation}
For the description of the boundary conditions on the boundary curve $\partial S$ it is also useful to introduce curve gradient and curve divergence. 

The curve gradient $\nabla_\parallel \bm{D}_c$ of a field $\bm{D}_c$ is defined analogously to its surface counterpart. For brevity, we assume the field $\bm{D}_c$ can be extended to the neighborhood of points $\bm{X} \in \partial S$. Consequently, we define the {\it curve gradient} via a projection of the surface gradient 
\bea
\nabla_\parallel \bm{D}_c = \left(\nabla_s \bm{D}_c \right)  \bm{I}_\parallel ,
\eea
where $\bm{I}_\parallel = \bm{I}_s - \bm{V}\otimes \bm{V}$ is the projection to the tangent space of the curve $\partial S$ and $\bm{V}$ denotes unit in-plane normal to the boundary curve $\partial S$. Definition of the curve divergence $\mathrm{Div}_\parallel$ is analogous to equation \eqref{eq:divs_def_Silhavy}, only replacing surface gradient with its curve counterpart. With this definition, the curve divergence of a tensor field $\bm{R}_c:\Gamma\to V\otimes\mathbb{R}^3$ is the field $\Div_\parallel(\bm{R}_c):\Gamma \to V$ given explicitly by
\begin{align}\label{eq:divc}
\Div_\parallel(\bm{R}_c)=\bm{R}_{c,s}\cdot \bm{T},
\end{align}
where $s$ is the arclength parameter of the boundary curve $\partial S$, $\bm{T}$  is its unit tangent vector and  “\({}_{,s}\)” denotes the differentiation with respect to variable $s$.

The curve counterpart to the surface Piola transformation \eqref{eq:piola_t} reads
\bea \label{eq:piola_curve}
\Div_\parallel(\bm R_c) = J_\parallel\ddiv_\parallel (J_\parallel^{-1} \bm R_c \bm{F}_\parallel^T ),
\eea
where $\bm R_c:\Gamma \to V\otimes\mathbb{R}^3 $  is a tensor field satisfying $\bm R_c = \bm R_c \bm{I}_\parallel$, $\ddiv_\parallel$ denotes divergence operator on the deformed curve $\bm{\phi}(\partial S)$ and $J_\parallel$ means the Jacobian of the curve transformation  given by $J_\parallel =|\bm{F}_s \bm{T}|= J_s| \bm{F}_s^{-T} \bm{V}|$.

\section{Variational formulation}\label{sec:VarFormulation}

\subsection{Generic formulation based on surface second gradient}

The total energy of the system is formulated within a general framework that accommodates both compressible and incompressible materials, with the latter treated via the penalty method. It consists of the bulk and surface energy, together with the work of prescribed external bulk forces ${\bm{f}}$ and surface forces ${\bm{q}}$, and is given by
\begin{equation}
E[\bm{y}]= \int_\Omega 
   W(\bm{F}) \,\rd V 
          + \int_S \varPhi(\bm{F}_s,\bm{H}_s) \,\rd \Gamma
          - \int_\Omega \bm{y}\cdot {\bm{f}} \,\rd  V
          - \int_{\partial\Omega} \bm{y}\cdot {\bm{q}} \,\rd \Gamma,
\end{equation}
where
\begin{equation}\label{eq:FHs}
\bm{F} = \nabla \bm{y}, 
\qquad 
\bm{F}_s = \nabla_s \bm{y}, 
\qquad 
\bm{H}_s = \nabla_s \bm{F}_s \circ (\bm{I}_s\times \bm{I}_s).
\end{equation}
Here, recall that the tensor $\bm{I}_s = \bm{I} - \bm{N}\otimes\bm{N}$ denotes the projector onto the tangent space of $S$, and $\nabla_s(\cdot) = \nabla(\cdot)\,\bm{I}_s$ is the surface gradient acting as a linear map on $\mathbb{R}^3$. 

The strain energy density function $W$ describes the stored bulk energy per unit reference volume, and the surface strain energy density $\varPhi$ characterizes the stored surface energy per unit reference surface area.  We recall that a surface strain energy function expressed directly as $\varPhi = \varPhi ({\bm{F}}_s, {\bm{H}}_s)$ is, in general, not objective, as it depends on the choice of observer.  
To ensure frame-indifference, the surface energy must be expressed in terms of objective quantities that remain invariant under superposed rigid-body motions.  
This requirement can be, e.g., satisfied by reformulating the energy as a function of the surface right Cauchy--Green tensor and its surface gradient,  
\begin{align}
\varPhi  = \tilde{\varPhi }({\bm{C}}_s, {\nabla}_s{\bm{C}}_s), 
\qquad \text{with} \quad {\bm{C}}_s = {\bm{F}}_s^{T}{\bm{F}}_s,
\end{align}
in direct analogy with the framework of generalized (or higher-gradient) elasticity.  However, in the present work, we do not pursue this reformulation explicitly and retain the dependence of the surface energy on ${\bm{F}}_s$ and ${\bm{H}}_s$ for notational convenience.

Define the first Piola–Kirchhoff stress $\bm{P}$, the referential surface stress $\bm{L}_s$, and the referential third-order surface couple-stress $\bm{A}_s$ by
\begin{equation}
\label{eq:constitutiveBulk}
\bm{P}   = \frac{\partial W}{\partial\bm{F}}, 
\qquad
\bm{L}_s = \frac{\partial \varPhi }{\partial \bm{F}_s}, 
\qquad
\bm{A}_s = \frac{\partial \varPhi }{\partial \bm{H}_s}.
\end{equation}
Using the definition \eqref{eq:FHs} for $\bm{F}_s$ and $\bm{H}_s$, the first variation of the energy functional can be written as
\begin{equation}\label{eq:deltaE-compact}
\delta E 
= \int_\Omega \bm{P} : \nabla\delta \bm{y}\,\mathrm{d}V 
 + \int_S (\bm{L}_s : \nabla_s \delta\bm{y} + \bm{A}_s \;\vdots\; \nabla^2_s\delta\bm{y})\, \mathrm{d}\Gamma
 - \int_\Omega \delta\bm{y}\cdot \bm{f} \,\mathrm{d}V
 - \int_{\partial\Omega} \delta\bm{y}\cdot \bm{q} \,\mathrm{d}\Gamma,
\end{equation}
where $\delta$ denotes variation, and $:$ and $\vdots$ represent the double and triple contraction of tensors, respectively.

Applying Green’s identity in the bulk yields
\begin{equation}
\int_\Omega \bm{P} :\nabla  \delta\bm{y} \,\mathrm{d}V
= - \int_\Omega \delta\bm{y}\cdot\Div (\bm{P}) \,\mathrm{d}V
  + \int_{\partial\Omega} \delta\bm{y}\cdot \bm{P}\bm{N} \,\mathrm{d}\Gamma.
\end{equation}
Since the surface stress $\bm{L}_s $ is superficial, the surface divergence theorem \citep{steinmann2008boundary} gives
\begin{equation}
\int_S \bm{L}_s  : \nabla_s\delta \bm{y} \,\mathrm{d}\Gamma
= - \int_S \delta \bm{y} \cdot \Div_s(\bm{L}_s ) \,\mathrm{d}\Gamma
  + \int_{\partial S} \delta \bm{y} \cdot \bm{L}_s  \bm{V} \,\mathrm{d}s,
\end{equation}
where $\bm{V}$ is the unit in-plane normal to the boundary curve $\partial S$.

For the second-gradient term, two successive surface integrations by parts yield
\begin{align}
\begin{split}
\int_S \bm{A}_s\;\vdots\; \nabla^2_s\delta\bm{y}\,\mathrm{d}\Gamma
&= - \int_S \nabla_s \delta\bm{y} : \Div_s(\bm{A}_s) \,\mathrm{d}\Gamma
   + \int_{\partial S} \nabla_s \delta\bm{y} : \bm{A}_s \bm{V} \,\mathrm{d}s \\
&= \int_S \delta \bm{y} \cdot \Div_s\left(\Div_s(\bm{A}_s)\bm{I}_s\right)\mathrm{d}\Gamma
  - \int_{\partial S} \delta \bm{y} \cdot \Div_s (\bm{A}_s)\bm{I}_s\bm{V} \,\mathrm{d}s
  + \int_{\partial S} \nabla_s \delta\bm{y} : \bm{A}_s \bm{V} \,\mathrm{d}s.
  \end{split}
\end{align}
Decomposing the last boundary term into parts normal and tangential to $\partial S$, and applying the line divergence theorem \citep{steinmann2008boundary} and noting that $\partial(\partial S)=\emptyset$, we obtain
\begin{equation}
\int_{\partial S} \nabla_s \delta\bm{y} : \bm{A}_s \bm{V} \,\mathrm{d}s
= \int_{\partial S}  \nabla_{s,\perp} \delta\bm{y}\cdot \bm{A}_{s,\perp}
  - \delta \bm{y} \cdot \Div_{\parallel}(\bm{A}_{s,\parallel})\,\mathrm{d}s.
\end{equation}
Here, the normal and tangential components are defined by 
\begin{align}
\nabla_{s,\perp} \delta\bm{y} = (\nabla_{s} \delta\bm{y} )\bm{V},\qquad \bm{A}_{s,\perp} = (\bm{A}_s \bm{V})\cdot \bm{V},\qquad \bm{A}_{s,\parallel} = (\bm{A}_s \bm{V})\cdot \bm{I}_\parallel,
\end{align}
where $\bm{I}_\parallel$ denotes  the projector onto the tangent of $\partial S$.

Substituting these identities into \eqref{eq:deltaE-compact}, we obtain
\begin{align}
\begin{split}
\delta E
&= - \int_\Omega \delta \bm{y}\cdot (\Div(\bm{P})+{\bm{f}})\,\mathrm{d}V
   + \int_{\partial\Omega} \delta \bm{y}\cdot(\bm{P}\bm{N} - {\bm{q}}) \,\mathrm{d}\Gamma  - \int_S \delta \bm{y} \cdot \Div_s\left(\bm{L}_s - \Div_s(\bm{A}_s)\bm{I}_s\right)\mathrm{d}\Gamma \\
&\quad + \int_{\partial S} \big[
     \delta \bm{y} \cdot \big((\bm{L}_s  - \Div_s(\bm{A}_s)\bm{I}_s)\bm{V} - \Div_{\parallel}(\bm{A}_{s,\parallel})\big)
   +  \nabla_{s,\perp}\delta\bm{y}\cdot \bm{A}_{s,\perp}
   \big]\,\mathrm{d}s.
   \end{split}
\end{align}
By the fundamental lemma of the calculus of variations, the equilibrium equations and boundary conditions take the coordinate-free form
\begin{align}
& \Div(\bm{P}) + {\bm{f}} = \bm{0} \hspace{11.1em} \text{in}\ \Omega, \label{eq:bulk-eq}\\
&\bm{P}\bm{N} = {\bm{q}}  \hspace{14.6em} \text{on } \partial\Omega\setminus S, \label{eq:bulk-bc}\\
&\Div_s \left(\bm{L}_s  - \Div_s(\bm{A}_s)\bm{I}_s\right) + \bm{q} - \bm{P}\bm{N} = \bm{0}  \qquad\text{on } S, \label{eq:surf-eq}\\
&\big(\bm{L}_s  - \Div_s (\bm{A}_s)\bm{I}_s\big)\bm{V} - \Div_{\parallel}(\bm{A}_{s,\parallel}) = \bm{0}\ \, \qquad \text{on}\ \partial S, \label{eq:edge-bc}\\
&\bm{A}_{s,\perp} = \bm{0}\hspace{14.4em} \text{on } \partial S. \label{eq:edge-bc-add}
\end{align}

\subsection{Curvature-based formulation}
In many surface elasticity formulations (e.g., \cite{steigmann1999elastic,gao2014curvature}), the surface energy depends on the curvature tensor rather than the full third-order tensor $\bm{H}_s$. Recall the definition of relative curvature
\begin{equation}
\bm{\kappa} = -\bm{n}(\bm{F}_s)\cdot \bm{H}_s,
\end{equation}
where we emphasize the dependence of the surface normal $\bm{n}$ on $\bm{F}_s$. Note that some authors, e.g., \cite{vsilhavy2013direct}, omit the minus sign; here we keep it for consistency with \cite{steigmann1999elastic}.

To obtain a reduced formulation in terms of curvature, we introduce the curvature-based surface energy density
\begin{align}
{\varPsi}(\bm{F}_s,\bm{\kappa}):= \varPhi (\bm{F}_s,\bm{H}_s).
\end{align}
Using the relation $\bm{H}_s = -{\partial \bm{\kappa}}/{\partial \bm{n}}$ together with the differentiation of the constraint $\bm{F}_s^T\bm{n}=\bm{0}$, we obtain
\begin{align}
\bm{L}_s 
&= \frac{\partial {\varPsi}}{\partial \bm{F}_s}
  + \frac{\partial {\varPsi}}{\partial \bm{\kappa}} : \frac{\partial \bm{\kappa}}{\partial \bm{F}_s}
= {\bm{P}}_s - (\bm{H}_s : {\bm{M}_s}) \frac{\partial \bm{n}}{\partial \bm{F}_s}
= {\bm{P}}_s + \bm{n}\otimes \bm{F}_s^{-1}\big(\bm{H}_s:{\bm{M}_s}\big),
\\
\bm{A}_s &= \frac{\partial {\varPsi}}{\partial \bm{H}_s}
= \frac{\partial {\varPsi}}{\partial \bm{\kappa}} \,\frac{\partial \bm{\kappa}}{\partial \bm{H}_s}
= -\,\bm{n}\otimes {\bm{M}_s},
\end{align}
where the surface first Piola–Kirchhoff  stress ${\bm{P}}_s$ and surface moment tensors $\bm{M}_s$ are defined by
\begin{equation}
{\bm{P}}_s= \frac{\partial {\varPsi}}{\partial \bm{F}_s},
\qquad
{\bm{M}}_s= \frac{\partial {\varPsi}}{\partial \bm{\kappa}}. \label{eq:const_PM}
\end{equation}
Consequently, the tensor whose surface divergence appears in \eqref{eq:surf-eq} can be rewritten as
\begin{equation}\label{eq:PminusDivA}
\begin{split}
\bm{L}_s  - \Div_s (\bm{A}_s)\bm{I}_s
&= {\bm{P}}_s + \bm{n}\otimes \bm{F}_s^{-1}(\bm{H}_s:{\bm{M}_s})
  + \Div_s(\bm{n}\otimes {\bm{M}_s})\,\bm{I}_s\\
&={\bm{P}}_s + \bm{F}_s^{-1}\Div_s (\bm{n}\otimes \bm{F}_s {\bm{M}_s}),
\end{split}
\end{equation}
where the last equality follows from the identity
\begin{equation}
\bm{n}\otimes \bm{F}_s^{-1}(\bm{H}_s:{\bm{M}_s})
 + \Div_s(\bm{n}\otimes {\bm{M}_s})\,\bm{I}_s
= \bm{F}_s^{-1}\Div_s (\bm{n}\otimes \bm{F}_s {\bm{M}_s}).
\end{equation}
Expanding the surface divergence, one can express the result using the curvature tensor $\bm{b}$ as
\begin{equation}\label{eq:DivnFsMs}
\bm{F}_s^{-1}\Div_s (\bm{n}\otimes \bm{F}_s {\bm{M}_s})
= -\bm{b}\bm{F}_s {\bm{M}_s} + \bm{n}\otimes \bm{F}_s^{-1}\Div_s(\bm{F}_s {\bm{M}_s}).
\end{equation}
In view of equations \eqref{eq:PminusDivA} and \eqref{eq:DivnFsMs}, the surface equilibrium equation \eqref{eq:surf-eq} can be written solely in terms of the second-order surface stress ${\bm{P}}_s$ and surface moment ${\bm{M}_s}$, thereby eliminating the need for third-order tensors, as
\begin{equation}\label{eq:SurfaceEqQ}
\Div_s ({\bm{P}}_s 
- \bm{b} \bm{F}_s {\bm{M}_s}
+ \bm{n}\otimes \bm{F}_s^{-1}\Div_s(\bm{F}_s {\bm{M}_s}))
+ \bm{q} - \bm{P}\bm{N} = \bm{0}
\qquad \text{on } S.
\end{equation}

Finally, by introducing an \emph{effective surface 
first Piola--Kirchhoff stress} $\bm{Q}_s$ by
\begin{align}\label{eq:Qsmy}
&\bm{Q}_s := {\bm{P}}_s 
- \bm{b} \bm{F}_s {\bm{M}_s}
+ \bm{n}\otimes \bm{F}_s^{-1}\Div_s(\bm{F}_s {\bm{M}_s})\,,
\end{align}
one can easily express the surface equilibrium equation \eqref{eq:surf-eq} and boundary conditions  \eqref{eq:edge-bc} and  \eqref{eq:edge-bc-add} as
\begin{align}
&\Div_s \left(\bm{Q}_s\right) + \bm{q} - \bm{P}\bm{N} = \bm{0}\hspace{3.05em} \text{on } S, \label{eq:surf-eq2} \\
&\bm{Q}_s\bm{V} + \Div_{\parallel}(\bm{n}\otimes \bm{M}_s\bm{V}) = \bm{0}\qquad \text{on}\ \partial S,\label{eq:edge-bc1}\\
& \bm{V}\cdot\bm{M}_s\bm{V} = 0 \hspace{8em} \text{on}\ \partial S. \label{eq:edge-bc2}
\end{align}
The projection operator $I_{\parallel}$ no longer appears in the boundary condition thanks to $\eqref{eq:edge-bc1}$ and $\bm{I}_s\bm{M}_s=\bm{M}_s$.
In this form, \eqref{eq:surf-eq2}--\eqref{eq:edge-bc2} mimics the classical equations 
from surface elasticity \citep{steigmann1999elastic,gao2014curvature}, except that curvature effects are embedded in $\bm{Q}_s$.

\section{Incremental theory}\label{sec:incremental}

In this section, we derive the incremental equations for curvature-dependent surface elasticity by linearizing the governing equations obtained in Section \ref{sec:VarFormulation}.

\subsection{Kinematics}
\begin{figure}[h!]
	\centering
	\includegraphics[width=0.7\linewidth]{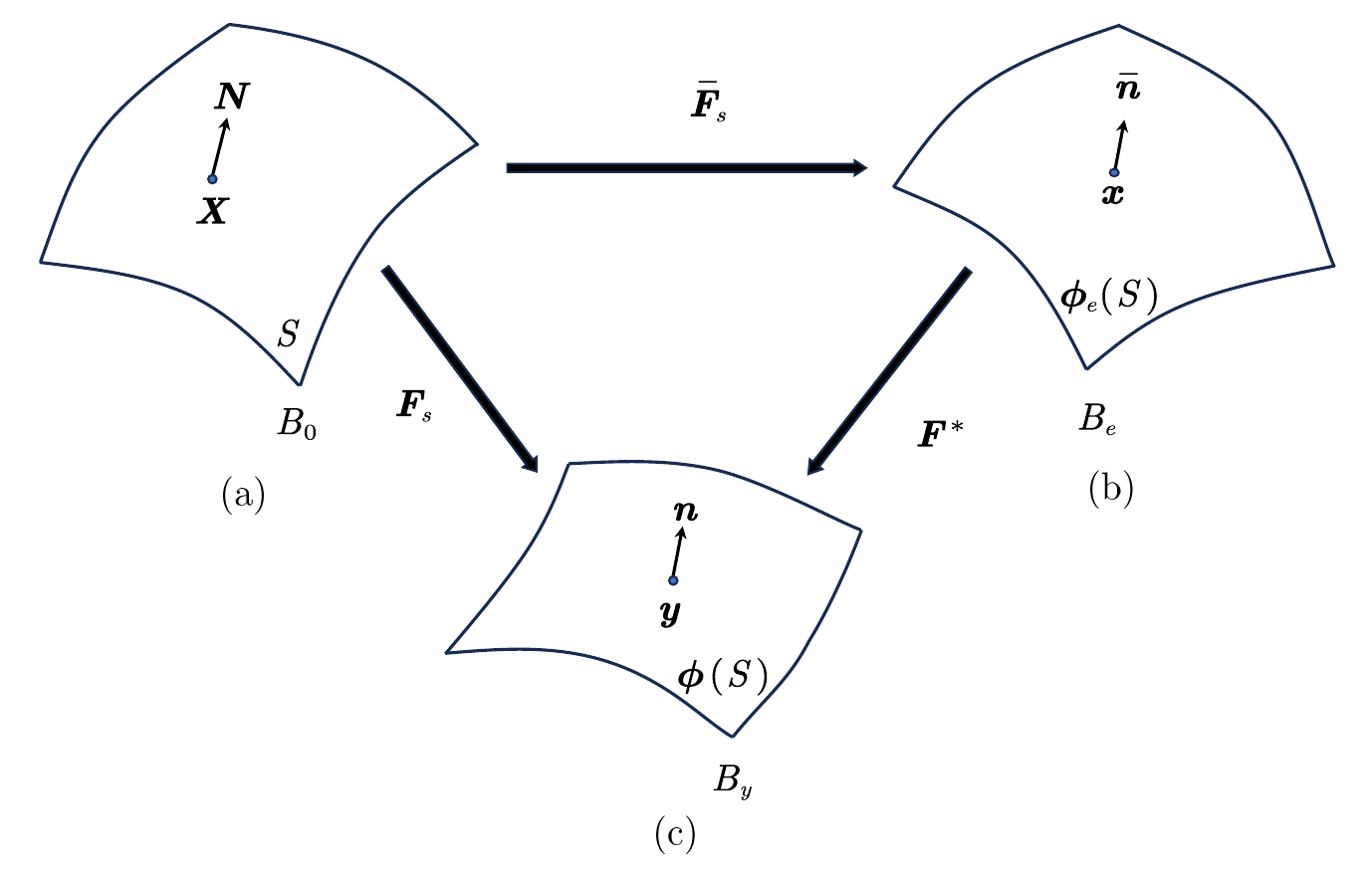}
	\caption{Three configurations of an elastic surface: (a) reference configuration, (b) finitely deformed configuration, and (c) current configuration. 
    }
	\label{fig:three-configurations}
\end{figure}

Following \cite{ogden1984non}, we consider three configurations of an elastic body arising from successive deformations. The reference configuration, in which the body is undeformed and occupies a region $\Omega \subset \mathbb{R}^3$, is denoted by $B_0$. Under external loading, the body undergoes a finite deformation into an equilibrium configuration $B_e$. A further deformation is then applied to $B_e$, producing the current configuration $B_y$, as shown in Fig. \ref{fig:three-configurations}. 

The position of a material point in these three configurations is written as
\begin{align}
  \bm{X} \in B_0, 
  \qquad \bm{x} = \bm{\phi}_e(\bm{X}) \in B_e, 
  \qquad {\bm{y}} = \bm{\phi}_y(\bm{x}) = \bm{\phi}(\bm{X}) \in B_y ,
\end{align}
where $\bm{\phi}_e : B_0 \to B_e$ and $\bm{\phi}_y : B_e \to B_y$ are the deformation maps and $\bm{\phi}:=\bm{\phi}_y\circ\bm{\phi}_e$ denotes their composition. The corresponding deformation gradients associated with the deformations $B_0 \to B_y$, $B_0 \to B_e$, and $B_e \to B_t$ are defined by
\begin{align}
  \bm{F} = \frac{\partial {\bm{y}}}{\partial \bm{X}}, 
  \qquad \bar{\bm{F}} = \frac{\partial \bm{x}}{\partial \bm{X}}, 
  \qquad \bm{F}^* = \frac{\partial {\bm{y}}}{\partial \bm{x}}. 
\end{align}
By the chain rule, these gradients are related through the multiplicative decomposition
\begin{align}\label{eq:FFbar}
  \bm{F} = \bm{F}^* \, \bar{\bm{F}} .
\end{align}
Restricting \eqref{eq:FFbar} to the surface yields
\begin{align}\label{eq:iFs}
  \bm{F}_s = \bm{F}^*_s \, \bar{\bm{F}}_s .
\end{align}

To analyze the stability of the equilibrium solution, we regard the superimposed deformation $\bm{\phi}_y$ as a small perturbation of the equilibrium configuration $B_e$. Specifically, the associated displacement field $\bm{y} -\bm{x}$ is written as $\epsilon \bm{u}$, where $\epsilon$ is a small parameter. 
Assuming $\epsilon$  arbitrarily small allows the equilibrium equations and boundary conditions to be linearized. 
A bifurcation is then identified at the configuration where the total deformation map  $\bm{\phi}$   loses injectivity, that is, when multiple solutions become admissible.
This procedure is commonly referred to as the \emph{incremental} or \emph{small-on-large} approach, see, e.g., \cite{ogden1984non}, \cite{bigoni2012nonlinear}, and \cite{goriely2017mathematics}.

\subsection{Equilibrium equations}

In this setting, we assume that the finitely deformed configuration $B_e$, obtained from $\bm{\phi}_e(\bm{X})$, is in equilibrium. That is, the associated first Piola--Kirchhoff stress $\bar{\bm{P}}$, and the effective first Piola-Kirchhoff surface stress $\bar{\bm{Q}}_s$ satisfy the equilibrium equations and boundary conditions. The complete set of governing equations describing this equilibrium 
state is given by:
\begin{align}
&\Div (\bar{\bm{P}}) + {\bm{f}} = \bm{0} \hspace{6.4em} \text{in } \Omega, \label{eq11bar} \\ 
&\bar{\bm{P}} {\bm{N}} = {\bm{q}} \hspace{9.9em} \text{on } \partial\Omega \setminus S, \label{eq12bar}\\ 
&\Div_s(\bar{\bm{Q}}_s) + {\bm{q}}-\bar{\bm{P}} {\bm{N}} = \bm{0} \hspace{3.1em} \text{on } S, \label{eq13bar}\\ 
&\bar{\bm{Q}}_s\bm{V} + \Div_{\parallel}(\bar{\bm{n}} \otimes \bar{\bm{M}}_s\bm{V}) = \bm{0}\qquad \text{on}\ \partial S, \label{eq14bar}\\
&\bm{V}\cdot\bar{\bm{M}}_s\bm{V} = {0}\hspace{8em}  \text{on }\ \partial S. \label{eq:16bar} 
\end{align}
On top of this equilibrium configuration, we superimpose an incremental deformation  parametrized by $\epsilon$, such that
\begin{align}
\bm{y}(\bm{x})=\bm{x}+\epsilon \bm{u}(\bm{x}).
\end{align}
This leads  to the perturbed deformation gradients
\begin{align}
 &\bm{F}^* = \bm{I} + \epsilon \bar{\nabla} \bm{u} = \bm{I} + \epsilon \bm{\eta},\\
 & \bm{F}_s^* = \bm{i}_s + \epsilon \bar{\nabla}_s \bm{u} = \bm{i}_s + \epsilon \bm{\eta}_s,  \label{eq:Fst_surf} 
\end{align}
where we introduce the incremental bulk and surface displacement gradient tensors $\bm{\eta} = \bar{\nabla} \bm{u}$ and $\bm{\eta}_s = \bar{\nabla}_s \bm{u}$, with $\bar{\nabla}$ and $\bar{\nabla}_s$ denoting the bulk and surface gradients with respect to the configuration $B_e$.

For the derivation of incremental equations, we expand the surface stress and moment tensors $\bm{P}$, $\bm{Q}_s$, $\bm{M}_s$ as well as the current unit normal $\bm{n}$ in terms of powers of the parameter~$\epsilon$ as
\begin{align}
&\bm{P}    = \bar{\bm{P}}    + \epsilon \bm{P}^{[1]}    + \mathcal{O}(\epsilon^2), \label{eq:exp1} \\
&\bm{Q}_s  = \bar{\bm{Q}}_s  + \epsilon \bm{Q}_s^{[1]}  + \mathcal{O}(\epsilon^2), \\
&\bm{M}_s  = \bar{\bm{M}}_s  + \epsilon \bm{M}_s^{[1]}  + \mathcal{O}(\epsilon^2), \label{eq:exp3} \\
&\bm{n}  = \bar{\bm{n}}  + \epsilon \bm{n}^{[1]}  + \mathcal{O}(\epsilon^2), 
\end{align}
 where barred quantities denote tensors related to the intermediate configuration, i.e., independent of the perturbation $\epsilon$, and superscript $[1]$ indicates the first-order perturbation (whose exact form is discussed in the next section).

 Note that one might worry that terms such as $\bm{P}$ and $\bar{\bm{P}}$ lie in different tangent spaces and hence should not be equated directly. This issue is resolved by noting that the formulas for bulk and surface deformation gradients  \eqref{eq:Fst_surf} should in fact be pre-multiplied by \emph{shifters}, which transport the quantities from the tangent space of $B_e$ to that of the perturbed configuration $B_y$ (see Chapter~1.3 of \cite{marsden1994mathematical}). Subsequently, shifters should also be included in the expansions \eqref{eq:exp1}--\eqref{eq:exp3}. However, since in our formulation we extended the domain and co-domain of $\bm{F}_s$ to $\mathbb{R}^3$, the tangent spaces of both deformed and undeformed configurations coincide with $\mathbb{R}^3$, and hence it is not necessary to include the shifters in the definitions.

 Subsequently, we substitute these relations into the equilibrium equations 
\eqref{eq:bulk-eq}, \eqref{eq:bulk-bc}, \eqref{eq:SurfaceEqQ}, and \eqref{eq:edge-bc2} and make use of the equilibrium of the intermediate configuration \eqref{eq11bar}--\eqref{eq:16bar}. Upon dividing the resulting equations by $\epsilon$ and taking the limit $\epsilon\to 0$, we obtain by  the incremental problem: 
\begin{align}
&\Div (\bm{P}^{[1]}) = \bm{0} \hspace{14.8em} \text{in } \Omega, \label{eq11} \\ 
&\bm{P}^{[1]} {\bm{N}} = \bm{0} \hspace{16.4em} \text{on } \partial\Omega \setminus S, \label{eq12}\\ 
&\Div_s (\bm{Q}_s^{[1]}) -\bm{P}^{[1]}{\bm{N}} = \bm{0} \hspace{10.8em} \text{on } S, \label{eq13}\\ 
&{\bm{Q}}^{[1]}_s\bm{V} + \Div_{\parallel}\big(\bar{\bm{n}}\otimes{\bm{M}}^{[1]}_s\bm{V}+{\bm{n}}^{[1]}\otimes\bar{\bm{M}}_s\bm{V}\big)= \bm{0}  \qquad \text{on } \partial S, \label{eq14}
\\ 
&\bm{V}\cdot{\bm{M}}^{[1]}_s\bm{V} = {0} \hspace{15em}\text{on } \partial S. \label{eq15}
\end{align}

\subsection{Linearization of kinematic and stress measures}

In order to formulate the incremental form of the bulk and surface equilibrium equations, 
we derive the linearizations of the relevant kinematic and stress quantities. 
We begin by recalling the general perturbation expansions for bulk and surface tensor-valued functions, 
and then proceed to the specific bulk and surface measures.

\subsubsection{General perturbation expansion}

For a bulk quantity, we consider a generic tensor-valued function
\begin{align}
\mathcal{Q} = \mathcal{Q}(\bm{F})
\end{align}
depending on the deformation gradient $\bm{F}$.  
The perturbed state is expressed as
\begin{align}
\bm{F}(\varepsilon) = \bar{\bm{F}} + \varepsilon \bm{\eta}\bar{\bm{F}} + \mathcal{O}(\varepsilon^2),
\end{align}
where $\bar{\bm{F}}$ denotes the deformation gradient in the finitely deformed configuration, and $\bm{\eta}$ is the incremental displacement gradient.  
Expanding $\mathcal{Q}$ in powers of $\varepsilon$ yields
\begin{align}
\mathcal{Q}(\bm{F}(\varepsilon))
= \mathcal{Q}(\bar{\bm{F}}) 
+ \varepsilon\,
\frac{\partial \mathcal{Q}}{\partial \bm{F}}\Big|_{\bar{\bm{F}}}:(\bm{\eta}\bar{\bm{F}})
+ \mathcal{O}(\varepsilon^2),
\end{align}
so that the first-order perturbation reads
\begin{align}\label{eq:Q1}
\mathcal{Q}^{[1]} 
= \frac{\partial \mathcal{Q}}{\partial \bm{F}}\Big|_{\bar{\bm{F}}}:(\bm{\eta}\bar{\bm{F}}).
\end{align}

For a surface quantity, the dependence extends to both the surface deformation gradient and its surface gradient.  Let
\begin{align}
\mathcal{Q}_s = \mathcal{Q}_s(\bm{F}_s, \bm{H}_s),
\end{align}
where $\bm{F}_s$ is the surface deformation gradient and $\bm{H}_s = \nabla_s \bm{F}_s \circ(\bm{I}_s\times \bm{I}_s)$ its surface gradient.  
The perturbed state is given by
\begin{align}
\bm{F}_s(\varepsilon) = \bar{\bm{F}}_s + \varepsilon \bm{\eta}_s \bar{\bm{F}}_s + \mathcal{O}(\varepsilon^2),
\qquad
\bm{H}_s(\varepsilon) = \bar{\bm{H}}_s + \varepsilon \nabla_s(\bm{\eta}_s\bar{\bm{F}}_s) + \mathcal{O}(\varepsilon^2).
\end{align}
Expanding $\mathcal{Q}_s$ gives
\begin{align}
\mathcal{Q}_s(\bm{F}_s(\varepsilon),\bm{H}_s(\varepsilon))
= \mathcal{Q}_s(\bar{\bm{F}}_s,\bar{\bm{H}}_s)
+ \varepsilon \Big(
\frac{\partial \mathcal{Q}_s}{\partial \bm{F}_s}\Big|_{(\bar{\bm{F}}_s,\bar{\bm{H}}_s)}:(\bm{\eta}_s\bar{\bm{F}}_s)
+ \frac{\partial \mathcal{Q}_s}{\partial \bm{H}_s}\Big|_{(\bar{\bm{F}}_s,\bar{\bm{H}}_s)}\;\vdots\;\nabla_s(\bm{\eta}_s\bar{\bm{F}}_s)
\Big) + \mathcal{O}(\varepsilon^2),
\end{align}
and consequently,
\begin{align}\label{eq:Qs1}
\mathcal{Q}_s^{[1]} 
= \frac{\partial \mathcal{Q}_s}{\partial \bm{F}_s}\Big|_{(\bar{\bm{F}}_s,\bar{\bm{H}}_s)}:(\bm{\eta}_s\bar{\bm{F}}_s)
+ \frac{\partial \mathcal{Q}_s}{\partial \bm{H}_s}\Big|_{(\bar{\bm{F}}_s,\bar{\bm{H}}_s)}\;\vdots\;\nabla_s(\bm{\eta}_s\bar{\bm{F}}_s).
\end{align}

The perturbation formulas \eqref{eq:Q1} and \eqref{eq:Qs1} provide the general structure for the linearization of all quantities depending on the deformation gradient (in the bulk) and on the surface deformation gradient and its curvature-related derivative (on the surface).

\subsubsection{Linearization of bulk quantities}

The bulk quantity that appears in the incremental formulation is the first Piola-Kirchhoff stress, which is written directly as
\begin{equation}
\bm P
= \frac{\partial W}{\partial \bm F}.
\label{eq:P_direct}
\end{equation}
Perturbing the first Piola-Kirchhoff stress \eqref{eq:P_direct} about \(\bar{\bm F}\) along \(\bm\eta \bar{\bm F}\) gives
\begin{equation}
\bm P^{[1]}
= \mathbb{A} : (\,\bm\eta \bar{\bm F}\,),
\label{eq:P1_master_direct}
\end{equation}
where $\mathbb A$ denotes the tangent stiffness
\begin{align}
\mathbb A
:= \frac{\partial^2 W}{\partial \bm F\,\partial \bm F}\Big|_{\bar{\bm F}}.
\end{align}

\subsubsection{ Linearization of surface quantities}
We now turn to the linearization of the relevant surface quantities, i.e., the first–order perturbations needed to formulate the incremental equations on the boundary surface.  Although the incremental surface elasticity relations were recently derived in \citet{yu2025incremental}, our setting is more complex due to the presence of the surface moment tensor and its explicit dependence on the surface relative curvature.

We first recall that the effective surface (Piola–type) stress is defined by
\begin{equation}\label{eq:Qs}
\bm{Q}_s={\bm{P}}_s-\bm{b}\bm{F}_s\bm{M}_s+\bm{n}\otimes \bm{F}_s^{-1}\Div_s(\bm{F}_s\bm{M}_s).
\end{equation}
A first–order perturbation of \eqref{eq:Qs} about the barred state yields
\begin{align}\label{eq:Qs1_master}
\begin{split}
\bm{Q}_s^{[1]}
&= {\bm{P}}_s^{[1]}-\bar{\bm{b}}\,\bar{\bm{F}}_s \bm{M}_s^{[1]}-(\bm{b}\bm{F}_s)^{[1]} \bar{\bm{M}}_s\\
&\quad+\bm{n}^{[1]} \otimes \bar{\bm{F}}_s^{-1}\Div_s(\bar{\bm{F}}_s\bar{\bm{M}}_s)+\bar{\bm{n}} \otimes (\bm{F}_s^{-1})^{[1]}\Div_s(\bar{\bm{F}}_s\bar{\bm{M}}_s)\\
&\quad+\bar{\bm{n}} \otimes \bar{\bm{F}}_s^{-1}\Div_s(\bm{F}_s^{[1]}\,\bar{\bm M}_s + \bar{\bm F}_s\,\bm{M}_s^{[1]}).
\end{split}
\end{align}
Therefore, we next derive the first-order perturbations of the required kinematic quantities.

\paragraph{\textbf{Unit normal}}

The first-order perturbation of the current unit normal $\bm{n}$ is obtained as the derivative of of $\bm{n}$ with respect to $\bm{F}_s$, applied to the perturbation 
$\bm{\eta}_s \bar{\bm{F}}_s$:
\begin{equation}\label{eq:n1}
\bm{n}^{[1]} 
= \frac{\partial \bm{n}}{\partial \bm{F}_s}\Big|_{\bar{\bm{F}}_s} 
: (\bm{\eta}_s \bar{\bm{F}}_s).
\end{equation}
Differentiation of the relation $\bm{F}_s^T\bm{n}=\bm{0}$ implies
\begin{align}\label{eq:dn}
\rd\bm{n} = -\bm{F}_s^{-T}(\rd \bm{F}_s)^T \bm{n}.
\end{align}
Thus, \eqref{eq:n1} simplifies to
\begin{equation}
\bm{n}^{[1]} = -\bm{F}_s^{-T}(\bm{\eta}_s \bar{\bm{F}}_s)^T \bm{n}= - \bm{\eta}_s^T \bar{\bm{n}}. \label{eq:norm_incr}
\end{equation}

\paragraph{\textbf{Product} $\bm{b}\bm{F}_s$}
In the incremental formulation, only the linearization of 
the product $\bm{b}\bm{F}_s$ is required, which allows us to avoid computing 
$\bm{b}^{[1]}$ explicitly. By the chain rule, we have the relation
\begin{align}
\bm{b}\bm{F}_s=-\nabla_s\bm{n}.
\end{align}
The first-order perturbation thus yields
\begin{align}
(\bm{b}\bm{F}_s)^{[1]}=-\nabla_s\bm{n}^{[1]}=\underbrace{\bar{\nabla}_s(\bm{\eta}^T_s\bar{\bm{n}})}_{\bm{\omega}_s}\bar{\bm{F}}_s.
\end{align}

\paragraph{\textbf{Relative curvature}}

The first-order perturbation of the relative curvature $\bm{\kappa}$ is given by
\begin{align}\label{eq:kappa-deriv-split}
\bm{\kappa}^{[1]}
= \frac{\partial\bm{\kappa}}{\partial\bm{F}_s}\Big|_{(\bar{\bm{F}}_s,\bar{\bm{H}}_s)} : (\bm{\eta}_s \bar{\bm{F}}_s)
+
\frac{\partial\bm{\kappa}}{\partial\bm{H}_s}\Big|_{(\bar{\bm{F}}_s,\bar{\bm{H}}_s)} \;\vdots\; \nabla_s(\bm{\eta}_s \bar{\bm{F}}_s) .
\end{align}
It follows from the definition $\bm{\kappa} = -\,\bm{n}\cdot \bm{H}_s$ that
\begin{align}
&\frac{\partial \bm{\kappa}}{\partial \bm{F}_s}\Big|_{(\bar{\bm{F}}_s,\bar{\bm{H}}_s)} 
: (\bm{\eta}_s \bar{\bm{F}}_s)=-\bm{n}^{[1]}\cdot\bar{\bm{H}}_s = (\bm{\eta}_s^T \bar{\bm{n}})\cdot \bar{\bm{H}}_s,
\\
&\frac{\partial \bm{\kappa}}{\partial \bm{H}_s}\Big|_{(\bar{\bm{F}}_s,\bar{\bm{H}}_s)} 
\;\vdots\; \nabla_s(\bm{\eta}_s \bar{\bm{F}}_s) 
= -\bm{I}_s(\bar{\bm{n}}\cdot \nabla_s(\bm{\eta}_s \bar{\bm{F}}_s)) =
-\bar{\bm{F}}_s^T( \bar{\bm{n}}\cdot \nabla_s\bm{\eta}_s)
-(\bm{\eta}^T_s\bm{n})\cdot \bar{\bm{H}}_s,
\end{align}
where we have used the Leibniz rule to expand the gradient $\nabla_s(\bm{\eta}_s\bar{\bm{F}}_s)$. Adding these contributions together yields
\begin{align}\label{eq:kappa-deriv-final}
\bm{\kappa}^{[1]}
=-\bar{\bm{F}}_s^T\left( \bar{\bm{n}}\cdot \nabla_s\bm{\eta}_s\right) = \bm{F}^T_s(\underbrace{-\bar{\bm{i}}_s(\bar{\bm{n}}\cdot \bar{\nabla}_s\bm{\eta}_s)}_{\bm{\rho}_s})\,\bar{\bm{F}}_s =  \bar{\bm{F}}^T_s {\bm{\rho}}_s\bar{\bm{F}}_s,
\end{align}
where  we passed from surface gradient $\nabla_s$ with respect to the reference configuration to the surface gradient ${\nabla}_s(\cdot) = \bar{\nabla}_s(\cdot)\bar{\bm{F}}_s$ with respect to the intermediate configuration. This reformulation is convenient, since the incremental field $\bm{\eta}_s$ represents the displacement gradient between the intermediate and current configurations. 
In addition, the surface identity $\bar{\bm{i}}_s$ was injected in the expression, since the action of $\bm{F}_s^T$ discards the normal component as $\bm{F}^T\bm{n} = \bm{0}$ and thus projects onto the tangent space. By the Leibniz rule, the second-order tensor $\bm{\rho}_s$ can be rewritten without involving third-order tensors as
\begin{align}
\bm{\rho}_s=-\bar{\bm{i}}_s\bar{\nabla}_s(\bm{\eta}_s^T\bar{\bm{n}})-\bm{\eta}_s^T\bar{\bm{b}}.
\end{align}

\paragraph{\textbf{Inverse surface deformation gradient}}

Starting from the identity
\begin{equation}
\bm{F}_s \bm{F}_s^{-1}=\bm{i}_s, \qquad 
\bm{i}_s = \bm{I} - \bm{n}\otimes \bm{n},
\end{equation}
its differentiation gives
\begin{equation}
(\rd \bm{F}_s)\,\bm{F}_s^{-1} + \bm{F}_s\,\rd\bm{F}_s^{-1} 
= -(\rd \bm{n})\otimes \bm{n} -\bm{n}\otimes (\rd \bm{n}).
\end{equation}
Moreover, using \eqref{eq:dn}, we obtain
\begin{equation}\label{eq:FsInv-linearization-is}
\rd\bm{F}_s^{-1} 
= -\,\bm{F}_s^{-1}(\rd\bm{F}_s)\bm{F}_s^{-1}
+ \bm{F}_s^{-1}\bm{F}_s^{-T}(\rd\bm{F}_s)^T(\bm{n}\otimes\bm{n}).
\end{equation}
Setting $\rd\bm{F}_s=\bm{\eta}_s\bar{\bm{F}}_s$ in \eqref{eq:FsInv-linearization-is}  and evaluating   at the barred state, we obtain the first-order perturbation of $\bm{F}^{-1}_s$  as 
\begin{align}
(\bm{F}^{-1}_s)^{[1]} =-\,\bar{\bm{F}}_s^{-1}(\bm{\eta}_s\bar{\bm{F}}_s)\bar{\bm{F}}_s^{-1}
+ \bar{\bm{F}}_s^{-1}\bar{\bm{F}}_s^{-T}(\bm{\eta}_s\bar{\bm{F}}_s)^T(\bar{\bm{n}}\otimes\bar{\bm{n}})=- \bar{{\bm{F}}}_s^{-1}{  \underbrace{(\bar{\bm{i}}_s{\bm{\eta}}_s - {\bm{\eta}}_s^T \bar{\bm{n}} \otimes\bar{\bm{n}}  )}_{{\bm{\xi}_s}} } =-\bar{{\bm{F}}}_s^{-1}{\bm{\xi}_s}.	
\end{align}

In summary, the first-order perturbations of the required kinematic quantities are given by
\begin{align}
\begin{split}
&\bm{n}^{[1]} = -\,\bm{\eta}_s^T \bar{\bm{n}},\qquad (\bm{b}\bm{F}_s)^{[1]}=\bm{\omega}_s\bar{\bm{F}}_s,\qquad \bm{\kappa}^{[1]}=\bar{\bm{F}}^T_s {\bm{\rho}}_s\bar{\bm{F}}_s,\qquad (\bm{F}^{-1}_s)^{[1]}=-\bar{{\bm{F}}}_s^{-1}{\bm{\xi}_s},
\end{split}
\end{align}
where
\begin{align}\label{eq:kinematic-q}
\begin{split}
\bm{\omega}_s=\bar{\nabla}_s(\bm{\eta}_s^T\bar{\bm{n}}),\qquad \bm{\rho}_s=-\bar{\bm{i}}_s\bar{\nabla}_s(\bm{\eta}_s^T\bar{\bm{n}})-\bm{\eta}_s^T\bar{\bm{b}},\qquad \bm{\xi}_s=\bar{\bm{i}}_s{\bm{\eta}}_s - {\bm{\eta}}_s^T \bar{\bm{n}}\otimes\bar{\bm{n}},\qquad 
\end{split}
\end{align}

Finally, the linearized increments of the surface stress-like quantities are given by
\begin{align}
&{\bm{P}}_s^{[1]} = 
\mathbb{A}_s 
   : \big( {\bm{\eta}_s} \, \bar{{\bm{F}}}_s \big)
 + \mathbb{B}_s
   : \big( \bar{{\bm{F}}}_s^{T} \, {\bm{\rho}_s} \, \bar{{\bm{F}}}_s \big), \label{eq:P_incr} \\[4pt]
&{\bm{M}}_s^{[1]} = 
\mathbb{ C}_s
   : \big( {\bm{\eta}}_s \, \bar{{\bm{F}}}_s \big)
 + \mathbb{D}_s
   : \big( \bar{{\bm{F}}}_s^{T} \, {\bm{\rho}_s} \, \bar{{\bm{F}}}_s \big), \label{eq:M_incr}
\end{align}
where the surface moduli are defined as
\begin{align}
   &\mathbb{A}_s  = \frac{\partial {\bm{P}}_s}{\partial \bm{F}_s}\Big|_{(\bar{\bm{F}}_s, \bar{\bm{\kappa}})}, 
     \qquad
      \, \mathbb{B}_s  = \frac{\partial {\bm{P}}_s}{\partial \bm{\kappa}}\Big|_{(\bar{\bm{F}}_s, \bar{\bm{\kappa}})},
    \\
   & \mathbb{C}_s = \frac{\partial \bm{M}_s}{\partial \bm{F}_s}\Big|_{(\bar{\bm{F}}_s, \bar{\bm{\kappa}})}, 
    \qquad  \mathbb{D}_s = \frac{\partial \bm{M}_s}{\partial \bm{\kappa}}\Big|_{(\bar{\bm{F}}_s, \bar{\bm{\kappa}})}.
\end{align}
Substituting all these expressions into \eqref{eq:Qs1_master}, we obtain the final form of the incremental stress $\bm{Q}_s^{[1]}$ as
\begin{align}
\begin{split}\label{eq:Q_s^{[1]}}
\bm{Q}_s^{[1]} =&\mathbb{A}_s:( \bm{\eta}_s  \bar{\bm{F}}_s )
+\mathbb{B}_s:( \bar{\bm{F}}_s^{T}  \bm{\rho}_s \bar{\bm{F}}_s )-\bar{\bm{b}}\bar{\bm{F}}_s\big(\mathbb{C}_s: (\bm{\eta}_s  \bar{\bm{F}}_s) + \mathbb{D}_s: ( \bar{\bm{F}}_s^{T}  \bm{\rho}_s  \bar{\bm{F}}_s ) \big)-\bm{\omega}_s \bar{\bm{F}}_s\bar{\bm{M}}_s\\
&-\bm{\eta}_s^T \bar{\bm{n}}\otimes \bar{\bm{F}}_s^{-1}\Div_s(\bar{\bm{F}_s}\bar{\bm{M}}_s)-\bar{\bm{n}} \otimes \bar{\bm{F}}_s^{-1} \bm{\xi}_s \Div_s(\bar{\bm{F}_s}\bar{\bm{M}}_s)\\
&+\bar{\bm{n}} \otimes \bar{\bm{F}}_s^{-1}\Div_s\big(\bm{\eta}_s  \bar{\bm{F}}_s \bar{\bm M}_s + \bar{\bm F}_s (
   \mathbb{C}_s:(\bm{\eta}_s  \bar{\bm{F}}_s)
 + \mathbb{D}_s: ( \bar{\bm{F}}_s^{T}  \bm{\rho}_s  \bar{\bm{F}}_s ) )\big).
\end{split}
\end{align}
This incremental constitutive equation for the surface, together with the corresponding one \eqref{eq:P1_master_direct} for the bulk, completes the formulation of the incremental theory.

\subsection{Push-forward  to the intermediate configuration}\label{sec:push-forward}

It is also convenient to formulate the equilibrium equations in the intermediate configuration. 
To this end, we introduce the incremental actual stress tensor, its surface counterpart, the actual moment tensor, and the actual incremental moment tensor as
\begin{align}\label{eq:icr}
&\bm{\chi} = \bar{J}^{-1}\bm{P}^{[1]}\bar{\bm{F}}^{T}, 
\qquad\quad  
\bm{\chi}_s = \bar{J}_s^{-1}\bm{Q}^{[1]}_s\bar{\bm{F}}_s^{T},\\
&\bar{\bm{\tau}}_s =\bar{J}_s^{-1} \bar{\bm{F}}_s \bar{\bm{M}}_s \bar{\bm{F}}_s^T,
\qquad 
\bm{\tau}_s^{[1]} = \bar{J}_s^{-1} \bar{\bm{F}}_s \bm{M}_s^{[1]} \bar{\bm{F}}_s^T.
\end{align}
Moreover, we introduce the in-plane unit normal $\bar{\bm{v}}$ to the boundary curve $\bm{\phi}(\partial S)$, which is obtained by
\bea\label{eq:v_act}
\bar{\bm{v}} =\frac{\bar{\bm{F}}_s^{-T} \bm{V}}{|\bar{\bm{F}}_s^{-T} \bm{V}|}.   
\eea

The push–forward of the reference governing system \eqref{eq11}--\eqref{eq15}, obtained using the standard bulk and surface Piola transformations \eqref{eq:piola_t} along with the relation \eqref{eq:v_act}, reads:
\begin{align}
&\overline{\ddiv}(\bm\chi) = \bm 0
\hspace{13.2em}  \text{in}\ \bm\phi_e(\Omega), \label{eq:inc1}\\
&\bm\chi \bar{\bm n} =\bm0 
\hspace{14.7em} \text{on } \bm\phi_e(\partial\Omega\setminus S), \label{eq:inc2}\\
&\overline{\ddiv}_s(\bm\chi_s)-\bm\chi \bar{\bm n}= \bm 0
\hspace{10.2em}  \text{on } \bm\phi_e(S), \label{eq:inc4}
\\
&\bm\chi_s \bar{\bm{v}}
+
\overline{\ddiv}_{\parallel} (
\bar{\bm n}\otimes \bm\tau_s^{[1]}\bar{\bm{v}}
+\bm n^{[1]}\otimes \bar{\bm\tau}_s\bar{\bm{v}})
= \bm 0
\qquad   \text{on } \bm\phi_e(\partial S), \label{eq:edge1_spatial}\\
&\bar{\bm{v}} \cdot \bm{\tau}_s^{[1]}\bar{\bm{v}}
=  0\hspace{12.9em} \text{on } \bm\phi_e(\partial S), \label{eq:edge2_spatial}
\end{align}
where $\overline{\ddiv}$, $\overline{\ddiv}_s$, and $\overline{\ddiv}_{\parallel}$ denote the bulk, surface divergence, and line divergence in the intermediate configuration. The only nontrivial task is to derive \eqref{eq:edge1_spatial} from  \eqref{eq14}. For this, we first substitute the definitions \eqref{eq:icr} and \eqref{eq:v_act} into \eqref{eq14} and then apply the Piola transform \eqref{eq:piola_curve}  on the boundary curve $\partial S$, which yields
\bea \label{eq:div_subder}
 \bm\chi_s\,\bar{\bm{v}}
+
 \overline{\ddiv}_{\parallel}\!\Big(\frac{1}{\bar{J}_s |\bar{\bm{F}}_s^{-T} \bm{V}|}(
\bar{\bm{n}}\otimes {\bm{M}}^{[1]}_s\bm{V}+{\bm{n}}^{[1]}\otimes \bar{\bm{M}}_s\bm{V}) \bar{\bm{F}}_s^T
\Big) = \bm{0}.
\eea
 The first term inside the divergence is reformulated using \eqref{eq:v_act} as
\bea
&&\frac{1}{\bar{J}_s |\bar{\bm{F}}_s^{-T} {\bm{V}}|} \bar{\bm{n}}\otimes({\bm{M}}^{[1]}_s{\bm{V}}) \bar{\bm{F}}_s^T =\frac{1}{\bar{J}_s}\bar{\bm{n}}\otimes\bar{\bm{F}}_s\bar{\bm{M}}^{[1]}\bar{\bm{F}}^T_s\bar{\bm{v}}= \bar{\bm{n}}\otimes {\bm{\tau}}^{[1]}_s\bar{\bm{v}}. 
\eea
 Applying similar operations to the second term in \eqref{eq:div_subder}, we obtain the boundary condition \eqref{eq:edge1_spatial}.

For applications, it is convenient to recast the incremental actual surface stress tensor $\bm{\chi}_s$ into a form analogous to the effective surface stress tensor (cf. \eqref{eq:Qsmy}). To this end, we further introduce the incremental  surface actual stress and moment measures
\begin{align}\label{eq:addIncrQuant}
\bm{\sigma}_s=\bar{J}_s^{-1}\bm{P}^{[1]}_s\bar{\bm{F}}_s^T,\qquad \bm{m}_s=\bar{J}_s^{-1}(\bm{F}_s\bm{M}_s)^{[1]}\bar{\bm{F}}_s^T=\bm{\tau}_s^{[1]}+\bm{\eta}_s\bar{\bm{\tau}}_s. 
\end{align}
In view of \eqref{eq:Q_s^{[1]}}, $\eqref{eq:icr}_2$ and the Piola transformation  \eqref{eq:piola_t}, the incremental actual surface stress $\bm{\chi}_s$ can be expressed in the compact form 
\begin{align}\label{eq:chis}
\bm{\chi}_s=\bm{\sigma}_s-\bar{\bm{b}}\bm{m}_s+\bar{\bm{n}}\otimes\bar{\bm{i}}_s\overline{\ddiv}_s(\bm{m}_s)+\bm{\theta}_s,
\end{align}
where the additional term $\bm{\theta}_s$ accounts for the geometric changes induced by the incremental deformation $B_e \to B_y$ and is given explicitly by
\begin{align}\label{eq:thetas}
\bm{\theta}_s=\bar{J}_s^{-1}[(\bar{\bm{b}}\bm{\eta}_s-\bm{\omega}_s)\bar{\bm{F}}_s\bar{\bm{M}}_s\bar{\bm{F}}_s^T-\bm{\eta}_s^T\bar{\bm{n}}\otimes\bar{\bm{i}}_s\Div_s(\bar{\bm{F}_s}\bar{\bm{M}}_s)-\bar{\bm{n}}\otimes{\bm{\xi}}_s\Div_s(\bar{\bm{F}_s}\bar{\bm{M}}_s)].
\end{align}
In the form \eqref{eq:chis}, the structure of incremental surface stress $\bm{\chi}_s$ stands out automatically. Moreover, this formulation makes the calculation of $\bm{\chi}_s$ straightforward: reducing it to evaluation of the stress and moment measures $\bm{\sigma}_s$ and $\bm{m}_s$, which are readily obtainable from linearization of the surface Piola-Kirchhoff stress and surface moment.

\section{Coordinate expressions}\label{sec:coords}

In this section, we express the incremental equations in a curvilinear coordinate system tailored to the boundary surface. We first introduce
the notation for covariant/contravariant bases, metric coefficients, and the
surface–adapted chart, and then write the bulk and surface incremental equations in component form. Throughout, Latin indices $i,j, \dots\in\{1,2,3\}$
refer to bulk coordinates and Greek indices $\alpha,\beta, \dots\in\{1,2\}$ to in–plane
(surface) coordinates.

\subsection{Bulk}

 We first briefly review the coordinate formulation of the bulk-related quantities and of incremental equations \eqref{eq11}--\eqref{eq12}.

The material position vector $\bm{X}$ is parametrized by fixed reference curvilinear coordinates $\theta^i$, 
\begin{align}
\bm{X} = \bm{X}(\theta^1,\theta^2,\theta^3).
\end{align}
The referential base vectors are then obtained by
\begin{align}
\bm G_i = \bm X_{,i} =\frac{\partial \bm{X}}{\partial \theta_i},
\end{align}
where the subscript “\({}_{,i}\)” denotes partial differentiation with respect to coordinate $\theta^i$. The undeformed metric coefficients, determinant of the associated matrix, and the contravariant basis then read 
\begin{align}
G_{ij}=\bm G_i \cdot\bm G_j,\qquad
G=\det(G_{ij}), \qquad(G^{ij})=(G_{ij})^{-1},\qquad 
\bm G^{\,i}=G^{ij}\,\bm G_j,
\end{align}
where $(G_{ij})$ denotes the $3\times 3$ matrix whose $ij$-th entry is $G_{ij}$ and analogously for $(G^{ij})$.

To allow for a concise description of the boundary surface, we adopt a \emph{surface-adapted} chart: points on
\(S\subset\partial\Omega\) are parameterized by
\begin{align} \bm X(\theta^1,\theta^2,\theta^3=\theta^s),
\end{align}
where $\theta^s$ denotes the constant value of the transverse coordinate $\theta^3$ identifying the surface. Consequently, the base vectors $\bm{G}_1$ and $\bm{G}_2$ are tangent to the boundary surface. The unit normal to the undeformed
surface is
\begin{align}
\bm N=\frac{\bm{G}_1 \times \bm{G}_2}{|\bm{G}_1 \times \bm{G}_2|} =\frac{\bm G^{3}}{\sqrt{G^{33}}}.
\end{align}

It is convenient to use the same coordinates to describe points in both the
reference and current (intermediate) configurations. The current position vector is
\begin{align}
\bm x(\theta^i) = \bm{\phi}_e\big(\bm X(\theta^i)\big),
\end{align}
and the current covariant base vectors are
\begin{align}
\bm g_i = \bm x_{,i} = \bar{\bm F} \bm G_i,
\end{align}
where \(\bar{\bm F}\) is the deformation gradient evaluated at the intermediate state. The current (deformed) metric coefficients, metric determinant, and contravariant basis are
\begin{align}
g_{ij}=\bm g_i \cdot \bm g_j,\qquad g=\det(g_{ij}),\qquad 
(g^{ij})=(g_{ij})^{-1},\qquad
\bm g^{\,i}=g^{ij}\,\bm g_j.
\end{align}
The Christoffel symbols of the deformed basis (i.e., of the Levi–Civita connection associated with \(g_{ij}\)) read
\begin{align}
\bar{\Gamma}^{\,k}_{\ ij} \;=\; \bm g_{i,j} \cdot \bm g^{\,k},
\qquad \text{with}\quad \bm g_{i,j}:=\frac{\partial \bm g_i}{\partial \theta^j}.
\end{align}
The coordinate expression of the material (reference) divergence of a 
second-order tensor $\bm{R} ={{R}}_i^{\ j} \bm{g}^i \otimes \bm{G}_j$ is given by (see, for instance, \cite{lee2012introduction}, p. 436)
\begin{equation}\label{eq:div_bulk_clean}
\Div(\bm{R})
= \frac{1}{\sqrt{G}}\big(\sqrt{G} \bm{R} \cdot \bm{G}^{\,j}\big)_{,j}
= \Big(
\frac{1}{\sqrt{G}}\big(\sqrt{G} {R}_{i}^{\ j}\big)_{,j}
- {R}_{j}^{\ k}\,\bar{\Gamma}^{\,j}_{\ ki}
\Big)\bm{g}^{\,i}.
\end{equation}

Using the same coordinates in the reference and current configurations, the deformation gradient at the intermediate configuration and its determinant  are 
\bea\label{eq:FsJbar}
{\bar{\bm{F}}}=\bm{g}_{i} \otimes \bm{G}^{i},\qquad \bar{J}= \det(\bar{\bm{F}})= \sqrt{g/G}.
\eea The inverse deformation gradient reads
\bea
{\bar{\bm{F}}^{-1}} = \bm{G}_{i} \otimes \bm{g}^{i}.
\eea
The unit outward normal to the deformed surface is obtained by Nanson's formula
\begin{align}
\bar{\bm n} = \frac{\bar{\bm{F}}^{-T}\bm{N}}{|\bar{\bm{F}}^{-T}\bm{N}|} = \frac{\bm{G}^{3}\cdot \bm{G}_{i}\otimes \bm{g}^{i}  }{|\bm G^{3}\cdot \bm{G}_{i}\otimes \bm{g}^{i}  |} = \frac{\bm{g}^{3}}{\sqrt{g^{33}}}.
\end{align}
The incremental displacement tensor is expressed in the deformed basis as
\bea
 {{\bm{\eta}}} = \eta^i_{\ j} \bm{g}_i\otimes\bm{g}^j .
\eea

From the requirement of frame indifference, the strain energy depends on the deformation gradient $\bm{F}$ only through 
the induced metric of the deformed configuration, i.e., 
$g_{ij} = \bm{g}_i \cdot \bm{g}_j$. 
Accordingly, the energy function $W(\bm{F})$ 
can be equivalently expressed as a function of the metric coefficients, 
\bea
W(\bm{F}) =\tilde{W}(\bm{F}\bm{G}_i \cdot \bm{F}\bm{G}_j) = \tilde{W}(g_{ij}).
\eea
In view of the identities $
{\partial g_{jk}}/{\partial \bm F}
= \bm g_j \otimes \bm G_k+\bm g_k\otimes \bm G_j $ and the symmetry $\partial\tilde{W}/\partial g_{jk}=\partial\tilde{W}/\partial g_{kj}$, the chain rule gives  the bulk first Piola-Kirchhoff stress in the form
\bea \label{eq:coordP_bulk}
\bm{P} = \frac{\partial W}{\partial \bm{F}}= \frac{\partial \tilde{W}}{\partial {g}_{jk}} \frac{\partial {g}_{jk}}{\partial  \bm{F}}=
\frac{\partial \tilde{W}}{\partial {g}_{jk}} (\bm g_j \otimes \bm G_k+\bm g_k\otimes \bm G_j)
= 
2\frac{\partial \tilde{W}}{\partial {g}_{kj}} \bm{g}_k \otimes \bm{G}_j = {2{g}_{ik} \frac{\partial \tilde{W}}{\partial {g}_{kj}} }\bm{g}^i \otimes \bm{G}_j.
\eea
In other words, the mixed components of \(\bm P\) in the basis \(\{\bm g^{\,i}\otimes \bm G_j\}\) are
\(P_{i}^{\ j}= 2\,g_{ik}\,\partial \tilde{W}/\partial g_{kj}\).

The tangent stiffness  in the intermediate configuration ${\mathbb A}$ is then obtained by 
\begin{equation}
{\mathbb A}
= \frac{\partial^2 W}{\partial \bm F\,\partial \bm F}\Big|_{\bar{\bm F}}
= 2\,\frac{\partial}{\partial \bm F}
\!\Big(\frac{\partial \tilde{W}}{\partial g_{ij}}\,\bm g_i\!\otimes\!\bm G_j\Big).
\end{equation}
Using further 
\( \partial \bm g_i/\partial \bm F = \bm g_p\!\otimes\!\bm g^{\,p} \!\otimes\!\bm G_i\),
we obtain
\begin{align}
\begin{split}
{\mathbb A}
&= 4\,\frac{\partial^2 \tilde{W}}{\partial g_{ij}\,\partial g_{kl}}\,
   \bm g_i\!\otimes\!\bm G_j\!\otimes\!\bm g_k\!\otimes\!\bm G_l
 + 2\,\frac{\partial \tilde{W}}{\partial g_{ij}}\,
   \bm g_p\!\otimes\!\bm G_j\!\otimes\!\bm g^{\,p}\!\otimes\!\bm G_i \\
&= \Big(4\,g_{im}\,g_{kn}\,\frac{\partial^2 \tilde{W}}{\partial g_{ij}\,\partial g_{kl}}
        + 2\,g_{nm}\,\frac{\partial \tilde{W}}{\partial g_{lj}}\Big)\,
   \bm g^{\,m}\!\otimes\!\bm G_j\!\otimes\!\bm g^{\,n}\!\otimes\!\bm G_l.
\end{split}
\end{align}
Thus, the components in the mixed basis \(\{\bm g^{\,m}\!\otimes\!\bm G_j\!\otimes\!\bm g^{\,n}\!\otimes\!\bm G_l\}\), are
\begin{equation}
{\mathbb A}^{\;\;j\;\;l}_{m\;n}
= 4\,g_{im}\,g_{kn}\,\frac{\partial^2 \tilde{W}}{\partial g_{ij}\,\partial g_{kl}}
  + 2\,g_{nm}\,\frac{\partial \tilde{W}}{\partial g_{lj}}. \label{eq:bulk_stiff_coeff}
\end{equation}
The incremental bulk stress then follows from the linearized law (cf.\ \eqref{eq:P1_master_direct})
\begin{align}\label{eq:PbulkIncr_clean}
\bm P^{[1]}
&= {\mathbb A} : (\bm\eta\bar{\bm F}) = {\mathbb A}^{\;\;j\;\;l}_{m\;n}\,\eta^{n}_{\ l} \,\bm g^{\,m}\!\otimes\!\bm G_j,\qquad \text{i.e.,}\quad {P^{[1]}}_{m}^{\ j}
= {\mathbb A}^{\;\;j\;\;l}_{m\;n}\,\eta^{n}_{\ l}.
\end{align}
The components of the pushed–forward incremental stress
\(\bm\chi=\chi_{i}^{\ j}\,\bm g^{\,i}\!\otimes\!\bm g_j\) defined in \eqref{eq:icr} satisfy
\begin{equation}\label{eq:chi_coords_clean}
\chi_{i}^{\ j}
= \bar{J}^{-1}\;{P^{[1]}}_{i}^{\ j}
\end{equation}
with $\bar{J}$ given in \eqref{eq:FsJbar}.

By the coordinate expression of the surface divergence \eqref{eq:div_bulk_clean}, the incremental bulk equilibrium \eqref{eq11} becomes
\bea
\frac{1}{\sqrt{G}}\big(\sqrt{G} {{P}^{[1]}}_{i}^{\ j}  \big)_{,j}-  {P^{[1]}}_{j}^{\ m} \bar{\Gamma}^j_{\ m i}=0. \label{eq:bulk_bs_coord}
\eea
In spatial form, using components of \(\bm\chi\) and \(g=\det(g_{ij})\), we obtain
\bea
\frac{1}{\sqrt{g}}\big(\sqrt{g} {\chi}_{i}^{\ j} \big)_{,j}-  {\chi}_{j}^{\ m} \bar{\Gamma}^j_{\ m i} = 0. \label{eq:bulk_bs_coord_actual}
\eea
The incremental boundary condition \eqref{eq12} on  part of the boundary in coordinate form reads 
\bea
{P^{[1]}}_{i}^{\ 3}= 0\qquad \text{on}\ \partial\Omega\setminus S,
\eea
and its spatial counterpart \eqref{eq:inc2} is
\bea
{\chi}_{i}^{\ 3}  = 0\qquad \text{on}\ \bm{\phi}_e(\partial\Omega\setminus S).
\eea

\subsection{Surface}
We now give a coordinate formulation of the surface objects and of the incremental boundary
condition \eqref{eq13}. With the \emph{surface–adapted} chart \(\{\theta^i\}\) introduced above,
the boundary surface \(S\subset\partial\Omega\) is parametrized by the in–plane coordinates
\(\theta^\alpha\) (\(\alpha=1,2\)) and \(\theta^3=\theta^s\).
The corresponding covariant surface bases and their reciprocals are the bulk bases
evaluated at the surface,
\begin{align}
\bm G^{\alpha}=G^{\alpha\beta}\bm G_\beta,\qquad \bm{g}_\alpha=\bar{\bm F}_s \bm G_\alpha,\qquad
\bm g^{\alpha}=g^{\alpha\beta}\bm g_\beta,
\end{align}
with surface metrics
\begin{align}
G_{\alpha\beta}=\bm G_\alpha\cdot \bm G_\beta,\qquad
g_{\alpha\beta}=\bm g_\alpha\cdot \bm g_\beta,\qquad
G_s=\det(G_{\alpha\beta}),\quad g_s=\det(g_{\alpha\beta}).
\end{align}
Here, $(G_{\alpha\beta})$ denotes the $2\times 2$ matrix whose $\alpha\beta$-th entry is $G_{\alpha\beta}$ and similarly for $(g_{\alpha\beta})$.  Moreover, certain quantities associated with the surface require an extension 
of the basis to three dimensions. In such cases, the third base vectors or 
one-forms, denoted by $\bm{G}_{3}$, $\bm{g}_{3}$ and 
$\bm{G}^{3}$, $\bm{g}^{3}$, are taken from the bulk basis and 
evaluated at the surface, i.e.,\ at $\theta^{3} = \theta^s$. 
With a slight abuse of notation, the same symbols are used to denote the base 
vectors and one-forms both in the bulk and on the surface. 
This does not lead to any ambiguity, as bulk quantities are only related to 
their surface counterparts through their evaluation at the surface.

In analogy with the bulk divergence, the
surface divergence of a superficial tensor field $\bm{R}_s = R^{\alpha\beta}_s\bm{g}_\alpha\otimes\bm{G}_\beta$ has the coordinate expression
\begin{align}\label{eq:surfDivCoord}
\Div_s (\bm{R}_s)= \frac{1}{\sqrt{G_s}}\big(\sqrt{G_s}\bm{R}_s \cdot \bm{G}^\beta\big)_{,\beta}= \Big(\frac{1}{\sqrt{G_s}}\big(\sqrt{G_s}\,R_s^{\alpha\beta}\big)_{,\beta}
+ R_s^{\gamma\delta}\bar{\Gamma}^{\alpha}_{\ \gamma\delta}\Big)\bm{g}_\alpha
+ R_s^{\alpha\beta}\bar{b}_{\alpha\beta}\,\bar{\bm{n}},
\end{align}
where the term with $\bar{\Gamma}^{\alpha}_{\ \gamma\delta}$ accounts for the
variation of the covariant basis $\bm{g}_\alpha$ along the surface, and the
term with $\bar{b}_{\alpha\beta}$ collects the normal projection generated by surface curvature.

The surface deformation gradient and its inverse read
\begin{align}
{\bar{\bm{F}}}_s = \bm{g}_{\alpha} \otimes \bm{G}^{\alpha},\qquad
{\bar{\bm{F}}}^{-1}_s = \bm{G}_{\alpha} \otimes \bm{g}^{\alpha}.
\end{align}
The surface Jacobian $\bar{J}_s = \sqrt{{g_s}/{G_s}}$ is connected to the bulk Jacobian by
\begin{align}
\bar{J}_s =  \bar{J} \sqrt{\frac{g^{33}}{G^{33}}}, 
\end{align}
which is due to the tangent property of the first two base vectors. The curvature tensor $\bar{\bm{b}}$ and the relative curvature tensor $\bm{\kappa}$ can be represented componentwise as
\begin{align}
\bar{\bm{b}}
= \bar b_{\alpha\beta}\,\bm g^{\alpha}\otimes\!\bm g^{\beta},
\qquad
\bar b_{\alpha\beta} = \,\bm n\cdot \bm g_{\alpha,\beta},
\qquad \bm\kappa
= \kappa_{\alpha\beta}\,\bm G^{\alpha} \otimes\!\bm G^{\beta}.
\end{align}
so that, in our sign convention, $\bar{\bm{b}} = -\,\kappa_{\alpha\beta}\,\bm g^{\alpha}\!\otimes\!\bm g^{\beta}$ and
$\bm\kappa = -\,\bar b_{\alpha\beta}\,\bm G^{\alpha}\!\otimes\!\bm G^{\beta}$.

The surface incremental displacement gradient (restricted to the surface and expressed in the current
basis) is written as
\begin{align}
 {{\bm{\eta}}_s}=\bm{\eta}\big|_{\theta^3=\theta^s} \bar{\bm{i}}_s={{{\eta}}_s} ^i_{\ \alpha} \bm{g}_i\otimes\bm{g}^\alpha = {{{\eta}}_s} ^\beta_{\ \alpha} \bm{g}_\beta\otimes\bm{g}^\alpha + {{{\eta}}_s} ^3_{\ \alpha} \bm{g}_3\otimes\bm{g}^\alpha.
\end{align}
We now turn to the coordinate expressions required for the evaluation of 
$\bm{Q}_s^{[1]}$. In particular, we will make use of the identities 
\begin{align}\label{eq:ident}
\bar{\bm{n}}\cdot {\bm{\eta}}_s 
= \frac{{{\eta}_s}^{3}{}_{\!\alpha}}{\sqrt{g^{33}}}\,\bm{g}^{\,\alpha},
\qquad
\bar{\bm{i}}_s{\bm{\eta}}_s{\bar{\bm{F}}}_s
= {{\eta}_s}^{\beta}{}_{\!\alpha}\,\bm{g}_{\beta}\!\otimes\!\bm{G}^{\,\alpha}.
\end{align}
Based on the above relations, the quantities associated with the gradient of the incremental displacement gradient can be expressed as
\begin{align}
\begin{split}
&{\bm{\omega}_s}{\bar{\bm{F}}}_s = \nabla_s( {\bm{\eta}^T_s} \bar{\bm{n}} )  =\nabla_s\Big( \frac{{\eta_s}^3_{\ \alpha}}{\sqrt{g^{33}}}\bm{g}^\alpha \Big) = \Big[\Big(\frac{{\eta_s}^3_{\ \alpha}}{\sqrt{g^{33}}}\Big)_{,\beta}- \frac{{\eta_s}^3_{\ \gamma}}{\sqrt{g^{33}}} \bar{\Gamma}^{\gamma}_{\ \alpha\beta} \Big]\bm{g}^\alpha \otimes \bm{G}^\beta + \frac{{\eta_s}^3_{\ \alpha}}{{g^{33}}} \bar{b}^\alpha_\beta \bm{g}^3 \otimes \bm{G}^\beta.
\end{split}\\
&\bm{\rho}_s =-\bar{\bm{i}}_s\bm{\omega}_s-\bm{\eta}_s^T\bar{\bm{b}} =  -\Big(\frac{{\eta_s} ^3_{\ \alpha, \beta}  }{\sqrt{g^{33}}} -\frac{{\eta_s} ^3_{\ \gamma}  }{\sqrt{g^{33}}}
\bar{\Gamma}^\gamma_{\ \alpha \beta }+{\eta_s} ^\gamma_{\ \alpha}  
\bar{b}_{\gamma \beta} \Big) \bm{g}^\alpha \otimes\bm{g}^\beta, \\ 
&{\bar{\bm{F}}}_s^T\bm{\rho}_s {\bar{\bm{F}}}_s =
-\Big( \frac{{\eta_s} ^3_{\ \alpha,\beta}}{\sqrt{g^{33}}}   -\frac{{\eta_s} ^3_{\ \gamma}  }{\sqrt{g^{33}}}
\bar{\Gamma}^\gamma_{\ \alpha \beta }+{\eta_s} ^\gamma_{\ \alpha}  
\bar{b}_{\gamma \beta} \Big) \bm{G}^\alpha \otimes\bm{G}^\beta.
\end{align}

We now focus on the coordinate formulation of the surface stress and moment tensors $\bm{P}_s$ and $\bm{M}_s$.  To satisfy the requirement of material frame indifference, it is convenient to express the surface energy in terms of the function 
\begin{align}
\varPsi(\bm{F}_s,\bm{\kappa})={\tilde{\varPsi}}(\bm{F}_s\bm{G}_\alpha\cdot\bm{F}_s\bm{G}_\beta\;,\;\bm{\kappa}\bm{g}_\delta\cdot\bm{g}_\gamma)={\tilde{\varPsi}}(g_{\alpha\beta}, {\kappa}_{\gamma\delta}),
\end{align}
where $g_{\alpha\beta}$ are the metric coefficients of the deformed surface basis and ${\kappa}_{\gamma\delta}$ denote the components of the relative curvature tensor. Accordingly, the surface stress $\bm{P}_s$ and surface  moment $\bm{M}_s$ are given by
\begin{align}\label{eq:coordP}
&\bm{P}_s = \frac{\partial {\varPsi}}{\partial \bm{F}_s} = \frac{\partial \tilde{\varPsi}}{\partial {g}_{\beta\nu}} \frac{\partial {g}_{\beta\nu}}{\partial  \bm{F}_s}=
\frac{\partial \tilde{\varPsi}}{\partial {g}_{\beta\nu}} ( \bm{g}_\nu \otimes \bm{G}_\beta + \bm{g}_\beta \otimes \bm{G}_\nu )
= 
2\frac{\partial \tilde{\varPsi}}{\partial {g}_{\beta\nu}} \bm{g}_\nu \otimes \bm{G}_\beta =\underbrace{2{g}_{\nu\alpha}\frac{\partial \tilde{\varPsi}}{\partial {g}_{\beta\nu}}}_{{P_s}_{\alpha}^{\ \beta}}  \bm{g}^\alpha \otimes \bm{G}_\beta,\\
&\bm{M}_s = \frac{\partial {\varPsi}}{\partial \bm{\kappa}} = \frac{\partial \tilde{\varPsi}}{\partial {\kappa}_{\alpha \beta}}\frac{\partial \kappa_{\alpha \beta} }{\partial \bm{\kappa}} = \frac{\partial \tilde{\varPsi}}{\partial {\kappa}_{\alpha \beta} } \bm{G}_\alpha  \otimes \bm{G}_\beta.\label{eq:coordM}
\end{align}

We next evaluate the second derivatives of the energy, required to compute the incremental stresses. We start with the derivatives of $\bm{M}_s$, whose calculation is straightforward since $\bm{M}_s$ is a referential tensor.
\begin{align}
&{\mathbb C}_s= \frac{\partial \bm{M}_s }{\partial \bm{F}_s}\Big|_{(\bar{\bm{F}}_s, \bar{\bm{\kappa}})} = \frac{\partial }{\partial \bm{F}_s} \Big( \frac{\partial \tilde{\varPsi}}{\partial {\kappa}_{\alpha\beta}} \bm{G}_\alpha \otimes \bm{G}_\beta\Big) = \underbrace{2 g_{\nu \gamma}\frac{\partial ^2\tilde{\varPsi}}{\partial {\kappa}_{\alpha\beta} \partial g_{\delta \nu}}  }_{{{\mathbb C}_s}^{\alpha \beta \ \delta}_{\ \ \gamma}}  \bm{G}_\alpha\otimes \bm{G}_\beta  \otimes \bm{g}^\gamma \otimes \bm{G}_\delta,
\\ 
&{\mathbb D}_s = \frac{\partial \bm{M}_s }{\partial \bm{\kappa}}\Big|_{(\bar{\bm{F}}_s, \bar{\bm{\kappa}})} =
\frac{\partial ^2\tilde{\varPsi}}{\partial {\kappa}_{\alpha \beta} \partial {\kappa}_{\gamma \delta} } \bm{G}_\alpha  \otimes \bm{G}_\beta \otimes \bm{G}_\gamma  \otimes \bm{G}_\delta.
\end{align}
The derivative of ${\bm{P}}_s$ with respect to $\bm{\kappa}$ is similarly straightforward, since neither the metric coefficients nor the basis vectors depend on $\bm{\kappa}$, that is,
\begin{align}
{\mathbb B}_s  = \frac{\partial \bm{P}_s }{\partial  \bm{\kappa}} \Big|_{(\bar{\bm{F}}_s, \tilde{\bm{\kappa}})}= 
\underbrace{2 {g}_{\nu\alpha} \frac{\partial^2 \bar{\varPsi}}{\partial {g}_{\beta\nu} \partial {\kappa}_{\gamma \delta}}}_{{{\mathbb B}_s}^{\  \beta \gamma \delta}_{\alpha\ \ }}\bm{g}^\alpha  \otimes \bm{G}_\beta \otimes \bm{G}_\gamma  \otimes \bm{G}_\delta.
\end{align}
In contrast, the derivative of $\bm{P}_s$ with respect to $\bm{F}_s$ also accounts for the basis variations. Using the derivative \(\partial\bm g_{\alpha}/\partial \bm F_s = \bm I\otimes\bm G_{\alpha}\) gives
\begin{align}
\begin{split}
{\mathbb A}_s & = \frac{\partial \bm{P}_s }{\partial  \bm{F}_s}\Big|_{(\bar{\bm{F}}_s, \bar{\bm{\kappa}})} = 2
\frac{\partial  }{\partial  \bm{F}_s} \Big( \frac{\partial \tilde{\varPsi}}{\partial {g}_{\alpha\beta}} \bm{g}_\alpha \otimes \bm{G}_\beta \Big)= 
4 \frac{\partial ^2\tilde{\varPsi}}{\partial {g}_{\alpha\beta} \partial {g}_{\gamma\delta}} \bm{g}_\alpha \otimes \bm{G}_\beta \otimes \bm{g}_\gamma \otimes \bm{G}_\delta + 
2 \frac{\partial \tilde{\varPsi}}{\partial {g}_{\alpha\beta}} 
\frac{\partial  \left( \bm{g}_\alpha \otimes \bm{G}_\beta \right)}{\partial  \bm{F}_s} \\
&=
4 g_{\alpha\omega} g_{\gamma\nu} \frac{\partial ^2\tilde{\varPsi}}{\partial {g}_{\alpha\beta} \partial {g}_{\gamma\delta}} \bm{g}^\omega \otimes \bm{G}_\beta \otimes \bm{g}^\nu \otimes \bm{G}_\delta + 
2 \frac{\partial \tilde{\varPsi}}{\partial {g}_{\alpha\beta}} 
\left( \bm{g}_\xi \otimes \bm{G}_\beta \otimes \bm{g}^\xi +   \bm{n} \otimes \bm{G}_\beta \otimes \bm{n}\right)\otimes \bm{G}_\alpha 
\\ 
&=
\underbrace{\Big(4 g_{\alpha\omega} g_{\gamma\nu} \frac{\partial ^2\tilde{\varPsi}}{\partial {g}_{\alpha\beta} \partial {g}_{\gamma\delta}} +2 {g}_{\nu\omega} \frac{\partial \tilde{\varPsi}}{\partial {g}_{\delta\beta}} \Big)}_{{{\mathbb A}_s}^{\ \beta \ \delta}_{\omega \ \nu}} \bm{g}^\omega \otimes \bm{G}_\beta \otimes \bm{g}^\nu \otimes \bm{G}_\delta + 
\underbrace{2\frac{1}{{g^{33}}} \frac{\partial \tilde{\varPsi}}{\partial {g}_{\alpha\beta}} }_{{{\mathbb A}_s}_{3 \ 3}^{\ \beta \ \alpha}}
  \bm{g}^3 \otimes \bm{G}_\beta \otimes \bm{g}^3\otimes \bm{G}_\alpha .
\end{split}
\end{align}
With the moduli at hand, the incremental surface stress ${\bm{P}}_s^{[1]}$ and moment ${\bm{M}}_s^{[1]}$ can be written as
\begin{align}
\begin{split}\label{eq:t0}
{\bm{P}}_s^{[1]}&= {\mathbb A}_s
   : \big( {\bm{\eta}_s}  \bar{{\bm{F}}}_s \big)
 + {\mathbb B}_s: \big( \bar{\bm{F}}_s^{T}  \bm{\rho}_s  \bar{\bm{F}}_s \big)  \\
 & =
 \big( {{\mathbb A}_s}^{\ \beta \ \omega}_{\alpha \ \gamma} {\eta_s}^\gamma_{\ \omega}  + {{\mathbb B}_s}^{\  \beta \gamma \delta}_{\alpha\ \ }{{\rho}_s}_{\gamma \delta} \big) \bm{g}^\alpha \otimes\bm{G}_\beta + 2 \frac{1}{{g^{33}}}\frac{\partial \tilde{\varPsi}}{\partial {g}_{\alpha\beta}} {\eta_s} ^3_{\ \alpha}
  \bm{g}^3  \otimes \bm{G}_\beta 
  \\
& =
\Big[ {{\mathbb A}_s}^{\ \beta \ \omega}_{\alpha \ \gamma} {\eta_s}^\gamma_{\ \omega}  - {{\mathbb B}_s}^{\  \beta \gamma \delta}_{\alpha\ \ }\Big(\frac{{\eta_s} ^3_{\ \gamma,\delta}  }{\sqrt{g^{33}}}  -\frac{{\eta_s} ^3_{\ \omega}  }{\sqrt{g^{33}}}
\bar{\Gamma}^\omega_{\ \gamma \delta }+{\eta_s} ^\nu_{\ \gamma}  
\bar{{b}}_{\nu \delta} \Big)\Big] \bm{g}^\alpha \otimes\bm{G}_\beta +2 \frac{1}{{g^{33}}}\frac{\partial \tilde{\varPsi}}{\partial {g}_{\alpha\beta}} {\eta_s} ^3_{\ \alpha}
  \bm{g}^3  \otimes \bm{G}_\beta,
 \end{split}\\
\begin{split}\label{eq:t1}
{\bm{M}}_s^{[1]}& = {\mathbb C}_s
: \big( {\bm{\eta}}_s  \bar{{\bm{F}}}_s \big)
+ {\mathbb D}_s
: \big( \bar{{\bm{F}}}_s^{T}  {\bm{\rho}_s}  \bar{{\bm{F}}}_s \big)\\ 
&=
\Big({{\mathbb C}_s}^{\alpha \beta \ \delta}_{\ \ \gamma} {{\eta}_s}_{\ \delta}^\gamma+
{\frac{\partial ^2\tilde{\varPsi}}{\partial {\kappa}_{\alpha\beta} \partial {\kappa}_{\omega \nu}}}  {{\rho}_s}_{\omega \nu} \Big) \bm{G}_\alpha \otimes\bm{G}_\beta   \\
&=
\Big[{{\mathbb C}_s}^{\alpha \beta \ \delta}_{\ \ \gamma} {{\eta}_s}_{\ \delta}^\gamma-
{\frac{\partial ^2\tilde{\varPsi}}{\partial {\kappa}_{\alpha\beta} \partial {\kappa}_{\omega \nu}}}  \Big(\frac{{{\eta}_s} ^3_{\ \omega,\nu}  }{\sqrt{g^{33}}}  -\frac{{{\eta}_s} ^3_{\ \gamma}  }{\sqrt{g^{33}}}
\bar{\Gamma}^\gamma_{\ \omega \nu }+{{\eta}_s} ^\delta_{\ \omega}  
\bar{{b}}_{\delta \nu} \Big) \Big] \bm{G}_\alpha \otimes\bm{G}_\beta.
\end{split}
\end{align}

Note that for isotropic surface materials, the strain energy density ${\varPsi}$ can be expressed as a function of scalar invariants constructed from the surface right Cauchy--Green tensor ${\bm{C}}_s = {\bm{F}}_s^{\!T}{\bm{F}}_s$ and the curvature tensor ${\bm{\kappa}}$.  
The complete set of independent invariants typically includes the classical invariants of ${\bm{C}}_s$, the curvature invariants of ${\bm{\kappa}}$, and a family of mixed invariants that couple the two.  
These invariants provide a natural framework for constructing isotropic surface energy functions and for systematically capturing the effects of stretch--curvature coupling.  
A discussion of these invariants is provided in ~\ref{ap:formulas_iso}. It is also worth emphasizing that throughout this work, we treat the reference coordinates as fixed and independent of deformation.  
Alternatively, one may adopt a formulation in which both the reference and deformed coordinates evolve with deformation, for instance, such that they remain aligned with the principal directions of the deformation gradient tensor.  
In such a setting, and under the assumption of isotropy, the components of the stiffness tensors ${\mathbb A}_s$, ${\mathbb B}_s$, ${\mathbb C}_s$, and ${\mathbb D}_s$ can be expressed explicitly in terms of the principal stretches, principal curvatures, and the relative orientation of the eigenvectors of the deformation gradient $\bar{\bm{F}}_s$ and of the curvature tensor $\bar{\bm{\kappa}}$ since these tensors are in general not coaxial. The expressions written in terms of the principal stretches for the special case in which the principal axes of $\bar{\bm{F}}_s$ and $\bar{\bm{\kappa}}$ are aligned are also discussed in~\ref{ap:formulas_iso}.

We are now ready to express the individual terms of $\bm{Q}_s^{[1]}$ from \eqref{eq:Q_s^{[1]}} in surface-adapted coordinates. For convenience, we repeat the full expression
\begin{align}\label{eq:repQ_s^{[1]}}
\begin{split}
\bm{Q}_s^{[1]}\;=\;&\bm{P}_s^{[1]}-\bar{\bm{b}} \bar{\bm{F}}_s \bm{M}_s^{[1]}-\bm{\omega}_s \bar{\bm{F}}_s\,\bar{\bm{M}}_s -\bm{\eta}_s^T \bar{\bm{n}} \otimes \bar{\bm{F}}_s^{-1}\bar{\bm{l}} -\bar{\bm{n}} \otimes \bar{\bm{F}}_s^{-1}\bm{\xi}_s \bar{\bm{l}}+\bar{\bm{n}} \otimes \bar{\bm{F}}_s^{-1}\Div_s\big(\bm{\eta}_s \, \bar{\bm{F}}_s\,\bar{\bm M}_s + \bar{\bm{F}}_s \bm{M}_s^{[1]}\big),
\end{split}
\end{align}
where, for further convenience, we have introduced the auxiliary vector
\begin{align}
\bar{\bm{l}} \;=\; \Div_s(\bar{\bm{F}}_s\bar{\bm{M}}_s).
\end{align}
Having already derived the coordinate representation of the first term in \eqref{eq:repQ_s^{[1]}}, we now proceed to evaluate the remaining contributions. 
The next two terms in \eqref{eq:repQ_s^{[1]}} read
\begin{align}
\begin{split}\label{eq:Q1_l21}
&-\bar{\bm{b}}{\bar{\bm{F}}}_s{\bm{M}}_s^{[1]} =  {\kappa}_{\xi \alpha } 
\Big[{{\mathbb C}_s}^{\alpha \beta \ \delta}_{\ \ \gamma} {{\eta}_s}_{\ \delta}^\gamma-
{\frac{\partial ^2\tilde{\varPsi}}{\partial \bm{\kappa}_{\alpha\beta} \partial \bm{\kappa}_{\omega \nu}}}  \Big(\frac{{{\eta}_s} ^3_{\ \omega,\nu}  }{\sqrt{g^{33}}}  -\frac{{{\eta}_s} ^3_{\ \gamma} }{\sqrt{g^{33}}} 
\bar{\Gamma}^\gamma_{\ \omega \nu }+{{\eta}_s} ^\delta_{\ \omega}  
\bar{{b}}_{\delta \nu} \Big) \Big] \bm{g}^\xi \otimes\bm{G}_\beta,\\
&-{\bm{\omega}_s}{\bar{\bm{F}}}_s \bar{\bm{M}}_s=-  \bar{M}_s^{\beta\delta} \Big[  \Big(\Big(\frac{{\eta_s}^3_{\ \alpha}}{\sqrt{g^{33}}}\Big)_{,\beta}- \frac{{{\eta}_s}^3_{\ \gamma} }{\sqrt{g^{33}}}\bar{\Gamma}^{\gamma}_{\ \alpha\beta} \Big)\bm{g}^\alpha\otimes \bm{G}_\delta +\frac{{{\eta}_s}^3_{\ \alpha} }{{g^{33}}} \bar{b}^\alpha_\beta \bm{g}^3\otimes \bm{G}_\delta \Big].
\end{split}
\end{align}

We next address the two terms involving $\bar{\bm{l}}$ in \eqref{eq:repQ_s^{[1]}}, namely $-\bm{\eta}_s^T \bar{\bm{n}} \otimes \bar{\bm{F}}_s^{-1}\bar{\bm{l}}$ and $-\bar{\bm{n}} \otimes \bar{\bm{F}}_s^{-1}\bm{\xi}_s\bar{\bm{l}}$.  Setting $\bm{R}_s=\bar{\bm{F}}_s\bar{\bm{M}}_s
= \bar{\bm{M}}_s^{\alpha\beta}\,\bm{g}_\alpha\otimes\bm{G}_\beta$ in \eqref{eq:surfDivCoord}, we obtain
\begin{align}
\bar{\bm{l}} = \Div_s (\bar{\bm{F}}_s\bar{\bm{M}}_s)
&=
\Big(\frac{1}{\sqrt{G_s}}\big(\sqrt{G_s} \bar{{M}}_s^{\alpha\beta}\big)_{,\beta}
+ \bar{{M}}_s^{\gamma\delta}\bar{\Gamma}^\alpha_{\ \gamma\delta}\Big)\bm{g}_\alpha
+ \bar{{M}}_s^{\alpha\beta}\bar{b}_{\alpha\beta}\,\bar{\bm{n}}.
\end{align}
 Inserting the surface divergence obtained above and using
$\bm{\eta}_s^{T}\bar{\bm{n}}={\eta_s}^3_{\ \omega}/{\sqrt{g^{33}}}\,\bm{g}^\omega$
then produces the coordinate expression
\begin{align}\label{eq:Q1_l3}
- {{\bm{\eta}}^T_s} \bar{\bm{n} } \otimes {\bar{\bm{F}}}_s^{-1} \bar{\bm{l}} = -\frac{{{\eta}_s}^3_{\ \omega}}{\sqrt{g^{33}}}\Big(\frac{1}{\sqrt{G_s}}\big(\sqrt{G_s}\bar{M}_s^{\alpha\beta}\big)_{,\beta} + \bar{M}_s^{\gamma \delta} \bar{\Gamma}^\alpha_{\ \gamma \delta}\Big)\bm{g}^\omega \otimes  \bm{G}_\alpha.
\end{align}
Note also that the last term involving $\bar{\bm{F}}_s^{-1}\bar{\bm{n}}$ disappears,
as $\bar{\bm{F}}_s^{-1}$ operates solely within the tangent space of the surface, so its action on the normal vector is identically zero, that is, $\bar{\bm{F}}_s^{-1}\bar{\bm{n}} = \bm{0}$. For the second term, we compute 
\begin{align}
\begin{split}
{\bar{\bm{F}}}_s^{-1} {{\bm{\xi}}_s} =&{\bar{\bm{F}}}_s^{-1} (  \bar{\bm{i}}_s{{\bm{\eta}}_s} - {{\bm{\eta}}^T_s} \bm{n} \otimes\bm{n}  ) ={{\eta}_s}^\alpha_{\ \beta} \bm{G}_\alpha \otimes  \bm{g}^\beta - \frac{{{\eta}_s}^3_{\ \omega}}{\sqrt{g^{33}}}  {\bar{\bm{F}}}_s^{-1} \bm{g}^\omega \otimes  \bar{\bm{n} } = 
{{\eta}_s}^\alpha_{\ \beta} \bm{G}_\alpha \otimes  \bm{g}^\beta -  \frac{{{\eta}_s}^3_{\ \omega} }{{g^{33}}}g^{\alpha \omega} \bm{G}_\alpha \otimes \bm{g}^3.
\end{split}
\end{align}
It then follows that
\begin{align}
-\bar{\bm{n} }\otimes {\bar{\bm{F}}}_s^{-1} {{\bm{\xi}}_s} \bar{\bm{l}} = -\frac{1}{\sqrt{g^{33}}}\Big[{{\eta}_s}^\alpha_{\ \beta} \Big(\frac{1}{\sqrt{G_s}}\big(\sqrt{G_s}\bar{M}_s^{\beta\omega}\big)_{,\omega} + \bar{M}_s^{\gamma \delta} \bar{\Gamma}^\beta_{\ \gamma \delta}\Big) -\frac{{{\eta}_s}^3_{\ \omega}}{\sqrt{g^{33}}} g^{\alpha \omega}\bar{M}_s^{\gamma\beta}\bar{b}_{\gamma\beta} \Big]\bm{g}^3\otimes  \bm{G}_\alpha.
\label{eq:Q1_l5}
\end{align}

For the last term in \eqref{eq:repQ_s^{[1]}}, we apply \eqref{eq:surfDivCoord} with
$\bm{R}_s=\bm{\eta}_s \bar{\bm{F}}_s \bar{\bm{M}}_s
+ \bar{\bm{F}}_s \bm{M}^{[1]}$,
and expand it term by term in the same fashion. 
Since
\bea
{{\bm{\eta}}_s}{\bar{\bm{F}}}_s \bar{\bm{M}}_s = {{\eta}_s}^i_{ \ \alpha} \bar{M}_s^{\alpha\beta}\bm{g}_i \otimes\bm{G}_\beta,\qquad \bar{\bm{F}}_s\bm{M}^{[1]}=\bar{M}^{[1]\alpha\beta}\bm{g}_\alpha\otimes \bm{G}_\beta,
\eea
the surface divergence of the  two contributions becomes
\begin{align}
&\Div_s \left({\bm{\eta}_s}{\bar{\bm{F}}}_s \bar{\bm{M}}_s\right) =\frac{1}{\sqrt{G_s}}\big(\sqrt{G_s}{{\eta}_s}^i_{ \ \alpha} \bar{M}_s^{\alpha\beta}  \big)_{,\beta} \bm{g}_i +  {{\eta}_s}^i_{ \ \omega} \bar{M}_s^{\omega\beta} \bar{\Gamma}^\alpha_{\ i \beta} \bm{g}_\alpha + {{\eta}_s}^i_{ \ \omega} \bar{M}_s^{\omega\beta} \delta_i^{\omega} \bar{b}_{\omega \beta} \bar{\bm{n}},\\
&\Div_s \big(\bar{\bm{F}}_s{\bm{M}}^{[1]}_s\big)
=
\Big(\frac{1}{\sqrt{G_s}}\big(\sqrt{G_s}{{M}}^{[1]\alpha\beta}_s\big)_{,\beta}
+ {{M}}^{[1]\gamma\delta}_s\bar{\Gamma}^\alpha_{\ \gamma\delta}\Big)\bm{g}_\alpha
+ {{M}}^{[1]\alpha\beta}_s\bar{b}_{\alpha\beta}\,\bar{\bm{n}}.
\end{align}
Hence, the last term reads
\begin{align}\label{eq:Q1_l42}
\begin{split}
\bar{\bm{n}}\otimes  \bar{\bm{F}}_s^{-1}\Div_s\big(\bm{\eta}_s \, \bar{\bm{F}}_s\,\bar{\bm M}_s + \bar{\bm{F}}_s \bm{M}_s^{[1]}\big)=&\frac{1}{\sqrt{g^{33}}} 
\Big(\frac{1}{\sqrt{G_s}}\big(\sqrt{G_s}{{\eta}_s}^\alpha_{ \ \omega} \bar{M}_s^{\omega\beta}  \big)_{,\beta} +  {{\eta}_s}^i_{ \ \omega} \bar{M}_s^{\omega\beta} \bar{\Gamma}^\alpha_{\ i \beta}\\
&+\frac{1}{\sqrt{G_s}}\big(\sqrt{G_s}{{{M}}_s^{[1]}}^{\alpha\beta} \big)_{,\beta} + {{{M}}_s^{[1]}}^{\gamma \delta} \bar{\Gamma}^\alpha_{\ \gamma \delta} \Big){\bm{g}^3 }\otimes \bm{G}_\alpha.
\end{split}
\end{align}
Consequently, adding together the formulas \eqref{eq:t0}, \eqref{eq:Q1_l21}, \eqref{eq:Q1_l3}, \eqref{eq:Q1_l5} and \eqref{eq:Q1_l42} yields the complete coordinate formulation of the incremental stress $\bm{Q}_s^{[1]}$ given in \eqref{eq:Q_s^{[1]}}.

Accordingly, the incremental boundary condition \eqref{eq13} can be expressed in coordinate form as
\begin{align}
&\frac{1}{\sqrt{G_s}}\big(\sqrt{G_s} {{Q}_s^{[1]}}_{\alpha}^{\ \beta}  \big)_{,\beta} -  {{Q}_s^{[1]}}_{i}^{\ \beta} \bar{\Gamma}^i_{\ \beta \alpha}-\frac{1}{\sqrt{G^{33}}} {{P}^{[1]}}_\alpha^{\ 3} =0\hspace{4.3em} \text{on}\ S,\\
&\frac{1}{\sqrt{G_s}}\big(\sqrt{G_s} {{Q}_s^{[1]}}_{3}^{\ \beta}  \big)_{,\beta}+\frac{1}{\sqrt{g^{33}}} {{Q}_s^{[1]}}_{\beta}^{\ \alpha} \bar{b}^{\beta}_{\ \alpha}-\frac{1}{\sqrt{G^{33}}} {{P}^{[1]}}_3^{\ 3}=0 \qquad \text{on}\ S. \label{eq:surf_bs_coord}
\end{align}
where the coordinates of the bulk incremental stress $\bm{P}^{[1]}$ are taken from equation \eqref{eq:PbulkIncr_clean}.

Having the coordinate form of the incremental stress tensor $\bm{Q}_s^{[1]} = {{Q}_s^{[1]}}_i^{\ \alpha} \bm{g}^i \otimes \bm{G}_{\alpha}$, the components of the \emph{actual} incremental surface stress tensor $\bm{\chi}_s = {{\chi}_s}_i^{\ \alpha} \bm{g}^i \otimes \bm{g}_{\alpha}$ follows from the transformation \eqref{eq:icr} as
\begin{align}\label{eq:chi_coords_surf}
{\chi_s}_i^{\ \alpha} =\bar{J}_s^{-1} {{{Q}}_s^{[1]}}_i^{\ \alpha}
\end{align}
with $\bar{J}_s=\sqrt{g_s/G_s}$.
Thus, the incremental boundary condition expressed in terms of the actual incremental stresses \eqref{eq:inc4} reads
\begin{align}
&\frac{1}{\sqrt{g_s}}\big(\sqrt{g_s} {\chi_s}_{\alpha}^{\ \beta}  \big)_{,\beta}  -  {\chi_s}_{i}^{\ \beta} \bar{\Gamma}^i_{\ \beta\alpha} -\frac{1}{\sqrt{g^{33}}} {\chi}_\alpha^{\ 3} = 0\hspace{4.3em} \text{on}\ \bm{\phi}_e(S), \label{eq:ibcc1}\\
&\frac{1}{\sqrt{g_s}}\big(\sqrt{g_s} {\chi_s}_{3}^{\ \beta}  \big)_{,\beta}  +\frac{1}{\sqrt{g^{33}}}{\chi_s}_{\beta}^{\ \alpha} \bar{b}^{\beta}_{\ \alpha} -\frac{1}{\sqrt{g^{33}}} {\chi}_3^{\ 3} = 0\qquad \text{on}\ \bm{\phi}_e(S). 
\label{eq:surf_bs_coord2}
\end{align}

Now let us focus on the remaining incremental boundary conditions \eqref{eq14} and \eqref{eq15} on $\partial S$. The unit boundary curve normal is expressed by
\begin{align}
\bm{V} = V_\alpha \bm{G}^\alpha ,
\end{align}
where its components are obtain by $V_\alpha = \bm{V} \cdot \bm{G}_\alpha$. Consequently, the boundary condition \eqref{eq15} reads
\begin{align}
{M_s^{[1]}}^{\alpha \beta} V_\alpha V_\beta  = 0 \qquad \text{on } \partial S.
\end{align}

In order to express boundary condition \eqref{eq14} in coordinates, it is useful to introduce a unit tangent $\bm{T} = T^\alpha \bm{G}_\alpha=T_\alpha \bm{G}^\alpha$ to the boundary curve $\partial S$. Subsequently, the projector on the tangent space of $\partial S$ reads $\bm{I}_\parallel = \bm{T} \otimes \bm{T}$. Let $s$ be the arclength variable of the curve $\partial S$. Then for a tensor field defined on $\partial S$ satisfying $\bm{R}_{\parallel}  \bm{I}_\parallel  = \bm{R}_{\parallel}$, its divergence on the boundary curve \eqref{eq:divc} can be written as
\bea
\Div_{\parallel} \bm{R}_{\parallel}  =( \bm{R}_{\parallel} \cdot \bm{T})_{,s}.
\eea
 Note that this formula can also be obtained from the surface divergence formula \eqref{eq:surfDivCoord}, realizing that $\beta=1$ and using $\theta^1=s$ and thus $\bm{G}^1=\bm{T}$. Choosing 
$\bm{R}_{\parallel} = \bar{\bm{n}}\otimes{\bm{M}}^{[1]}_s\bm{V}+{\bm{n}}^{[1]}\otimes\bar{\bm{M}}_s\bm{V}$ we have
\bea
\bm{R}_{\parallel} \cdot \bm{T} = ( \bm{T} \cdot {\bm{M}}^{[1]}_s\bm{V}) \bar{\bm{n}}  - ( \bm{T}\cdot\bar{\bm{M}}_s\bm{V}) \bm{\eta}_s^T{\bm{n}},
\eea
and equation \eqref{eq14} yields
\bea
{{{Q}}^{[1]}_s}_i{}^{ \alpha}{V}_{\alpha} \bm{g}^i
+ \Big( \frac{{{{M}}^{[1]}_s}^{\alpha\beta} T_\alpha {V}_\beta}{\sqrt{g^{33}}} \bm{g}^3   - {{\eta}_s}^{3}{}_{\!\gamma}\,\frac{{\bar{{M}}_s}^{\alpha\beta}T_\alpha{V}_\beta}{\sqrt{g^{33}}}\,\bm{g}^{\,\gamma} \Big)_{,s} = \bm{0}  \ &&\text{on } \partial S.\label{eq:bc_dS_coordsA}
\eea
The derivative with respect to arc length on $\partial S$ can be further recast in terms of derivatives with respect to surface coordinates $\theta^\alpha = \theta^\alpha(s)$ using the directional derivative relation $\frac{\partial (\cdot)}{\partial s}=\frac{\partial (\cdot)}{\partial\theta^\alpha} T^\alpha$. Consequently,  equation \eqref{eq:bc_dS_coordsA} is split into in-plane part and out-of-plane part components as
\begin{align}
&{{{Q}}^{[1]}_s}_\alpha{}^{ \beta}{V}_{\beta} 
- 
\frac{{{{M}}^{[1]}_s}^{\beta\gamma} T_\beta {V}_\gamma }{\sqrt{g^{33}}} \bar{\Gamma}^3_{\ \alpha \delta} T^\delta-
\Big({{\eta}_s}^{3}{}_{\!\alpha}\,\frac{{\bar{{M}}_s}^{\beta\gamma}T_\beta{V}_\gamma}{\sqrt{g^{33}}}\Big)_{,\nu} T^{\nu}
+
{{\eta}_s}^{3}{}_{\!\delta}\,\frac{{\bar{{M}}_s}^{\beta\gamma}T_\beta{V}_\gamma }{\sqrt{g^{33}}}\bar{\Gamma}^{\delta}_{\alpha \nu } T^\nu
= 0 \qquad \text{on} \ \partial S,\\
& {{{Q}}^{[1]}_s}_3{}^{ \alpha}{V}_{\alpha} +\Big( \frac{{{{M}}^{[1]}_s}^{\alpha\beta} T_\alpha {V}_\beta}{\sqrt{g^{33}}} \Big)_{,\gamma} T^\gamma 
- 
{{\eta}_s}^{3}{}_{\!\gamma}\,\frac{{\bar{{M}}_s}^{\alpha\beta}T_\alpha{V}_\beta}{g^{33}} \bar{b}^\gamma_\nu T^\nu 
=0 \hspace{11.7em} \text{on} \ \partial S.
\end{align}

For completeness, we also provide coordinate formulation for the actual version of the boundary conditions \eqref{eq:edge1_spatial} and \eqref{eq:edge2_spatial}. However, since the derivation is analogous, we only state the final result. equation \eqref{eq:edge1_spatial} is again decomposed into in-plane and out-of-plane components and reads 
\begin{align}
&{{{\chi}}^{[1]}_s}_\alpha{}^{ \beta}\bar{v}_{\beta} 
- 
\frac{{{{\tau}}^{[1]}_s}^{\beta\gamma} \bar{t}_\beta \bar{v}_\gamma }{\sqrt{g^{33}}} \bar{\Gamma}^3_{\ \alpha \delta} \bar{t}^\delta -
\Big({{\eta}_s}^{3}{}_{\!\alpha}\,\frac{{\bar{{\tau}}_s}^{\beta\gamma}\bar{t}_\beta\bar{v}_\gamma}{\sqrt{g^{33}}}\Big)_{,\nu} \bar{t}^{\nu}
+
{{\eta}_s}^{3}{}_{\!\delta}\,\frac{{\bar{{\tau}}_s}^{\beta\gamma}\bar{t}_\beta\bar{v}_\gamma }{\sqrt{g^{33}}}\bar{\Gamma}^{\delta}_{\ \alpha \nu } \bar{t}^\nu
=0\qquad \text{on} \ \bm{\phi}_e(\partial S),\\
& {{{\chi}}^{[1]}_s}_3{}^{ \alpha}\bar{v}_{\alpha} +\Big( \frac{{{{\tau}}^{[1]}_s}^{\alpha\beta} \bar{t}_\alpha \bar{v}_\beta}{\sqrt{g^{33}}} \Big)_{,\gamma} \bar{t}^\gamma 
- 
{{\eta}_s}^{3}{}_{\!\gamma}\,\frac{{\bar{{\tau}}_s}^{\alpha\beta}\bar{t}_\alpha\bar{v}_\beta  }{g^{33}} \bar{b}^\gamma_\nu   \bar{t}^\nu
 =0\hspace{10.7em}\text{on} \ \bm{\phi}_e(\partial S),
\end{align}
where we expressed the actual unit normal and tangent to the boundary curve in coordinates as 
\begin{align}
\bar{\bm{v}} = \bar{v}_\alpha \bm{g}^\alpha, \qquad \bar{\bm{t}} = \bar{t}_\alpha \bm{g}^\alpha = \bar{t}^\alpha \bm{g}_\alpha.
\end{align}
The coordinate version of \eqref{eq:edge1_spatial} is immediately obtained as
\begin{align}
{\tau_s^{[1]}}^{\alpha \beta} \bar v_\alpha \bar  v_\beta  = 0 \qquad \text{on } \bm{\phi}_e(\partial S).
\end{align}

\section{Application to elasto-capillary beading of soft cylinders}\label{sec:app}

The capillarity-induced beading of soft cylinders has attracted significant research interest in recent years, owing to a persistent discrepancy between theoretical predictions and experimental observations. A linear bifurcation analysis assuming constant (liquid-like) surface tension predicts that the instability occurs at zero wavenumber \citep{taffetani2015beading,fu2016localized,lestringant2020one,xuan2017plateau,fu2021necking,bakiler2023surface,ciarletta2012peristaltic}, corresponding to localized necking and beading. This prediction, however, is not directly related to experimental observations, which typically exhibit a periodic beading pattern. The discrepancy has been attributed to the fact that surface tension in solids is not constant, but varies with surface stretch and surface curvature \citep{magni2025elastic,taffetani2024curvature}, stemming from the intrinsic differences between solids and liquids at the molecular level. In this section, we reexamine the periodic beading in soft cylinders using the general incremental theory derived in Section \ref{sec:incremental}. We show that a bifurcation at finite wavenumber can emerge when surface stretching and bending are taken into account. As a byproduct, we recover the results in \cite{taffetani2024curvature}, thereby validating our incremental theory and highlighting its effectiveness.

\subsection{Homogeneous deformations}
Let us consider an infinite, soft cylinder composed of an elastic material, coated with an elastic surface described by the classical surface-tension model augmented by stretching or bending resistance. The cylinder is uniformly stretched along its axis and subsequently loaded by surface stresses acting on its outer surface. Since the cylinder is assumed to be infinite and the surface loading is uniform in the axial direction, the geometry and data are translationally invariant in the axial coordinate \(z\). Consequently, the problem has no physical ends, and no boundary conditions need to be prescribed at the axial extremities. Well-posedness is ensured by admissibility conditions on the unbounded domain, requiring that all fields remain bounded and of finite energy as \(|z| \to \infty\). In practice, this requirement is satisfied by the adopted harmonic ansatz \(e^{ikz}\). The only boundary condition to be enforced is that on the lateral surface, where the traction is balanced by the surface stresses. Together, these conditions fully determine the solution, and no additional axial boundary conditions are necessary.

For a cylindrical geometry, it is natural to adopt cylindrical coordinates $\theta^{1} = \Theta$, $\theta^2 = Z$, and $\theta^3 = R$, where $\Theta$, $Z$, and $R$ denote the angular, longitudinal, and radial coordinates, respectively. The reference position vector can then be expressed in terms of the orthonormal cylindrical basis $(\bm{e}_\theta(\Theta),\bm{e}_z,\bm{e}_r(\Theta))$ as 
\bea
\bm{X} = Z \bm{e}_z + R \bm{e}_r.
\eea
Realizing that $\rd\bm{e}_\theta/\rd \Theta = -\bm{e}_r$ and $\rd\bm{e}_r/\rd \Theta = \bm{e}_\Theta$, the covariant and contravariant  base vectors in the reference configuration are given by 
\begin{align}\label{eq:G123}
\bm{G}_{1} = R \bm{e}_\theta=R^2\bm{G}^1,\qquad  \bm{G}_{2} = \bm{e}_z=\bm{G}^2, \qquad    \bm{G}_{3} = \bm{e}_r=\bm{G}^3.
\end{align}
 Therefore, the determinant of the metric coefficients reads $G=R^2$.

Assuming that the cylinder undergoes a homogeneous deformation, the deformation is fully characterized by the axial stretch $\lambda$ and azimuthal stretch $a$. The deformed position vector is therefore
\begin{align}
\bm{x} =z\bm{e}_z + r\bm{e}_r
\end{align}
with the deformed cylindrical coordinates  given by
\begin{align}\label{eq:hom}
 z=\lambda Z,\qquad r=aR.
\end{align}
 This yields the covariant and  contravariant base vectors in the deformed configuration as
\begin{align}\label{eq:g123}
\bm{g}_{1} = r \bm{e}_\theta=r^2\bm{g}^1, \qquad\bm{g}_{2} =\lambda \bm{e}_z=\lambda^2\bm{g}^2, \qquad  \bm{g}_{3} =a \bm{e}_r=a^2\bm{g}^3.
\end{align} 
The corresponding nonzero metric coefficients and metric determinant are
\begin{align}
g_{11}=a^2 R^2,\qquad g_{22}=\lambda^2,\qquad g_{33}=a^2,\qquad g=a^4\lambda^2R^2.
\end{align}
The nonzero Christoffel symbols are then 
\[\bar{\Gamma}_{31}^1=\bar{\Gamma}_{13}^1 = 1/R, \qquad \bar{\Gamma}_{11}^3=-R.\]

Due to the choice of adapted coordinates, points on the surface are obtained by evaluating the bulk position at $R=A$, i.e., 
\begin{align}
\bm{X}_s=\bm{X}|_{R=A}=Z\bm{e}_z+A\bm{e}_r.
\end{align}
Consequently, the surface base vectors,  contravariant base vectors, metric coefficients, and Christoffel symbols follow directly from the bulk quantities evaluated at $R = A$. Hence, 
\begin{align}
\bm{G}_{1} = A \bm{e}_\theta=A^2\bm{G}^1, \qquad\bm{G}_{2} = \bm{e}_z=\bm{G}^2,
\end{align}
and $G_s = \det(G_{\alpha\beta})=  A$.
The same procedure applies to the corresponding quantities in the deformed configuration, and thus
\begin{align}
\bm{g}_{1} = aA \bm{e}_\theta=a^2A^2\bm{g}^1, \qquad \bm{g}_{2} = \lambda \bm{e}_z=\lambda^2 \bm{g}^2, 
\end{align}
The nonzero covariant and covariant metric coefficients and metric determinant are
\begin{align}
g_{11}=a^2A^2,\qquad g_{22}=\lambda^2,\qquad g^{11}=1/(a^2A^2),\qquad g^{22}=1/\lambda^2,\qquad g_s=\det(g_{\alpha\beta})=a^2A^2\lambda^2.
\end{align}
Note that all metric components are constant; therefore, their coordinate derivatives vanish. The reference and deformed normal vectors coincide and are given by
\begin{align}
\bm{N} = \bar{\bm{n}} = \bm{e}_r.
\end{align}
The nonzero Christoffel symbols for the surface are
\begin{align}
\bar{\Gamma}_{31}^1=\bar{\Gamma}_{13}^1 = 1/A, \qquad \bar{\Gamma}_{11}^3=-A.
\end{align}

 In view of the relations \eqref{eq:G123} and \eqref{eq:g123}, it is sufficient and more convenient to work with the orthonormal basis $(\bm{e}_\theta,\bm{e}_z,\bm{e}_r)$ rather than the curvilinear basis. This provides the advantage of clarifying the physical meanings of specific components and eliminates the need to distinguish between covariant and contravariant indices, since they coincide in an orthonormal basis. With this convention, the bulk and surface deformation gradients of the homogeneous deformation are then given by
 \begin{align}\label{eq:homgrad}
 &\bar{\bm{F}} =\bm{g}_i \otimes \bm{G}^i = a \bm{e}_\theta \otimes \bm{e}_\theta + \lambda \bm{e}_z \otimes \bm{e}_z  + a \bm{e}_r \otimes \bm{e}_r,\\ 
 &\bar{\bm{F}}_s = \bm{g}_\alpha \otimes \bm{G}^\alpha =a \bm{e}_\theta \otimes \bm{e}_\theta + \lambda \bm{e}_z \otimes \bm{e}_z.
 \end{align}
The curvature and relative curvature of the deformed surface read
\begin{align}
&\bar{\bm{b}}=-\bar{\nabla}_s\bar{\bm{n}}=-\frac{\partial\bar{\bm{n}}}{\partial\Theta}\otimes\bm{g}^1=-\frac{1}{aA}\bm{e}_\theta\otimes\bm{e}_\theta,\\
&\bar{\bm{\kappa}}=-\bar{\bm{F}}_s^T\bar{\bm{b}}\bar{\bm{F}}_s=\frac{a}{A}\bm{e}_\theta\otimes\bm{e}_\theta.
\end{align}
The projection onto the tangent space of the deformed surface is simply 
\begin{align}
\bar{\bm{i}}_s=\bm{I}-\bar{\bm{n}}\otimes\bar{\bm{n}}=\bm{e}_\theta\otimes\bm{e}_\theta+\bm{e}_z\otimes\bm{e}_z.
\end{align}

To maintain generality of the deformation,  we assume that the bulk solid is made of a compressible neo-Hookean hyperelastic material with the strain energy function \citep{bakiler2023surface,yu2025incremental}
\begin{align} \label{eq:bulkEn}
W=\frac{\mu}{2}(I_1-3-2\ln J)+ \frac{D}{2}\Big(\frac{J^2-1}{2}-\ln J\Big),
\end{align}
where $\mu$ and $D$ are the Lam\'{e} parameters,  $I_1=\tr(\bm{C})=G^{ij}g_{ij}$ is the first invariant of the right Cauchy-Green tensor $\bm{C}=\bm{F}^T\bm{F}$ and $J=\det(\bm{F})=\sqrt{g/G}$ is Jacobian determinant. The Poisson ration $\nu$ measuring the compressiblity is given by $\nu={D}/{(2(D+\mu))}$.
The incompressible deformation is recovered by letting the modulus parameter $D\to \infty$. For the homogeneous deformation \eqref{eq:hom},  the first Piola-Kirchhoff stress is obtained from the constitutive relation~\eqref{eq:coordP_bulk} as
\begin{align}\label{eq:Pbar}
\begin{split}
\bar{\bm{P}}&=\Big[\mu\Big(a-\frac{1}{a}\Big)+\frac{D}{2}\Big(a^3\lambda^2-\frac{1}{a}\Big)\Big]\bm{e}_\theta\otimes\bm{e}_\theta+\Big[\mu\Big(\lambda-\frac{1}{\lambda}\Big)+\frac{D}{2}\Big(a^4\lambda-\frac{1}{\lambda}\Big)\Big]\bm{e}_z\otimes\bm{e}_z\\
&\quad+\Big[\mu\Big(a-\frac{1}{a}\Big)+\frac{D}{2}\Big(a^3\lambda^2-\frac{1}{a}\Big)\Big]\bm{e}_r\otimes\bm{e}_r.
\end{split}
\end{align}

For the surface, we consider two classes of constitutive models that account for the surface stretching and bending resistance, respectively. The corresponding surface energy functions take the form
\begin{align}
&\varPsi=\gamma J_s+\frac{\alpha_s}{2}(J_s-1)^2\qquad\ \ \ \ \ (\text{stretching resistance}), \label{eq:suradd}\\
&\varPsi=\gamma J_s +\frac{\beta_s}{2}(H-H_0)^2J_s\qquad (\text{bending resistance}),  \label{eq:sur}
\end{align}
where $J_s=\sqrt{g_s/G_s}$ is the surface Jacobian determinant, $\gamma$ denotes the surface tension parameter, $\alpha_s$ and $\beta_s$ are the surface stretching and bending rigidity, respectively, $H=\frac{1}{2}\tr(\bm{b})$ is the mean curvature of the current surface, and $H_0$ is the spontaneous curvature. Here, multiplication by $J_s$ ensures that surface energy is measured with respect to the reference area. The first surface energy function \eqref{eq:suradd} has no dependence on the curvature, and can thus be analyzed using the framework developed in our previous work \citep{yu2025incremental}. Accordingly, we shall focus primarily on the second surface energy model \eqref{eq:sur}, also known as the Helfrich energy \citep{helfrich1973elastic}, while also presenting results for the first model for completeness.

Expressing the surface energy \eqref{eq:sur} directly in terms of the components of the metric tensor   and relative curvature yields
\bea \label{eq:psi_s}
\varPsi= \sqrt{\frac{{g_{11}g_{22}-g_{12}^2}}{G}}\Big[\gamma +\frac{\beta_s}{2}\Big(-\frac{1}{2} \kappa_{\alpha\beta}g^{\alpha\beta} -H_0\Big)^2\Big].
\eea
Substituting \eqref{eq:psi_s} into equations \eqref{eq:coordP} and \eqref{eq:coordM}, we obtain the surface stress and moment in the intermediate configuration
\begin{align}
&\bar{\bm{P}}_s = \Big[\gamma \lambda+\frac{\beta_s \lambda }{8 a^2 A^2}(4 a^2 A^2 H_0^2-4 a A H_0-3)\Big]\bm{e}_\theta\otimes\bm{e}_\theta+\Big[\gamma a+\frac{\beta_s}{8 a A^2}(2 a A H_0+1)^2\Big]\bm{e}_z\otimes\bm{e}_z,\label{eq:Psbar}\\
&\bar{\bm{M}}_s =\frac{\beta_s\lambda}{4 a^2 A}(2 a A H_0+1)\bm{e}_\theta\otimes\bm{e}_\theta+\frac{\beta_s}{4 \lambda A}(2 a AH_0+1)\bm{e}_z\otimes\bm{e}_z.\label{eq:Msbar}
\end{align}

The boundary condition \eqref{eq:SurfaceEqQ} in coordinate form is obtained analogously from only non-trivial equation \eqref{eq:surf_bs_coord}, which reads
\bea\label{eq:bccc}
\bar{P}_{33}= -\frac{1}{A}\Big(\bar{P}_{s11}+\frac{\bar{M}_{s11}}{A}\Big),
\eea
where the components are  referred  to  the ordered orthonormal basis $(\bm{e}_\theta,\bm{e}_z,\bm{e}_r)$.
Substituting \eqref{eq:Pbar}, \eqref{eq:Psbar} and \eqref{eq:Msbar} into \eqref{eq:bccc} yields a relation for stretches $a$ and $\lambda$:
\begin{align} \label{eq:alambda}
8 \mu a(a^2-1)A^3+4 D a (a^4\lambda^2-1)A^3+8\gamma \lambda a^2 A^2+\beta_s\lambda(4 a^2 A^2 H_0^2-1)=0.
\end{align}
By the implicit function theorem, this equation defines $a=a(\lambda)$ as a function of $\lambda$.  It is clear that, in the incompressible limit $D\to\infty$, this function simplifies to $a=\lambda^{-1/2}$. Finally, the resultant axial force is calculated as
\begin{align}
F_z \;&= \pi A^2 \bar{P}_{22} + 2 \pi A \bar{P}_{s22}= \pi A^2\Big( \mu \frac{\lambda^2-1}{\lambda}+\frac{D}{2}\frac{a^4\lambda^2-1}{\lambda}+\frac{2\gamma a}{A}+\frac{\beta_s}{4a A^3}(2 a A H_0+1)^2
\Big).
\end{align}

\subsection{Linear bifurcation analysis}
Having determined all the necessary quantities in the intermediate configuration, we now proceed to examine the incremental equations and analyze the onset of bifurcation.

Following the approach in Section \ref{sec:incremental}, we consider an axisymmetric perturbation of the form
\begin{align}
\bm{u}=v(z,r)\bm{e}_z+u(z,r)\bm{e}_r,
\end{align}
where the functions $v(z,r)$ and $u(z,r)$ describe the displacements in the longitudinal and radial directions. The incremental displacement gradient is then given by
\begin{align}
\bm{\eta}=&\bar{\nabla}_s\bm{u}=\frac{\partial \bm{u}}{\partial\theta^i}\otimes \bm{g}^i=\frac{u}{r}\bm{e}_\theta\otimes\bm{e}_\theta+\frac{\partial v}{\partial z}\bm{e}_z\otimes\bm{e}_z+\frac{\partial v}{\partial r}\bm{e}_z\otimes \bm{e}_r+\frac{\partial u}{\partial z}\bm{e}_r\otimes\bm{e}_z+\frac{\partial u}{\partial r}\bm{e}_r\otimes\bm{e}_r,
\end{align}
where the derivatives $\partial/{\partial\theta^i}$ are calculated using the chain rule and the relation \eqref{eq:hom}.

To obtain the components of the bulk incremental stress, we first substitute the particular energy form \eqref{eq:bulkEn} into equation \eqref{eq:bulk_stiff_coeff} to obtain the individual stiffness coefficients. These coefficients are then inserted into equations \eqref{eq:PbulkIncr_clean} and \eqref{eq:chi_coords_clean}, yielding the components of the incremental actual stress tensor $\bm{\chi}$ defined in \eqref{eq:icr}. Alternatively, since the bulk material is isotropic, these components can be computed via contracting the instantaneous modulus—expressed in terms of the principal stretches—with components of incremental displacement gradient  \citep{ogden1984non}. The nonzero components of the actual stress tensor $\bm{\chi}$ in the orthonormal basis, denoted by $\chi_{ij}$, are then given by
\begin{align}
\begin{split}\label{eq:chicomp}
&\chi_{11} = \frac{2\mu(a^2+1)+D(a^4\lambda^2+1)}{2a^2\lambda }\frac{u}{r}+ D a^2\lambda \Big(\frac{\partial u}{\partial r}+\frac{\partial v}{\partial z}\Big) ,\\ 
&\chi_{22} = \frac{2\mu(\lambda^2+1)+D(a^4\lambda^2+1)}{2a^2\lambda}\frac{\partial v}{\partial z}+Da^2\lambda\Big(\frac{\partial u}{\partial r}+\frac{u}{r}\Big),
\\ 
&\chi_{33} = \frac{2\mu(a^2+1)+D(a^4\lambda^2+1)}{2a^2\lambda}\frac{\partial u}{\partial r}+Da^2\lambda\Big(\frac{\partial v}{\partial z}+\frac{u}{r}\Big),
\\ 
&\chi_{23} = \frac{2\mu+D(1-a^4\lambda^2)}{2a^2\lambda}\frac{\partial u}{\partial z}+\frac{\mu}{\lambda}\frac{\partial v}{\partial r},
\\ 
&\chi_{32} = \frac{2\mu+D(1-a^4\lambda^2)}{2a^2\lambda}\frac{\partial v}{\partial r}+\frac{\mu\lambda }{a^2}\frac{\partial u}{\partial z}. 
\end{split}
\end{align}

Specializing the incremental equilibrium equations \eqref{eq:bulk_bs_coord_actual} to cylindrical coordinates, we obtain
\begin{align}
&\frac{\partial\chi_{22}}{\partial z}+\frac{\partial\chi_{23}}{\partial r}+\frac{\chi_{23}}{r}=0, \label{eq:iequi1_rev}\\
&\frac{\partial\chi_{32}}{\partial z}+\frac{\partial\chi_{33}}{\partial r}+\frac{\chi_{33}-\chi_{11}}{r}=0. \label{eq:iequi2_rev}
\end{align}
We then substitute the expressions for the incremental bulk stresses \eqref{eq:chicomp} into the equilibrium equations \eqref{eq:iequi1_rev} and \eqref{eq:iequi2_rev}, obtaining a system of  differential equations for the unknown displacement fields $v(z,r)$ and $u(z,r)$:
\begin{align}
\begin{split}\label{eq:equiuv1}
&\Big(2\mu(1+\lambda^2)+D(a^4\lambda^2+1)\Big)\frac{\partial^2 v}{\partial z^2}+2\mu a^2\frac{\partial^2 v}{\partial r^2}+\Big(2\mu+D(a^4\lambda^2+1)\Big)\frac{\partial^2 u}{\partial z\partial r}\\
&+\frac{2\mu a^2}{r}\frac{\partial v}{\partial r}
+\frac{2\mu+D(a^4\lambda^2+1)}{r}\frac{\partial u}{\partial z}= 0,
\end{split}
\\
&2 \mu \lambda^2 \frac{\partial^2 u}{\partial z^2}+
\Big(2\mu+D(a^4\lambda^2+1)\Big)\frac{\partial^2 v}{\partial z\partial r}+\Big(2\mu(a^2+1)+D(a^4\lambda^2+1)\Big)\Big(\frac{\partial^2 u}{\partial r^2}+\frac{1}{r}\frac{\partial u}{\partial r}-\frac{u}{r^2}\Big)=0. \label{eq:equiuv2}
\end{align}

Next, we analyze the surface deformation. The increment surface displacement gradient is obtained from its bulk counterpart as
\begin{align}
\bm{\eta}_s=\bm{\eta}|_{r=aA}\bar{\bm{i}}_s=\frac{u}{aA}\bm{e}_\theta\otimes\bm{e}_\theta+\frac{\partial v}{\partial z}\bm{e}_z\otimes\bm{e}_z+\frac{\partial u}{\partial z}\bm{e}_r\otimes\bm{e}_z,
\end{align}
where the displacement functions $u$ and $v$ are evaluated at $r=aA$. It follows that the intermediate quantities defined in \eqref{eq:kinematic-q}, which are used for calculating the first-order perturbation of relevant kinematic quantities, are given by
\begin{align}
&\bm{\omega}_s=\frac{\partial^2 u}{\partial z^2}\bm{e}_z\otimes\bm{e}_z, \label{eq:omegas}\\
&\bm{\xi}_s=\frac{u}{aA}\bm{e}_\theta\otimes\bm{e}_\theta+\frac{\partial v}{\partial z}\bm{e}_z\otimes\bm{e}_z-\frac{\partial u}{\partial z}\bm{e}_z\otimes\bm{e}_r, \label{eq:xis}\\
&\bm{\rho}_s=\frac{u}{a^2A^2}\bm{e}_\theta\otimes\bm{e}_\theta-\frac{\partial^2 u}{\partial z^2}\bm{e}_z\otimes\bm{e}_z. \label{eq:rhos}
\end{align}

To derive components of the incremental surface stress tensor, we may first substitute the specific form of the surface energy \eqref{eq:psi_s} into the general relations \eqref{eq:t0}, \eqref{eq:Q1_l21}, \eqref{eq:Q1_l3}, \eqref{eq:Q1_l5}, and \eqref{eq:Q1_l42}. The components of the incremental actual surface stress $\bm{\chi}_s$, defined in \eqref{eq:icr},  can then be obtained by means of equation \eqref{eq:chi_coords_surf}. This approach is straightforward but involves lengthy calculations. A more efficient method is to recognize that the principal directions of surface deformation tensor $\bm{C}_s=\bm{F}_s^T\bm{F}$ and relative curvature $\bm{\kappa}$ are aligned even for the axisymmetric deformation, as they both coincide with the $\bm{e}_\theta$ and $\bm{e}_z$ axes. Consequently, the surface energy \eqref{eq:sur} can be viewed as a function of the principal surface stretches $\lambda^s_1$, $\lambda^s_2$ and principal relative curvatures $\kappa_1$, $\kappa_2$, taking the form
 \begin{align}
 \varPsi=\varPsi^\text{p}(\lambda^s_1,\lambda_2^s,\kappa_1,\kappa_2)=\lambda_1^s\lambda^s_2\Big[\gamma+\frac{\beta_s}{2}\Big(\frac{\kappa_1}{2(\lambda^s_1)^2}+\frac{\kappa_2}{2(\lambda^s_2)^2}+H_0\Big)^2\Big].
 \end{align}
The simple expressions for the surface moduli given in \ref{ap:formulas_iso} (cf. \eqref{eq:moduli}) can then be employed to calculate the surface stress and moment measures $\bm{\sigma}_s$ and $\bm{m}_s$. Substituting these expressions together with \eqref{eq:omegas}--\eqref{eq:rhos} into \eqref{eq:chis}, we obtain the components $(\chi_{sij})$ of incremental actual surface stress tensor $\bm{\chi}_s$ in the orthonormal basis as follows:
\begin{align}
\begin{split}\label{eq:chiscomp}
&\chi_{s11}=\Big(\gamma+\frac{\beta_s}{8a^2A^2}(4H_0^2a^2A^2-1)\Big)\frac{\partial v}{\partial z}+\frac{\beta_s}{4a^3A^3}u-\frac{\beta_s H_0}{2}\frac{\partial^2 u}{\partial z^2} ,\\
&\chi_{s22}=\Big(\frac{\gamma}{a A}+\frac{\beta_s}{8a^3A^3}(4 H_0^2a^2A^2-1)\Big)u,\\
&\chi_{s32}=\Big(\gamma+\frac{\beta_s}{8a^2 A^2}(4H_0^2 a^2A^2+4H_0a A-1)\Big)\frac{\partial u}{\partial z}  -\frac{\beta_s}{4} \frac{\partial^3 u}{\partial z^3}. 
\end{split}
\end{align}
The boundary conditions on the surface are obtained from \eqref{eq:ibcc1} and \eqref{eq:surf_bs_coord2}, which  take  the form
\begin{align}
&\frac{\partial \chi_{s22 }  }{\partial z}  = \chi_{23} \hspace{5em} \text{at}\ r=aA \label{eq:bc_final1_rev},\\ 
&\frac{\partial \chi_{s32 }  }{\partial z}  -\frac{\chi_{s11}}{aA}  =  \chi_{33} \qquad \text{at}\ r=aA. \label{eq:bc_final2_rev}
\end{align}
Substituting the expressions for the incremental bulk and surface stress  \eqref{eq:chicomp} and \eqref{eq:chiscomp} into \eqref{eq:bc_final1_rev} and \eqref{eq:bc_final2_rev} yields two boundary conditions at $r=aA$ for the unknown displacements $u(z,r)$ and $v(z,r)$:
\begin{align}
&\frac{\mu}{\lambda}\frac{\partial v}{\partial r}+\Big(
\frac{2\mu+D(1-a^4\lambda^2)}{2a^2\lambda}-\frac{\gamma}{aA}
+\frac{\beta_s}{8a^3A^3}(1-4H_0^2a^2 A^2)\Big)\frac{\partial u}{\partial z}= 0\hspace{7.8em} \text{at}\ r=aA,\label{eq:bs_subs1} \\
\label{eq:bs_subs2}
\begin{split}
&\frac{\beta_s}{4}\frac{\partial^4 u}{\partial z^4}-\Big(\gamma+\frac{\beta_s}{8 a^2 A^2}(4 H_0^2 a^2 A^2+8 H_0 aA-1)\Big)\frac{\partial^2 u}{\partial z^2}+\Big(\frac{D\lambda a}{A}+\frac{\beta_s}{4 a^4 A^4}\Big)u\\
&+\frac{2\mu(a^2+1)+D(a^4\lambda^2+1)}{2a^2\lambda}\frac{\partial u}{\partial r}+\Big(D a^2\lambda+\frac{\gamma}{aA}+\frac{\beta_s}{8a^3 A^3}(4H_0^2a^2 A^2-1)\Big)\frac{\partial v}{\partial z}=0 \qquad \text{at}\ r=aA.
\end{split}
\end{align}
The partial differential equations \eqref{eq:equiuv1} and \eqref{eq:equiuv2}, supplemented by the boundary conditions \eqref{eq:bs_subs1} and \eqref{eq:bs_subs2}, constitute the governing equations for the linear bifurcation analysis.

To determine the critical wavenumber of the bifurcation, we look for a normal mode solution of the form
\begin{align}
u(r,z)=f(r)e^{\ri k z},\quad v(r,z)=g(r)e^{\ri k z},
\end{align}
where $k$ is the axial wavenumber. On substituting this solution into the incremental equilibrium equations  \eqref{eq:equiuv1}, \eqref{eq:equiuv2},  and boundary conditions \eqref{eq:bs_subs1}, \eqref{eq:bs_subs2}, and then eliminating  $g(r)$ in terms of $f(r)$, we obtain a linear boundary-value problem for $f(r)$:
\begin{align}
&\Big(\frac{\rd^2}{\rd r^2}+\frac{1}{r}\frac{\rd}{\rd r}-\Big(\frac{1}{r^2}+k^2 q_1^2\Big)\Big)\Big(\frac{\rd^2}{\rd r^2}+\frac{1}{r}\frac{\rd}{\rd r}-\Big(\frac{1}{r^2}+k^2q_2^2\Big)\Big)f(r)=0,\label{eq:diff1}\\
&f''(r)+\frac{1}{r}f'(r)+l_1(r) f(r)=0\hspace{7.5em}\text{at} \ r=a A,\label{eq:bcff1}\\
&f'''(r)+\frac{2}{r}f''(r)+l_2(r)f'(r)+l_3(r)f(r)=0\qquad \text{at} \ r=a A,\label{eq:bcff2}
\end{align}
where
\begin{align}
q_1=\frac{\lambda}{a},\quad q_2=\frac{2\mu(1+\lambda^2)+D(a^4\lambda^2+1)}{2\mu(1+a^2)+D(a^4\lambda^2+1)},
\end{align}
and the functions $l_1$, $l_2$, and $l_3$ are available but are too long to be presented here.

One can observe that the general solution to \eqref{eq:diff1} bounded at $r=0$ is given by
\begin{align}
f(r)=c_1 I_1(k q_1r)+c_2 I_1(k q_2 r),
\end{align}
where $I_1(x)$ denotes the modified Bessel function of the first kind, and $c_1$ and $c_2$ are arbitrary constants. By imposing the boundary conditions \eqref{eq:bcff1} and \eqref{eq:bcff2}, we obtain a system of homogeneous linear equations in the unknowns $c_1$ and $c_2$. A nontrivial solution exists only if the determinant of the coefficient matrix vanishes, which leads to the dispersion equation
\begin{align}\label{eq:Omega}
\Omega(k,a,\lambda)=0.
\end{align}
The complete expression of $\Omega$ is rather lengthy but is readily obtained using the symbolic computation software {\it Mathematica}. As a consistency check, we have verified that in the long-wavelength limit $k \to 0$, equation \eqref{eq:Omega} reduces to the limiting-point instability criterion ${\rd F_z}/{\rd\lambda} = 0$. This connection is expected, as proved in \cite{yu2022analytic}.  By eliminating the azimuthal stretch $a$ in \eqref{eq:Omega}  using  \eqref{eq:alambda}, we obtain a relation between $k$ and $\lambda$ that determines the dispersion relation of the bifurcation. The dispersion relation for the surface energy \eqref{eq:suradd}, which captures the surface stretching effects,  can be derived by a similar analysis; the details are omitted here for the sake of space.

For a numerical illustration, we adopt the parameter values $A=1$\ and $\mu=1$, which is equivalent to scaling all length variables by $A$ and stress variables by $\mu$. When the surface strain dependence is absent (i.e., $\alpha_s = \beta_s = 0$), the bifurcation occurs at zero wavenumber. However, by exploring the parameter space of $\gamma$, $\alpha_s$, $\beta_s$, $H_0$, and the bulk Poisson's ratio, it is found that bifurcation modes with nonzero wavenumber can emerge for suitable choices of these parameters. The corresponding bifurcation curves are shown in  Fig. \ref{fig:dispersion-curve} for the surface energy models capturing surface stretching and bending effects, respectively. It is observed that both surface stretching and bending resistance can induce bifurcation modes with a nonzero wavenumber. Therefore, further quantitative experiments are required to elucidate the wavenumber-selection mechanism underlying the periodic beading observed in previous studies.

\begin{figure}[h!]
	\centering
	\subfloat[]{\includegraphics[width=0.43\textwidth]{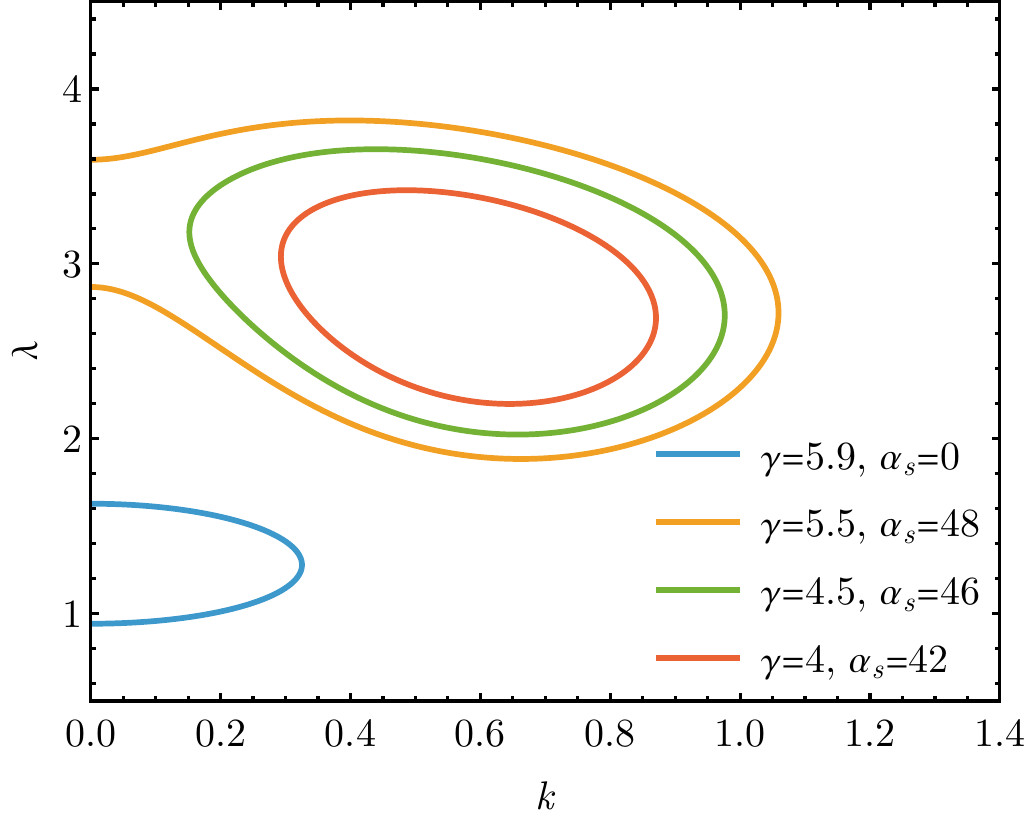}
	}\qquad
	\subfloat[]{\includegraphics[width=0.445\textwidth]{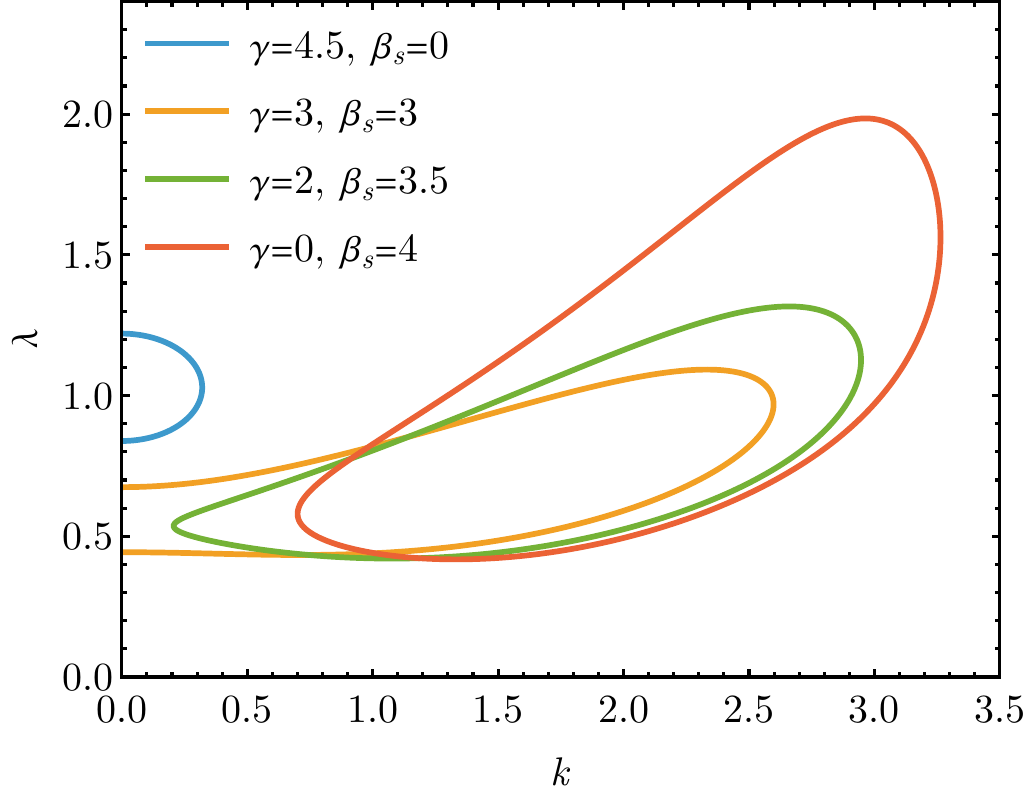}}
	\caption{(a) Bifurcation curve $\lambda$ versus $k$ at different values of $(\gamma,\alpha_s)$,  corresponding to the surface energy \eqref{eq:suradd} with stretching resistance. The cylinder is compressible with Poisson's ratio $\nu=0.49$. (b) Bifurcation curve $\lambda$ versus $k$ at different values of $(\gamma,\beta_s)$, corresponding to the surface energy \eqref{eq:sur} with bending resistance. The cylinder is compressible with Poisson's ratio $\nu=0.4$ and the spontaneous curvature $H_0=-2$. }
	\label{fig:dispersion-curve}
\end{figure}

\subsection{The incompressible limit $D\to\infty$}
In the incompressible limit $D \to \infty$, the expressions derived in the previous subsection simplify considerably. The corresponding incompressible quantities can be obtained by solving \eqref{eq:alambda} for $D$ and subsequently taking the limit $a \to \lambda^{-1/2}$. In this limit, the general dispersion equation \eqref{eq:Omega} simplifies to
\begin{align}\label{eq:bif}
\begin{split}
&-16\sqrt{\lambda}\Big(\lambda^3-1+2k\lambda\frac{I_0(k\lambda)}{I_1(k\lambda)}\Big)
+8k(\lambda^3+1)^2\frac{I_0(k\lambda^{-1/2})}{I_1(k\lambda^{-1/2})}+8\gamma\lambda(\lambda^3-1)(k^2-\lambda)\\
&+\beta_s \lambda(\lambda^3-1)(4H_0^2(k^2-\lambda)+8 H_0k^2\sqrt{\lambda}+2k^4+3\lambda^2-k^2\lambda)=0,
\end{split}
\end{align}
where we have set $A=1$ and $\mu=1$ to simplify the expression. It is straightforward to verify that the bifurcation condition \eqref{eq:bif} is identical to equation (32) in \cite{taffetani2024curvature} which was derived using stream-function formulation, upon identifying our parameters $(\gamma,\beta_s,H_0)$ with their notation $(l_{ec},l_{eb},-C)$.

To reproduce the results obtained by \citet{taffetani2024curvature}, we adopt the same input parameters. As an illustrative example, we set \(H_0 = -1.45\) and determine the dispersion relation for the bifurcation from \eqref{eq:bif}. By suitably balancing \(\gamma\) and \(\beta_s\), it is shown in Fig.~\ref{fig:dispersion-relation} that bifurcation modes with finite wavenumbers can appear as a result of the competition between bulk elasticity, surface tension, and bending stiffness. This reproduces the results presented in Fig.~1 of \citet{taffetani2024curvature}, providing quantitative validation of the present incremental theory. The reader is referred to that paper for a detailed discussion of the beading instability arising from curvature effects.

\begin{figure}[h!]
	\centering
	\includegraphics[width=0.5\linewidth]{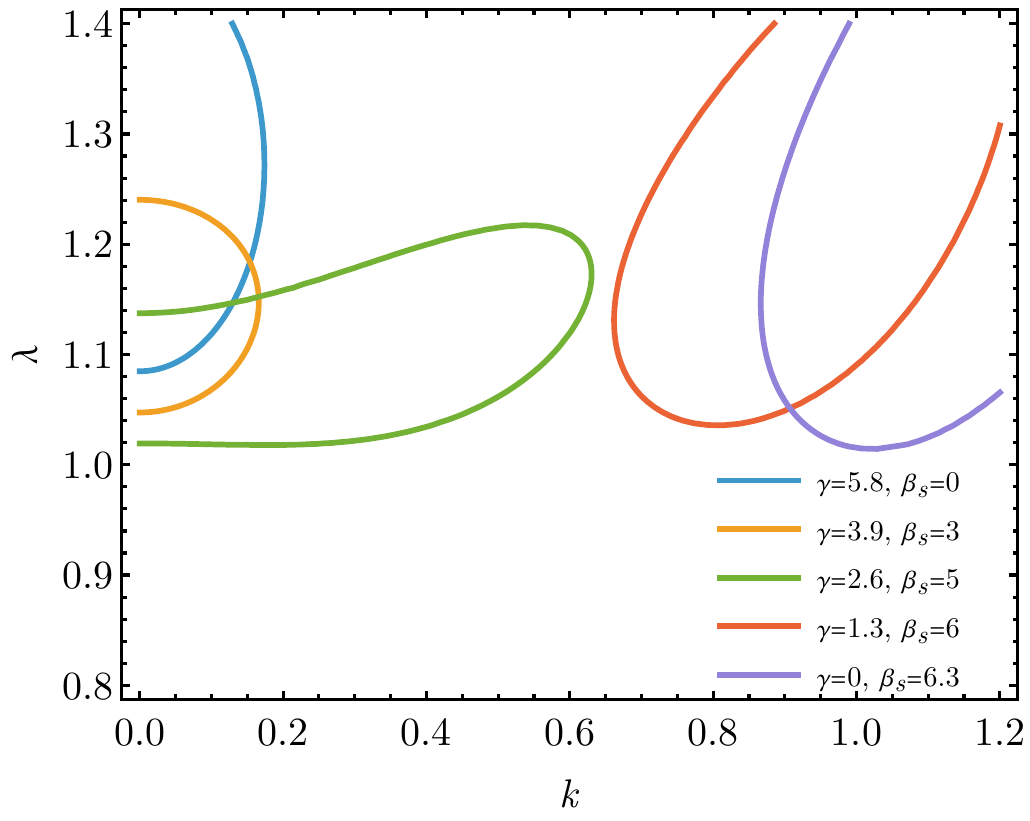}
	\caption{Bifurcation curve of $\lambda$ versus $k$ from \eqref{eq:bif} at different values of $(\gamma,\beta_s)$. The cylinder is incompressible and the spontaneous curvature  $H_0=-1.45$, consistent with \cite{taffetani2024curvature}.}
	\label{fig:dispersion-relation}
\end{figure}

\section{Conclusion}\label{sec:con}
Deriving surface equilibrium equations and their incremental counterparts 
directly from a variational principle is a challenging task when the surface 
possesses a general geometry. Such derivations typically require an explicit 
evaluation of surface curvature and its variation, demanding substantial 
familiarity with differential geometry, which often results in cumbersome 
expressions. The present work addresses this difficulty by establishing a 
compact and systematic incremental framework for surface–bulk systems that 
naturally incorporates curvature-dependent effects while remaining 
straightforward to use.

Building on a coordinate-free variational formulation, we derived the governing 
equations in a form that is both geometrically consistent and algebraically 
tractable. The resulting expressions can be written entirely in terms of abstract tensor notation, avoiding coordinate-dependent complexity. A key advantage of this formulation is that it involves only the curvature of the homogeneously deformed configuration, which can be determined directly from simple geometric considerations. As a result, the incremental equations are obtained without 
computing the curvature of the current (deformed) surface, thereby removing one 
of the major technical barriers that often hinder analytical or computational 
treatments of surface mechanics.

The presented formulation not only simplifies the derivation process but 
also enhances reusability and generality. Once the incremental equations are 
established, they can be readily applied to a broad class of elasto-capillary 
problems without rederiving the full variational equations. For example, the same set of incremental relations can be employed to study bifurcations in cylinders, spheres, or tori endowed with different surface energy laws, 
requiring only geometric specialization.

Therefore, the incremental framework developed here provides a unified setting for the stability analysis of elasto-capillary systems that account for surface curvature, enabling bifurcation problems to be addressed with the same clarity and efficiency as classical stability problems.
Extending our framework to higher orders will naturally enable the 
investigation of weakly and fully nonlinear regimes, including post-bifurcation 
behavior and pattern evolution in soft materials with surface energy.

\clearpage
\section*{Acknowledgements}
The work of Xiang Yu was supported by the National Natural Science Foundation of China (Grant No 12402068) and Guangdong Basic and Applied Basic Research Foundation (Grant No 2023A1515111141). Michal Šmejkal and Martin Horák gratefully acknowledge the financial support of the Czech Ministry of Education, Youth and Sports through the ERC CZ project SOFFA (No. LL2310), and M\v{S}MT-WTZ project 8J24AT004.

\appendix

\section{Constitutive laws of isotropic elastic surfaces} \label{ap:formulas_iso}

For isotropic surfaces, the surface energy $\varPsi$ can be based on the representation theorem expressed as a function of generalized principal invariants, see \cite{steigmann1999elastic}. Namely, it can be expressed as a function of six invariants of the right surface Cauchy-Green deformation tensor ${\bm{C}_s} = {\bm{F}^T_s}{\bm{F}_s}$ and the relative curvature tensor $\bm{\kappa}$ as
\begin{align}\label{eq:SE}
 \varPsi= \varPsi^\text{i}(I^s_1,I^s_2,I^s_3,I^s_4,I^s_5,I^s_6),
\end{align}
where the six invariants $I^s_1,I^s_2,\dots, I^s_6$ are defined by
\begin{align}
I^s_1=\tr(\bm{C}_s),\ I^s_2=\frac{1}{2}[ (\tr \bm{C}_s)^2 
-\tr(\bm{C}_s^2)],\ I^s_3=\tr(\bm{\kappa}),\ I^s_4=\frac{1}{2}[ (\tr\bm{\kappa})^2 
-\tr(\bm{\kappa}^2) ],\ I^s_5=\tr(\bm{C}_s\bm{\kappa}),\ I^s_6=\tr(\bm{C}_s\bm{\kappa}\bm{e}),
\end{align}
and $\bm{e}={e^{\alpha \beta}}\bm{G}_\alpha\otimes\bm{G}_\beta/\sqrt{G}$ is the two-dimensional permutation tensor. A direct calculation using \eqref{eq:const_PM} shows that the surface first Piola-Kirchhoff stress and moment tensor can be expressed as
\begin{align}
&\bm{P}_s=2\frac{\partial \varPsi^\text{i}}{\partial I_1^s}\bm{F}_s+2I_2^s\frac{\partial \varPsi^\text{i}}{\partial I_2^s}\bm{F}_s^{-T}+2\frac{\partial \varPsi^\text{i}}{\partial I_5^s}\bm{F}_s\bm{\kappa}+\frac{\partial \varPsi^\text{i}}{\partial I_6^s}\bm{F}_s(\bm{\kappa}\bm{e}-\bm{e}\bm{\kappa}),\label{eq:Ps}\\
&\bm{M}_s=\frac{\partial \varPsi^\text{i}}{\partial I_3^s}\bm{1}+\frac{\partial \varPsi^\text{i}}{\partial I_4^s}\bm{\kappa}^{*}+\frac{\partial \varPsi^\text{i}}{\partial I_5^s}\bm{C}_s+\frac{1}{2}\frac{\partial \varPsi^\text{i}}{\partial I_6^s}(\bm{e}\bm{C}_s-\bm{C}_s\bm{e}),\label{eq:Ms}
\end{align} 
where $\bm{\kappa}^*=\tr(\bm{\kappa})\bm{I}_s-\bm{\kappa}$ signifies the adjugate of $\bm{\kappa}$, and we have used the anti-symmetry of $\bm{e}$.

In the generic isotropic scenario, where the coupling terms $I_5^s$ and $I_6^s$ are present and $\bm{C}_s$ is not aligned with $\bm{\kappa}$ for general deformations (meaning their principal directions do not coincide), the first-order surface moduli can only be expressed in terms of the derivatives of the surface energy $\varPsi_s$ with respect to the six invariants.  In practical applications, the finite deformation $B_0\to B_e$ is typically homogeneous and both $\bm{C}_s$ and $\bm{\kappa}$ are aligned with the principal directions of this homogeneous deformation. In view of this, we shall assume that $\bm{C}_s$ and $\bm{\kappa}$ are aligned for the finite deformation $B_0\to B_e$, but may become misaligned once the incremental deformation is superimposed. 

For the finite homogeneous deformation  $B_0\to B_e$, let $\bar{\lambda}_1$ and $\bar{\lambda}_2$ denote the two principal stretches, whose principal directions coincide with those of $\bm{C}_s$. The corresponding  principal relative curvatures  are represented by $\bar{\kappa}_1$  and $\bar{\kappa}_2$. The six invariants of the surface associated with this finite deformation, denoted by $I_i$ ( $i=1,2,\dots,6$) with a slight abuse of notation, are given by
\begin{align}
\begin{split}
&I_1=\bar\lambda_1^2+\bar\lambda_2^2,\quad I_2=\bar\lambda_1^2\bar\lambda_2^2,\quad I_3=\bar{\kappa}_1+\bar{\kappa}_2,\quad I_4=\bar{\kappa}_1\bar{\kappa}_2,\quad I_5=\bar\lambda_1^2\bar{\kappa}_1+\bar\lambda_2^2\bar{\kappa}_2,\quad I_6=0.
\end{split}
\end{align}
Moreover let $\varPsi_i$ and $\varPsi_{ij}$ ($i,j=1,2,\dots, 6$) denote the derivatives of $\varPsi=\varPsi^\text{i}(I_1^s,I_2^s,I_3^s,I_4^s,I_5^s,I_6^s)$ evaluated at these values,
\begin{align}
\varPsi_i=\frac{\partial\varPsi^\text{i}}{\partial I_i^s}|_{(I_1,I_2,I_3,I_4,I_5,I_6)},\quad \varPsi_{ij}=\frac{\partial^2\varPsi^\text{i}}{\partial I_i^s\partial I_j^s}|_{(I_1,I_2,I_3,I_4,I_5,I_6)}.
\end{align}

As discussed in Section \ref{sec:incremental}, it is more convenient to write the equilibrium equations in the intermediate configuration. In this setting, the problem reduces to calculating the incremental actual stress tensor $\bm{\chi}_s=\bar{J}_s^{-1}\bm{Q}_s^{[1]}\bm{F}_s^T$, which can be recast  into the form (cf. \eqref{eq:chis})
\begin{align}
\bm{\chi}_s=\bm{\sigma}_s-\bar{\bm{b}}\bm{m}_s+\bar{\bm{n}}\otimes\bar{\bm{i}}_s\overline{\ddiv}_s(\bm{m}_s)+\bm{\theta}_s,
\end{align}
where $\bm{\sigma}_s$, $\bm{m}_s$ and $\bm{\theta}_s$ are defined in \eqref{eq:addIncrQuant} and \eqref{eq:thetas}. The quantity $\bm{\theta}_s$ is straightforward to calculate, as it is expressed directly in terms of the incremental surface deformation gradient. To compute the incremental surface stress and moment measures $\bm{\sigma}_s$ and $\bm{m}_s$, we recast them in the form
\bea
\bm{\sigma}_s=\mathcal{A}_s :{\bm{\eta}_s} + \mathcal{B}_s : {\bm{\rho}_s},\qquad \bm{m}_s= \mathcal{C}_s :{\bm{\eta}_s} + \mathcal{D}_s : {\bm{\rho}_s},
\eea
where we recall that $\bm{\rho}_s=-\bar{\bm{i}}_s\bar{\nabla}_s(\bm{\eta}_s^T\bar{\bm{n}})-\bm{\eta}_s^T\bar{\bm{b}}$ represents the increment of the relative curvature. In what follows,  the components of the surface stiffness tensors $\mathcal{A}_s$, $\mathcal{B}_s$, $\mathcal{C}_s$, and $\mathcal{D}_s$ are expressed with respect to an orthonormal basis $(\bm{e}_1, \bm{e}_2, \bm{e}_3)$ using the chain rule, where $\bm{e}_1$ and $\bm{e}_2$ coincide with the principal directions of $\bm{C}_s$ and $\bm{e}_3=\bm{n}$ is the unit outward normal.

The nonzero-components of the surface  stiffness tensor $\mathcal{A}_s$ are 
\begin{align}\label{eq:ap:As}
\begin{split}
&\bar{J}_s\mathcal{A}_{s\alpha\alpha\alpha\alpha}=2\bar{\lambda}_\alpha^2\varPsi_1+2I_2\varPsi_2+2\bar{\lambda}_\alpha^2\bar{\kappa}_\alpha\varPsi_5+4\bar{\lambda}_\alpha^4\varPsi_{11}+8\bar{\lambda}_\alpha^2 I_2\varPsi_{12}\\
&\hspace{4.5em}+8\bar{\lambda}^4_\alpha\bar{\kappa}_\alpha \varPsi_{15}+4 I_2^2\varPsi_{22}+8\bar{\lambda}_\alpha^2\bar{\kappa}_\alpha I_2\varPsi_{25}+4\bar{\lambda}_\alpha^4\bar{\kappa}_\alpha^2\varPsi_{55},\\
&\bar{J}_s\mathcal{A}_{s\alpha\alpha\beta\beta}=4I_2 (\varPsi_2+\varPsi_{11}+I_1\varPsi_{12}+I_3\varPsi_{15}+ I_2\varPsi_{22}+I_5\varPsi_{25}+I_4\varPsi_{55}),\\
&\bar{J}_s\mathcal{A}_{s\alpha\alpha\alpha\beta}=\bar{\lambda}_\alpha\bar{\lambda}_\beta(\bar{\kappa}_\alpha-\bar{\kappa}_\beta)[\varPsi_6+2\bar{\lambda}_\alpha^2(\varPsi_{16}+\bar{\lambda}_\beta^2\varPsi_{26}+\bar{\kappa}_\alpha\varPsi_{56})],\\
&\bar{J}_s\mathcal{A}_{s\alpha\alpha\beta\alpha}=2\bar{\lambda}_\alpha^3\bar{\lambda}_\beta(\bar{\kappa}_\alpha-\bar{\kappa}_\beta)(\varPsi_{16}+\bar{\lambda}_\beta^2\varPsi_{26}+\bar{\kappa}_\alpha\varPsi_{56}),\\
&\bar{J}_s\mathcal{A}_{s\alpha\beta\alpha\alpha}=-\bar{\lambda}_\alpha \bar{\lambda}_\beta(\bar{\kappa}_\alpha-\bar{\kappa}_\beta)[\varPsi_6+2\bar{\lambda}_\alpha^2(\varPsi_{16}+\bar{\lambda}_\beta^2\varPsi_{26}+\bar{\kappa}_\alpha\varPsi_{56})],\\
&\bar{J}_s\mathcal{A}_{s\alpha\beta\beta\beta}=-2\bar{\lambda}_\alpha\bar{\lambda}_\beta^3(\bar{\kappa}_\alpha-\bar{\kappa}_\beta)(\varPsi_{16}+\bar{\lambda}_\alpha^2\varPsi_{26}+\bar{\kappa}_\beta\varPsi_{56}),\\
&\bar{J}_s\mathcal{A}_{s\alpha\beta\alpha\beta}=2\bar{\lambda}_\beta^2(\varPsi_1+\bar{\kappa}_\beta\varPsi_5)+(\bar{\kappa}_\alpha-\bar{\kappa}_\beta)^2I_2\varPsi_{66}, \\
&\bar{J}_s\mathcal{A}_{s\alpha\beta\beta\alpha}=-I_2[2\varPsi_2+(\bar{\kappa}_\alpha-\bar{\kappa}_\beta)\varPsi_{66}],\\
&\bar{J}_s\mathcal{A}_{s3\alpha3 \alpha }=2\bar{\lambda}_\alpha^2\varPsi_1+2I_2\varPsi_2+2\bar{\lambda}_\alpha^2\bar{\kappa}_\alpha\varPsi_5,\\
&\bar{J}_s\mathcal{A}_{s3\alpha 3\beta}=\bar{\lambda}_\alpha\bar{\lambda}_\beta(\bar{\kappa}_\alpha-\bar{\kappa}_\beta)\varPsi_6.
\end{split}
\end{align}
The nonzero-components of the surface  stiffness tensor $\mathcal{B}_s$ are 
\begin{align}\label{eq:ap:Bs}
\begin{split}
&\bar{J}_s\mathcal{B}_{s\alpha\alpha\alpha\alpha}=2\bar{\lambda}_\alpha^2\varPsi_5+2\bar{\lambda}_\alpha^4\varPsi_{13}+2\bar{\lambda}_\alpha^4\bar{\kappa}_\alpha^{-1}I_4\varPsi_{14}+2\bar{\lambda}_\alpha^6\varPsi_{15}\\
&\hspace{5em}+2\bar{\lambda}_\alpha^2 I_2\varPsi_{23}+2\bar{\lambda}_\alpha^2\bar{\kappa}_\alpha^{-1}I_2I_4\varPsi_{24}+2\bar{\lambda}_\alpha^4 I_2\varPsi_{25}\\
&\hspace{5em}+2\bar{\lambda}_\alpha^4\bar{\kappa}_\alpha\varPsi_{35}+2\bar{\lambda}_\alpha^4 I_4\varPsi_{45}+2\bar{\lambda}_\alpha^6 \bar{\kappa}_\alpha\varPsi_{55},\\
&\bar{J}_s\mathcal{B}_{s\alpha\alpha\beta\beta}=2 I_2(\varPsi_{13}+\bar{\kappa}_\alpha \varPsi_{14}+\bar{\lambda}_\beta^2\varPsi_{15}+\bar{\lambda}_\beta^2\varPsi_{23}+\bar{\kappa}_\beta\bar{\lambda}_\beta^2 \varPsi_{24}\\
&\bar{J}_s\mathcal{B}_{s\alpha\alpha\alpha\beta}=-2\bar{\lambda}_\alpha^3\bar{\lambda}_\beta[\varPsi_6+\bar{\lambda}_\alpha^2(\varPsi_{16}+\bar{\lambda}_\beta^2\varPsi_{26}+\bar{\kappa}_\alpha\varPsi_{56})],\\
&\hspace{5em}+\bar{\lambda}_\beta^4 \varPsi_{25}+\bar{\kappa}_\alpha \varPsi_{35}+\bar{\kappa}^2_\alpha \varPsi_{45}+\bar{\kappa}_\alpha \bar{\lambda}_\beta^2 \varPsi_{55}),\\
&\bar{J}_s\mathcal{B}_{s\alpha\alpha\beta\alpha}=2\bar{\lambda}_\alpha^3\bar{\lambda}_\beta^3(\varPsi_{16}+\bar{\lambda}_\beta^2\varPsi_{26}+\bar{\kappa}_\alpha\varPsi_{56}),\\
&\bar{J}_s\mathcal{B}_{s\alpha\beta\alpha\alpha}=-\bar{\lambda}_\alpha^3\bar{\lambda}_\beta[\varPsi_6+(\bar{\kappa}_\alpha-\bar{\kappa}_\beta)(\varPsi_{36}+\bar{\kappa}_\beta\varPsi_{46}+\bar{\lambda}_\alpha^2\varPsi_{56})],\\
&\bar{J}_s\mathcal{B}_{s\alpha\beta\beta\beta}=\bar{\lambda}_\alpha\bar{\lambda}_\beta^3[\varPsi_6-(\bar{\kappa}_\alpha-\bar{\kappa}_\beta)(\varPsi_{36}+\bar{\kappa}_\alpha\varPsi_{46}+\bar{\lambda}_\beta^2\varPsi_{56})],\\
&\bar{J}_s\mathcal{B}_{s\alpha\beta\alpha\beta}=I_2[\varPsi_5+(\bar{\kappa}_\alpha-\bar{\kappa}_\beta)\bar{\lambda}_\alpha^2\varPsi_{66}].
\end{split}
\end{align}
The nonzero-components of the surface  stiffness tensor $\mathcal{C}_s$ are 
\begin{align}\label{eq:ap:Cs}
\begin{split}
&\bar{J}_s\mathcal{C}_{s\alpha\alpha\alpha\alpha}=\bar{\lambda}_\alpha^2\varPsi_3+\bar{\lambda}_\alpha^2\bar{\kappa}_\alpha^{-1}I_4+3\bar{\lambda}_\alpha^4 \varPsi_5+2\bar{\lambda}_\alpha^4\varPsi_{13}+2\bar{\lambda}_\alpha^4\bar{\kappa}_\alpha^{-1}I_4\varPsi_{14}\\
&\hspace{5em}+2\bar{\lambda}_\alpha^6\varPsi_{15}+2\bar{\lambda}_\alpha^2 I_2 \varPsi_{23}+2\bar{\lambda}_\alpha^2\bar{\kappa}_\alpha^{-1}I_2 I_4\varPsi_{24}+2\bar{\lambda}_\alpha^4 I_2\varPsi_{25}\\
&\hspace{5em}+2\bar{\lambda}_\alpha^4\bar{\kappa}_\alpha\varPsi_{35}+2\bar{\lambda}_\alpha^4 I_4\varPsi_{45}+2\bar{\lambda}_\alpha^6\bar{\kappa}_\alpha\varPsi_{55},\\
&\bar{J}_s\mathcal{C}_{s\alpha\alpha\beta\beta}=2 I_2(\varPsi_{13}+\bar{\kappa}_\beta \varPsi_{14}+\bar{\lambda}_\alpha^2 \varPsi_{15}+\bar{\lambda}_\alpha^2\varPsi_{23}+\bar{\lambda}_\alpha^2\bar{\kappa}_2 \varPsi_{24}\\
&\hspace{5em}+\bar{\lambda}_\alpha^4 \varPsi_{25}+\bar{\kappa}_\beta\varPsi_{35}+\bar{\kappa}_\beta^2\varPsi_{45}+\bar{\lambda}_\alpha^2\bar{\kappa}_\beta \varPsi_{55}),\\
&\bar{J}_s\mathcal{C}_{s\alpha\alpha\alpha\beta}=\frac{1}{2}\bar{\lambda}_\alpha\bar{\lambda}_\beta I_1\varPsi_6+\bar{\lambda}_\alpha^3\bar{\lambda}_\beta(\bar{\kappa}_\alpha-\bar{\kappa}_\beta)(\varPsi_{36}+\bar{\kappa}_\beta\varPsi_{46}+\bar{\lambda}_\alpha^2\varPsi_{56}),\\
&\bar{J}_s\mathcal{C}_{s\alpha\alpha\beta\alpha}=\bar{\lambda}_\alpha^3\bar{\lambda}_\beta\varPsi_6+\bar{\lambda}_\alpha^3\bar{\lambda}_\beta(\bar{\kappa}_\alpha-\bar{\kappa}_\beta)(\varPsi_{36}+\bar{\kappa}_\beta\varPsi_{46}+\bar{\lambda}_\alpha^2\varPsi_{56}),\\
&\bar{J}_s\mathcal{C}_{s\alpha\beta\alpha\alpha}=\frac{1}{2}\bar{\lambda}_\alpha\bar{\lambda}_\beta(3\bar{\lambda}_\alpha^2-\bar{\lambda}_\beta^2)\varPsi_6-\bar{\lambda}_\alpha^3\bar{\lambda}_\beta(\bar{\lambda}_\beta^2-\bar{\lambda}_\alpha^2)(\varPsi_{16}+\bar{\lambda}_\beta^2\varPsi_{26}+\bar{\kappa}_\alpha\varPsi_{56}),\\
&\bar{J}_s\mathcal{C}_{s\alpha\beta\beta\beta}=\bar{\lambda}_\alpha\bar{\lambda}_\beta^3[-\varPsi_6+(\bar{\lambda}_\alpha^2-\bar{\lambda}_\beta^2)(\varPsi_{16}+\bar{\lambda}_\alpha^2\varPsi_{26}+\bar{\kappa}_1\varPsi_{56})],\\
&\bar{J}_s\mathcal{C}_{s\alpha\beta\alpha\beta}=\bar{\lambda}_\beta^2(\varPsi_3+\bar{\kappa}_2\varPsi_4+I_1\varPsi_5)+\frac{1}{2}(\bar{\lambda}_\alpha^2-\bar{\lambda}_\beta^2)(\bar{\kappa}_\beta-\bar{\kappa}_\alpha)I_2\varPsi_{66},\\
&\bar{J}_s\mathcal{C}_{s\alpha\beta\beta\alpha}=I_2[\varPsi_5-\frac{1}{2}(\bar{\lambda}_\alpha^2-\bar{\lambda}_\beta^2)(\bar{\kappa}_\alpha-\bar{\kappa}_\beta)\varPsi_{66}],\\
&\bar{J}_s\mathcal{C}_{s 3\alpha 3\alpha}=\bar{\lambda}_\alpha^2\varPsi_3+\bar{\lambda}_\alpha^2\bar{\kappa}_2\varPsi_4+\bar{\lambda}_\alpha^4\varPsi_5,\\
&\bar{J}_s\mathcal{C}_{s3\alpha3\beta}=-\frac{1}{2}\bar{\lambda}_\alpha\bar{\lambda}_\beta(\bar{\lambda}_\alpha^2-\bar{\lambda}_\beta^2)\varPsi_6.
\end{split}
\end{align}
The nonzero-components of the surface  stiffness tensor $\mathcal{D}_s$ are 
\begin{align}\label{eq:ap:Ds}
\begin{split}
&\bar{J}_s\mathcal{D}_{s\alpha\alpha\alpha\alpha}=\bar{\lambda}_\alpha^4\varPsi_{33}+2\bar{\lambda}_\alpha^4\bar{\kappa}_\alpha^{-1}I_4\varPsi_{34}+2\bar{\lambda}_\alpha^6\varPsi_{35}+\bar{\lambda}_\alpha^4\bar{\kappa}_\alpha^{-2}I_4^2\varPsi_{44}\\
&\hspace{5em}+2\bar{\lambda}_\alpha^6\bar{\kappa}_\alpha^{-1}I_4\varPsi_{45}+\bar{\lambda}_\alpha^8\varPsi_{55},\\
&\bar{J}_s\mathcal{D}_{s\alpha\alpha\beta\beta}=I_2(\varPsi_4+\varPsi_{33}+I_3\varPsi_{34}+I_1\varPsi_{35}+I_4\varPsi_{44}+I_5\varPsi_{45}+I_2\varPsi_{55}),\\
&\bar{J}_s\mathcal{D}_{s\alpha\alpha\alpha\beta}=-\bar{\lambda}_\alpha^5\bar{\lambda}_\beta(\varPsi_{36}+\bar{\kappa}_\beta\varPsi_{46}+\bar{\lambda}_\alpha^2\varPsi_{56}),\\
&\bar{J}_s\mathcal{D}_{s\alpha\alpha\beta\alpha}=\bar{\lambda}_\alpha^3\bar{\lambda}_\beta^3(\varPsi_{36}+\bar{\kappa}_\beta \varPsi_{46}+\bar{\lambda}_\alpha^2\varPsi_{56}),\\
&\bar{J}_s\mathcal{D}_{s\alpha\beta\alpha\alpha}=\frac{1}{2}\bar{\lambda}_\alpha^3\bar{\lambda}_\beta(\bar{\lambda}_\alpha^2-\bar{\lambda}_\beta^2-)(\varPsi_{36}+\bar{\kappa}_\beta\varPsi_{46}+\bar{\lambda}_\alpha^2\varPsi_{56}),\\
&\bar{J}_s\mathcal{D}_{s\alpha\beta\beta\beta}=\frac{1}{2}\bar{\lambda}_\alpha\bar{\lambda}_\beta^3(\bar{\lambda}_\alpha^2-\bar{\lambda}_\beta^2-)(\varPsi_{36}+\bar{\kappa}_\alpha\varPsi_{46}+\bar{\lambda}_\beta^2\varPsi_{56}),\\
&\bar{J}_s\mathcal{D}_{s\alpha\beta\alpha\beta}=\frac{1}{2}I_2[-\varPsi_4+\bar{\lambda}_\alpha^2(\bar{\lambda}_\alpha^2-\bar{\lambda}_\beta^2)\varPsi_{66}],\\
&\bar{J}_s\mathcal{D}_{s\alpha\beta\beta\alpha}=-\frac{1}{2}I_2[\varPsi_4+\bar{\lambda}_\beta^2(\bar{\lambda}_\alpha^2-\bar{\lambda}_\beta^2)\varPsi_{66}].
\end{split}
\end{align}
In the above expressions, $\bar{J}_s=\bar{\lambda}_1\bar{\lambda}_2$, $\alpha,\beta=1,2$ and $\alpha\neq\beta$.  

In the special cases when the tensors $\bm{C}_s$ and $\bm{\kappa}$ are aligned (i.e.,\ $\bm{e}_\alpha$ correspond also to principal direction of $\bm{\kappa}$), or the coupling terms $I_5^s$ and $I_6^s$ are absent, the previous formulas can be significantly simplified. In these cases, it is convenient to introduce another functional form for the surface energy density depending on the principal stretches $\lambda_1^s$, $\lambda_2^s$ and principal relative curvatures $\kappa_1$, $\kappa_2$:
$$\varPsi=\varPsi^\text{p}(\lambda^s_1,\lambda^s_2,\kappa_1,\kappa_2).$$
Subsequently, the components of the surface  stiffness tensors expressed in terms of the orthonormal basis $(\bm{e}_1,\bm{e}_2,\bm{e}_3)$ reduce to
\begin{align}
\begin{split}\label{eq:moduli}
&\bar{J}_s\mathcal{A}_{s\alpha\alpha\beta\beta}=\bar{\lambda}_\alpha\bar{\lambda}_\beta \frac{\partial^2 \varPsi^\text{p}}{\partial\lambda^s_\alpha\partial\lambda^s_\beta},\\
&\bar{J}_s\mathcal{A}_{s\alpha\beta\alpha\beta}=\frac{\bar{\lambda}_\beta^2}{\bar{\lambda}_\alpha^2-\bar{\lambda}_\beta^2}\Big(\bar{\lambda}_\alpha\frac{\partial\varPsi^\text{p}}{\partial\lambda^s_\alpha}-\bar{\lambda}_\beta\frac{\partial\varPsi^\text{p}}{\partial\lambda_\beta^s}\Big),\quad \alpha\neq \beta,\\
&\bar{J}_s\mathcal{A}_{s\alpha\beta\beta\alpha}=\frac{\bar{\lambda}_\alpha\bar{\lambda}_\beta}{\bar{\lambda}_\beta^2-\bar{\lambda}_\alpha^2}\Big(\bar{\lambda}_\alpha\frac{\partial\varPsi^\text{p}}{\partial\lambda_\beta^s}-\bar{\lambda}_\beta\frac{\partial\varPsi^\text{p}}{\partial\lambda^s_\alpha}\Big),\quad \alpha\neq \beta,\\
&\bar{J}_s\mathcal{A}_{s3\alpha3\alpha}=\bar{\lambda}_\alpha\frac{\partial\varPsi^\text{p}}{\partial\lambda_\alpha^s},\\
& \bar{J}_s\mathcal{B}_{s\alpha\alpha\beta\beta}=\bar{\lambda}_\alpha\bar{\lambda}_\beta^2\frac{\partial^2\varPsi^\text{p}}{\partial\lambda^s_\alpha\partial\kappa_\beta},\\
&\bar{J}_s\mathcal{C}_{s\alpha\alpha\beta\beta}=\bar{\lambda}_\alpha^2\Big(\delta_{\alpha\beta}\frac{\partial\varPsi^\text{p}}{\partial\kappa_\alpha}+\bar{\lambda}_\beta\frac{\partial^2\varPsi^\text{p}}{\partial\lambda^s_\beta\partial\kappa_\alpha}\Big),\\
&\bar{J}_s\mathcal{C}_{s\alpha\beta\alpha\beta}=\bar{\lambda}_\beta^2\frac{\partial\varPsi^\text{p}}{\partial\kappa_\beta},\quad \alpha\neq \beta,\\
&\bar{J}_s\mathcal{C}_{s3\alpha3\alpha}=\bar{\lambda}_\alpha^2\frac{\partial\varPsi^\text{p}}{\partial\kappa_\alpha},\\
&\bar{J}_s\mathcal{D}_{s\alpha\alpha\beta\beta}=\bar{\lambda}_\alpha^2\bar{\lambda}_\beta^2\frac{\partial^2\varPsi^\text{p}}{\partial\kappa_\alpha\partial\kappa_\beta},\\
&\bar{J}_s\mathcal{D}_{s\alpha\beta\alpha\beta}=\frac{\bar{\lambda}_\alpha^2\bar{\lambda}_\beta^2}{2(\bar{\kappa}_\alpha-\bar{\kappa}_\beta)}\Big(\frac{\partial\varPsi^\text{p}}{\partial\kappa_\alpha}-\frac{\partial\varPsi^\text{p}}{\partial\kappa_\beta}\Big),\quad  \alpha\neq \beta,\\
&\bar{J}_s\mathcal{D}_{s\alpha\beta\beta\alpha}=\frac{\bar{\lambda}_\alpha^2\bar{\lambda}_\beta^2}{2(\bar{\kappa}_\alpha-\bar{\kappa}_\beta)}\Big(\frac{\partial\varPsi^\text{p}}{\partial\kappa_\alpha}-\frac{\partial\varPsi^\text{p}}{\partial\kappa_\beta}\Big),\quad \alpha\neq  \beta,
\end{split}
\end{align}
where all the derivatives are evaluated at $(\lambda^s_1,\lambda^s_2,\kappa_1,\kappa_2)=(\bar{\lambda}_1,\bar{\lambda}_2,\bar{\kappa}_1,\bar{\kappa}_2)$.  When the two principal stretches coincide, i.e., $\bar{\lambda}_1 =\bar{\lambda}_2$, or the principal relative curvatures are equal, $\bar{\kappa}_1 = \bar{\kappa}_2$, the corresponding expressions are to be understood as the limiting case $\bar{\lambda}_1 \to \bar{\lambda}_2$ (respectively $\bar{\kappa}_1 \to \bar{\kappa}_2$).

\bibliographystyle{model5-names}
\bibliography{mybibfile}

\end{document}